\pdfoutput=1
\documentclass[%
 aip,
 jmp,%
 amsmath,amssymb,
 reprint,%
]{revtex4-1}

\usepackage{graphicx}
\usepackage{dcolumn}
\usepackage{bm}
\usepackage{rotating}
\usepackage{colordvi}
\usepackage{color}

\input babarsym

\def\CP                {\ensuremath{{C\!P}}\xspace}
\def\CPT               {\ensuremath{{C\!PT}}\xspace} 
\def\T       {\ensuremath{T}\xspace}

\def\C       {\ensuremath{C}\xspace}
\def\P       {\ensuremath{P}\xspace}

\def\edm       {\ensuremath{\mathrm{e.d.m.}}\xspace}
\def\cms       {\ensuremath{\mathrm{c.m.}}\xspace}

\def\im         {\ensuremath{\mathrm{Im}}\xspace}

\def\invps   {\ensuremath{\rm \,ps^{-1}}\xspace}

\newcommand{\na}[2]{\mbox{\ensuremath{^{#2}{\rm #1}}}}

\def\D       {\ensuremath{D}\xspace}
\def\K       {\ensuremath{K}\xspace}

\def\P    {\ensuremath{P}\xspace}

\def\Pbar  {\kern 0.2em\overline{\kern -0.2em P}{}\xspace}

\def\X    {\kern 0.18em{\kern -0.18em X}{}\xspace}
\def\Xbar    {\kern 0.18em\overline{\kern -0.18em X}{}\xspace}

\def\A    {\kern 0.18em{\kern -0.18em A}{}\xspace}
\def\Abar    {\kern 0.18em\overline{\kern -0.18em A}{}\xspace}

\def\Bminus {\ensuremath{ B_{-} }\xspace}
\def\Bplus {\ensuremath{ B_{+} }\xspace}
\def\Bplusminus {\ensuremath{ B_{\pm} }\xspace}
\def\BminusT {\ensuremath{ B_{-}^\perp }\xspace}
\def\BplusT {\ensuremath{ B_{+}^\perp }\xspace}

\def\dGd {\ensuremath{\Delta \Gamma_d}\xspace}

\def\dmd {\ensuremath{\Delta m_d}\xspace}
\def\dt {\ensuremath{ \Delta t }\xspace}
\def\dz {\ensuremath{ \Delta z }\xspace}
\def\dttrue {\ensuremath{ \Delta t_{\rm true} }\xspace}
\def\de {\ensuremath{ \Delta E }\xspace}

\def\SpLpKs  {\ensuremath{ S^+_{\ellp,\KS} }\xspace}

\def\SpLmKs {\ensuremath{ S^+_{\ellm,\KS} }\xspace}
\def\SmLpKs  {\ensuremath{ S^-_{\ellp,\KS} }\xspace}
\def\SmLmKs {\ensuremath{ S^-_{\ellm,\KS} }\xspace}
\def\SpLpKl  {\ensuremath{ S^+_{\ellp,\KL} }\xspace}
\def\SpLmKl {\ensuremath{ S^+_{\ellm,\KL} }\xspace}
\def\SmLpKl  {\ensuremath{ S^-_{\ellp,\KL} }\xspace}
\def\SmLmKl {\ensuremath{ S^-_{\ellm,\KL} }\xspace}
\def\CpLpKs  {\ensuremath{ C^+_{\ellp,\KS} }\xspace}

\def\CpLmKs {\ensuremath{ C^+_{\ellm,\KS} }\xspace}
\def\CmLpKs  {\ensuremath{ C^-_{\ellp,\KS} }\xspace}
\def\CmLmKs {\ensuremath{ C^-_{\ellm,\KS} }\xspace}
\def\CpLpKl  {\ensuremath{ C^+_{\ellp,\KL} }\xspace}
\def\CpLmKl {\ensuremath{ C^+_{\ellm,\KL} }\xspace}
\def\CmLpKl  {\ensuremath{ C^-_{\ellp,\KL} }\xspace}
\def\CmLmKl {\ensuremath{ C^-_{\ellm,\KL} }\xspace}
\def\SpBzKs  {\SpLpKs}

\def\SpBzbKs {\SpLmKs}
\def\SmBzKs  {\SmLpKs}
\def\SmBzbKs {\SmLmKs}
\def\SpBzKl  {\SpLpKl}
\def\SpBzbKl {\SpLmKl}
\def\SmBzKl  {\SmLpKl}
\def\SmBzbKl {\SmLmKl}
\def\CpBzKs  {\CpLpKs}

\def\CpBzbKs {\CpLmKs}
\def\CmBzKs  {\CmLpKs}
\def\CmBzbKs {\CmLmKs}
\def\CpBzKl  {\CpLpKl}
\def\CpBzbKl {\CpLmKl}
\def\CmBzKl  {\CmLpKl}
\def\CmBzbKl {\CmLmKl}

\def\DeltaSpmT {\ensuremath{ \Delta S^\pm_{ \T} }\xspace}
\def\DeltaCpmT {\ensuremath{ \Delta C^\pm_{ \T} }\xspace}

\def\DeltaSpmCP {\ensuremath{ \Delta S^\pm_{ \CP} }\xspace}
\def\DeltaCpmCP {\ensuremath{ \Delta C^\pm_{ \CP} }\xspace}

\def\DeltaSpmCPT {\ensuremath{ \Delta S^\pm_{ \CPT} }\xspace}
\def\DeltaCpmCPT {\ensuremath{ \Delta C^\pm_{ \CPT} }\xspace}

\def\DeltaSmT {\ensuremath{ \Delta S^-_{ \T} }\xspace}
\def\DeltaCmT {\ensuremath{ \Delta C^-_{ \T} }\xspace}

\def\DeltaSmCP {\ensuremath{ \Delta S^-_{ \CP} }\xspace}
\def\DeltaCmCP {\ensuremath{ \Delta C^-_{ \CP} }\xspace}

\def\DeltaSmCPT {\ensuremath{ \Delta S^-_{ \CPT} }\xspace}
\def\DeltaCmCPT {\ensuremath{ \Delta C^-_{ \CPT} }\xspace}

\def\DeltaSpT {\ensuremath{ \Delta S^+_{ \T} }\xspace}
\def\DeltaCpT {\ensuremath{ \Delta C^+_{ \T} }\xspace}

\def\DeltaSpCP {\ensuremath{ \Delta S^+_{ \CP} }\xspace}
\def\DeltaCpCP {\ensuremath{ \Delta C^+_{ \CP} }\xspace}

\def\DeltaSpCPT {\ensuremath{ \Delta S^+_{ \CPT} }\xspace}
\def\DeltaCpCPT {\ensuremath{ \Delta C^+_{ \CPT} }\xspace}

\newcommand{\phm}{\ensuremath{\phantom{-}}}

\def\beq{\begin{equation}}
\def\eeq{\end{equation}}
\def\bea{\begin{eqnarray}}
\def\eea{\end{eqnarray}}
\def\bq{\begin{quote}}
\def\eq{\end{quote}}
\def\ben{\begin{enumerate}}
\def\een{\end{enumerate}}
\def\nn{\nonumber}

\def\twoDLL  {\ensuremath{-2\Delta\ln{\cal L}}\xspace}

\def\textSize {\normalsize}
\def\captionSize {\normalsize}

\definecolor{myBlue}{rgb}{0.0, 0.0, 1.0}
\def\Vera{}
\def\Refs{}
\definecolor{myGreen}{rgb}{0.0, 0.8, 0.0}
\def\Auth{}

\begin{document}

\title[Colloquium: Time-reversal violation with quantum-entangled \B mesons]
{Colloquium: Time-reversal violation with quantum-entangled \B mesons}

\author{J. Bernab\'eu}
 \altaffiliation[Also at ]{Department of Theoretical Physics, Universitat de Val\`encia, E-46100 Burjassot, Spain}
\email{Jose.Bernabeu@uv.es.}
\author{F.~Mart\'inez-Vidal}%
\email{Fernando.Martinez@ific.uv.es.}
\affiliation{ 
IFIC, Universitat de Val\`encia-CSIC, E-46071 Val\`encia, Spain
}%

\date{\today}

\begin{abstract}
\textSize 
Symmetry transformations have been proven a bedrock tool for understanding the nature of particle interactions, formulating and 
testing fundamental theories. Based on the up to now unbroken \CPT symmetry,
the violation of the \CP symmetry between matter and antimatter by weak interactions, discovered in the decay of kaons in 1964 
and observed more recently in 2001 in \B mesons, 
strongly suggests that the behavior of these particles under weak interactions must also be asymmetric under time reversal \T.
However, until the recent years there has not been a direct detection of the expected time-reversal violation
in the time evolution of any system.
This Colloquium examines the field of time-reversal symmetry breaking in the fundamental laws of physics. 
For transitions, its observation requires an asymmetry with exchange of initial and
final states. We discuss the conceptual basis for such an exchange with
unstable particles, using the quantum properties of
Einstein-Podolsky-Rosen (EPR) entanglement available at \B meson factories 
combined with the decay as a filtering measurement.
The method allows a clear-cut separation of different transitions
between flavor and \CP eigenstates in the decay of neutral \B mesons.
These ideas have been implemented 
for the experiment by the \babar\ Collaboration at SLAC's \B factory. The results, presented in 2012, 
\Vera{prove beyond any doubt}
the violation of time-reversal invariance in the time evolution between these two states of the neutral \B \Vera{meson}.
\end{abstract}

\maketitle

\textSize 

\tableofcontents

\section{Introduction}
\label{sec:Intro}

In particle physics not all processes are expected to run in the same way with time in one sense as they do in the opposite sense,
a symmetry transformation known as time reversal \T.
The direct observation of this phenomenon in neutral \B mesons was 
reported by the \babar\ Collaboration at SLAC in November 2012~\cite{Lees:2012kn}, 
and was echoed in other journals and magazines~\cite{Zeller:2012,PhysToday:2012,Nature:2012,PhysWorld:2012nov,PhysWorld:2012dec,TheEconomist:2012}.
The so-called \CPT theorem, applicable to phenomena described by a
local quantum field theory with Lorentz invariance and Hermiticity,
implies that the \CP violation observed in 1964 with neutral kaons~\cite{Christenson:1964,Christenson:1965}
and in 2001 with neutral \B mesons~\cite{Aubert:2001nu,Abe:2001xe} should 
\Vera{also reveal independently a}
\T violation for those systems.
Why 48 years after the discovery of \CP violation?
The conceptual basis for solving the problem of how to probe time reversal with unstable systems was 
proposed~\cite{Banuls:1999aj,Banuls:2000ki,Wolfenstein:1999re} in 1999 through the 
Einstein-Podolsky-Rosen (EPR) quantum entanglement of the two neutral \B mesons produced at the so-called \B factories,
in addition to using their two decays as filtering measurements to project definite states of the neutral \B meson.
In this Colloquium we review the fundamentals of time-reversal symmetry, 
its implication for transitions between two quantum states of the neutral \B meson, 
the implementation of 
genuine \T asymmetries by means of specific
decay channels, and the experimental analysis leading to the direct detection performed by \babar\ at a significance of $14\sigma$.

The outline of the article is as follows. We first introduce the basics of time-reversal physics,
presenting briefly the role of the time-reversal symmetry in the fundamental laws of classical and quantum mechanics, 
the different scenarios to search for \T violation and 
to 
\Vera{prove}
it experimentally, 
in particular when considering unstable systems. 
In Sec.~\ref{sec:BFACTORIES} we introduce the \B factories and 
discuss how the quantum entanglement 
\Vera{in decays of the \FourS resonance}
has been employed
during the last decade 
to perform flavor tagging for exploring \CP violation in neutral \B mesons.
Section~\ref{sec:CONCEPT} discusses how the lack of 
definite states of the two mesons in the entangled system before their decay can 
lead to either flavor tagging or ``\CP tagging'' for the preparation of neutral \B meson states
\Vera{required}
to directly 
\Vera{test}
time-reversal symmetry. 
Section~\ref{sec:ANALYSIS} briefly presents the \babar\ detector and data sample, describes
how the time-reversal physics is extracted from the data, and summarizes the results and their interpretation. 
\Vera{We} conclude and discuss some perspectives in Sec.~\ref{sec:CONCLUSION}.

\section{Time-reversal symmetry in physics}
\label{sec:TRINPHYSICS}

Time enters at the most elementary level as a parameter in the description of physical phenomena, serving to identify the order
of a sequence of events
in the evolution of a physical system.
Quantitatively it can be constructed in terms of a well established and 
continuing sequence of repetitive 
events. If the period of repetition is constant it may be used as a unit of time. 
An accurate definition of this unit is a prerequisite to reach a good 
precision in time measurements to observe the details of the evolution.

The symmetry transformation that changes a physical system with
a given sense of the time evolution into another with the opposite sense is called time reversal \T. 
It corresponds to changing the sign of the velocity vector ${\bf v}$ or the 
momentum ${\bf p}$, without changing the position ${\bf r}$. In the dynamical
equations of motions, or their solutions, such a transformation corresponds
formally to replacing $t$ by 
\Auth{$-t$.} 
The \T transformation changes the sign
of other dynamical variables \Vera{such} as angular \Vera{momentum}. 
For fields, the magnetic field 
\Vera{changes} 
its sign under time reversal, whereas the electric field does not.

\subsection{Classical and quantum mechanics}
\label{sec:TRINPHYSICS-cm-qm}

\begin{figure}[htb!]
\begin{center}
\begin{tabular}{cc}
 \includegraphics[width=0.23\textwidth]{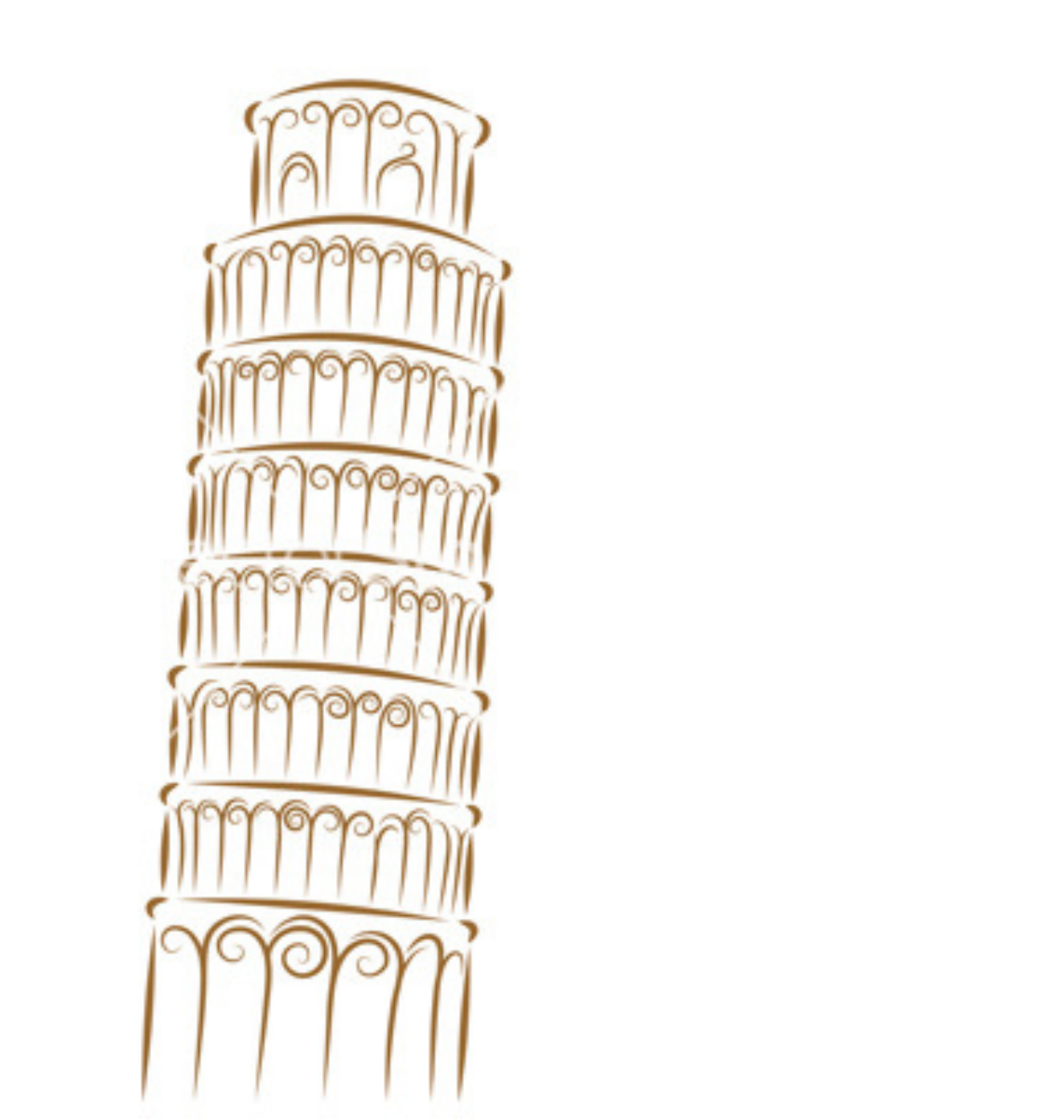} &
 \includegraphics[width=0.23\textwidth]{pisa-tower.pdf} \\
\end{tabular}
 \put(-240,50){(a)}
 \put(-177,30){\large $\Downarrow \bf{v_i} = 0$}
 \put(-177,-10){\large $\Downarrow \bf{v}$}
 \put(-177,-54){\large $\Downarrow \bf{v_f}$}
 \put(-121,50){(b)}
 \put(-58,30){\large $\Uparrow \bf{-v_i} = 0$}
 \put(-58,-10){\large $\Uparrow \bf{-v}$}
 \put(-58,-54){\large $\Uparrow \bf{-v_f}$}
\caption{\label{fig:tower}\captionSize (a) Trajectory of a stone falling from the leaning tower. (b) Trajectory after time-reversal transformation.
}
\end{center}
\end{figure}

The time-reversal transformation in classical mechanics corresponds to substitute for each trajectory ${\bf r}(t)$
the trajectory ${\bf r}(-t)$, i.e. to moving along the given trajectory with the opposite velocity at 
each point, as illustrated in Fig.~\ref{fig:tower}. 
It is not obvious that the dynamics remains invariant under this \T transformation.
If the original trajectory is dynamically possible, 
$d{\bf p}/dt = {\bf F}$ with a force ${\bf F}$ depending on the sense (sign) of the velocity leads 
to a violation of \T invariance.

In quantum mechanics, 
Wigner's time-reversal transformation~\cite{Wigner:1932}, 
\bea
\Auth{\psi(t) \to} \T\psi(t) \equiv \psi_\T(t) & = & \psi^*(-t), 
\label{eq:WignerT}
\eea
keeps the Schr\"odinger equation, $i\hbar \partial \psi(t)/\partial t = {\cal H}\psi(t)$, 
invariant under 
a \T transformation 
if the Hamiltonian ${\cal H}$ is real.
\Vera{This has} \Auth{three}
fundamental \Vera{consequences}~\cite{Henley:2013sja,Sachs:1987gp,Lee:1990,Branco:1999fs}. 
\Vera{First}, 
the \T operator 
\Auth{is antiunitary.} This property can be seen, for example,
evaluating the scalar product of two states,
\bea
\langle \psi_\T(t) | \phi_\T(t) \rangle & = & \langle \psi(-t) | \phi(-t) \rangle^* = \langle \phi(-t) | \psi(-t) \rangle.\ \ \
\eea
\Auth{Thus,} time reversal has to do with interchange of bra and ket states.
\Vera{Second,}
the complex conjugation implies that time reversal does not have observable and conserved
eigenvalues, 
\Auth{unless $\psi(t)$ is purely real.}
\Vera{Third,} 
for a plane wave with momentum $\bf{p}$, 
$\psi({\bf r},t)=\exp [i({\bf p}\cdot{\bf r} -Et)/\hbar]$,
the time-reversed wavefunction is 
$\psi^*({\bf r},-t)= \exp [i(-{\bf p}\cdot{\bf r} -Et)/\hbar]$, i.e. 
the \T-transformed function describes a particle with momentum $-{\bf p}$ \Auth{and energy $E$},
thus it is not necessary to interpret the transformed function as a particle going backwards in time.
For \Auth{this reason} the \T transformation is often referred to as ``motion reversal'' rather
than ``time reversal''.

\Refs{The \T transformation is implemented in the space of states by the antiunitary operator $U_\T$ in such a way that, 
for spinless particles,
\bea                
U_\T {\bf r} U_\T^\dagger = {\bf r} & , \ & U_\T {\bf p} U_\T^\dagger = -{\bf p},
\label{eq:rpTransform}
\eea
and $\psi_\T(t) = U_\T \psi(-t)$.}
\Auth{Equation~(\ref{eq:rpTransform})}
guarantees the invariance of the commutation rule 
between ${\bf r}$ and ${\bf p}$, 
thus 
we might say that \T transforms quantum mechanics into quantum mechanics.
\Refs{For a Hamiltonian ${\cal H}$ invariant under time reversal, $[{\cal H},U_\T]=0$, the time-evolution operator 
${\cal U}(t,t_0)$ transforms as 
\bea
U_\T {\cal U}(t,t_0) U_\T^\dagger & = & {\cal U}^\dagger(t,t_0).
\label{eq:Utt0}
\eea
The antiunitary character of $U_\T$ allows to write  $U_\T = U K$,
where $U$ is unitary ($U^{-1} = U^\dagger$) and $K$ is an operator which complex
conjugates all complex numbers. 
For the matrix elements of time-dependent transitions we have
}
\Auth{
\bea
& & \langle f | {\cal U}(t,t_0) | i \rangle = \langle f | U_\T^\dagger U_\T {\cal U}(t,t_0) U_\T^\dagger U_\T | i \rangle \nn \\
& & \ \ \ \ \ \ = \langle U_\T f | {\cal U}^\dagger(t,t_0) | U_\T i \rangle^* = \langle U_\T i | {\cal U}(t,t_0) | U_\T f \rangle,\ \ \
\label{eq:ME}
\eea
where time-reversal invariance is assumed when passing from the first to the second lines of Eq.~(\ref{eq:ME}). 
As a consequence, the comparison between $i \to f$ and $U_\T f \to U_\T i$ transitions
is a genuine test of this invariance.}
It is because of these special properties 
that the role of time reversal is distinct from that of any other 
symmetry operation in physics, and makes its experimental investigation 
significantly more difficult than other symmetries.

Therefore, time-reversal
\Auth{in classical mechanics as well as in quantum mechanics}
is related to the following fundamental \Auth{question} 
(see Fig.~\ref{fig:tower}): 
consider a point over a trajectory (a state in quantum mechanics),
invert the velocities of all particles in that point, and let it evolve; 
shall we obtain the former initial point of the trajectory with all velocities reversed?
Obviously, for a fair comparison the experiment should be repeated in the laboratory with 
exactly the same boundary conditions as in the \T mirror, since the motion is not only determined by the equations
of motion but also by the boundary conditions, and the symmetries of the motion cannot be greater than those of the
latter.
From Newton's mechanics to electrodynamics, the dynamical laws of physics are symmetric under \T transformation. 

Motion-reversal symmetry
implies, 
for a given configuration of energy-momenta and spin, 
\Auth{the reciprocity relation~\cite{Henley:2013sja,Sachs:1987gp,Lee:1990,Branco:1999fs}:}
the probability of an initial state $i$ being transformed into a final state $f$ is the same as the probability that
an initial state identical to $f$, but with momenta ${\rm \bf p}$ and spins ${\rm \bf s}$ reversed, transforms into the state $i$ 
with momenta and spins reversed,
\begin{eqnarray}
|\langle f |{\cal S}| i\rangle|^2 & = & \Auth{|\langle U_\T i |{\cal S} | U_\T f \rangle|^2},
\label{Eq:TRV-detailedBalance}
\end{eqnarray}
where
${\cal S}$ 
is the transition matrix determined by the Hamiltonian ${\cal H}$.
\Auth{Here,}
$| i \rangle \equiv | {\rm \bf p}_i,{\rm \bf s}_i \rangle$ and
      $\langle f | \equiv \langle {\rm \bf p}_f,{\rm \bf s}_f |$ are
the initial and final states,
\Auth{$\langle U_\T i |$} and \Auth{$| U_\T f \rangle$} are the \Auth{transformed} states of $| i \rangle$ and $\langle f |$,
respectively,
$\Auth{\langle U_\T i |} \equiv \langle -{\rm \bf p}_i,-{\rm \bf s}_i | $
and
$\Auth{| U_\T f \rangle} \equiv | -{\rm \bf p}_f,-{\rm \bf s}_f \rangle $.
It should be noted that \T invariance is a sufficient, but not necessary, condition
for 
\Auth{Eq.~(\ref{Eq:TRV-detailedBalance}).}
Therefore,
\Auth{a breaking of reciprocity}
is an unambiguous signal for \T violation.
If ${\cal S}$ is Hermitian, 
$|\langle f|{\cal S}|i\rangle| = 
\Auth{|\langle U_\T i |{\cal S}|U_\T f \rangle| =
 |\langle U_\T f |{\cal S}|U_\T i \rangle|}$; in this case,
\T invariance implies 
\T-odd
invariance, and vice versa,
where the 
\T-odd
transformation only refers to changing the sign of all odd variables under $t \to -t$ in ${\cal H}$, without exchanging 
initial and final states.
This occurs, for instance, to first order in weak interactions when final state interactions (FSI) \Auth{can} 
be neglected~\cite{Branco:1999fs}.

The 
\Auth{reciprocity relation}
establishes a connection between the differential cross-sections for reactions
$a+b\to c+d$ and $c+d\to a+b$~\Auth{(detailed balance).} 
It has been verified by experiment in nuclear reactions due to strong or weaker interactions, 
for example~\cite{Blanke:1983zz} 
\bea
\alpha(0^+)+\na{Mg}{24}(0^+) \rightarrow \na{Al}{27}(5/2^+)+ p(1/2^+). 
\eea
Here, if is there any \T violation it cannot exceed a half per mil.

\subsection{Complex systems and the arrow of time}
\label{sec:TRINPHYSICS-arrow}

When \Vera{discussing} 
\T violation we should clearly \Vera{distinguish} 
$t$ asymmetry of complex systems.
For example, our daily experience shows us that when a vase falls and breaks into pieces it is not possible that the pieces 
of the group fly 
\Refs{back in reverse order,}
forming the vase.
This macroscopic $t$ asymmetry, also known as ``arrow of time'', 
is in the nature of thermodynamics. As discussed by Eddington~\cite{Eddington:1928}, 
the arrow of time is a property of entropy alone, a measurement of disorder: 
the arrow gives the direction of progressive increase of disorder in isolated systems.
How it is then possible to generate thermodynamic irreversibility from fundamental laws that are \T symmetric? 
The answer to this question is that the thermodynamic irreversibility is associated to the irreversibility of the initial conditions for 
systems with large number of degrees of freedom, larger for more disordered \Vera{states}, 
making very unlikely (although not forbidden) for a system to evolve 
from a disordered to a more ordered state.
\Vera{In the example} of the falling vase, with ${\cal O}(10^{24})$ of particles and collisions, it is not possible in practice to \Refs{set up} 
the initial conditions for the reversed process (positions and velocities).

There is no doubt that the Universe is expanding, even accelerating at \Refs{the} present 
\Refs{times,} 
thus we have a time-asymmetric behavior. But this Universe $t$ asymmetry is perfectly compatible with \T-symmetric laws of physics. 
If we think of the initial condition of our Universe (likely inflation~\cite{Guth:1980zm,Guth:1982ec,Ade:2013ov,Ade:2014xna}) 
as an improbable state, i.e. more ordered,
the cosmological $t$ asymmetry is probably connected with the arrow of time for complex systems.
However, none of these time asymmetries is fundamental \T violation.

In particle physics, as 
\Refs{will} be discussed further in detail in Sec.~\ref{sec:TRINPHYSICS-unstable}, decays are 
an example of 
\Refs{thermodynamically} 
asymmetric processes: the number of variables or degrees of freedom
describing the final state
is larger than the number of variables needed to describe the initial state. 

\subsection{\Refs{Discrete symmetries broken} by weak interactions}
\label{sec:TRINPHYSICS-ckm}

It is well known since 1957 that weak interactions have little respect for symmetries. That year space inversion (parity, \P) symmetry was
discovered to be broken in 
\Vera{beta decays~\cite{LeeYang:1956,Wu:1957,Garwin:1957hc}}. Then, there was the hope that the combination 
of \P with charge conjugation (\CP) was a
good symmetry. But just a few years later, in 1964, 
\Refs{there}
was discovered a small but unambiguous violation of the \CP symmetry in
\K meson decays~\cite{Christenson:1964,Christenson:1965}. More recently, in 2001, the \B factory experiments \babar\ and Belle, 
observed that \CP is 
\Vera{violated} in \B mesons~\cite{Aubert:2001nu,Abe:2001xe}.

\begin{figure}[htb!]
\begin{center}
 \includegraphics[width=0.4\textwidth]{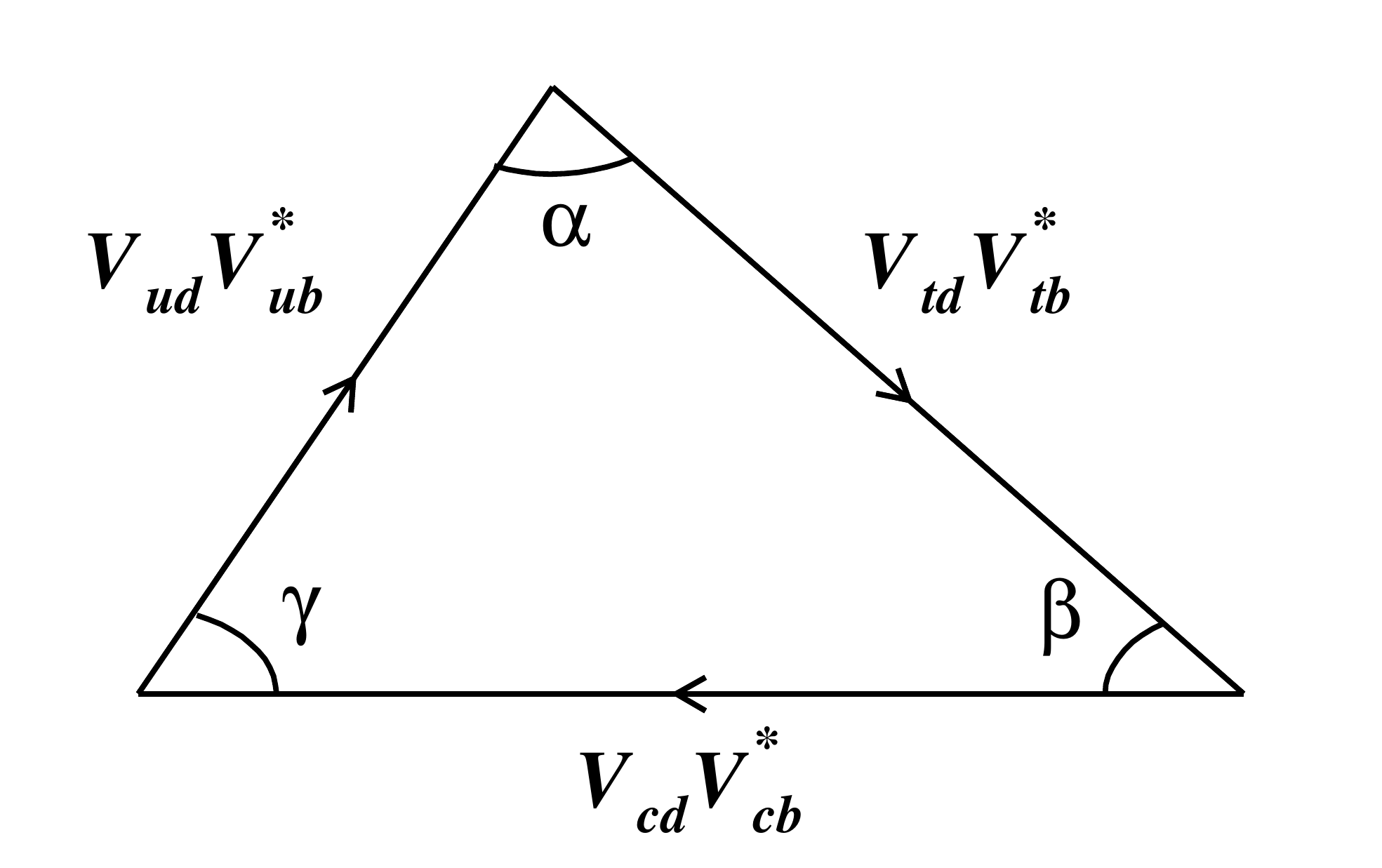}
\caption{\label{fig:bdUT}\captionSize
The $\b\d$ unitarity triangle representing the CKM unitarity conditions. 
The three sides 
\Vera{are determined from}
semileptonic and non-leptonic \B decays, including \Bz-\Bzb oscillations.
Since they are of comparable
length the angles are sizeable and one expects large \CP asymmetries in \B decays in the SM.
There are other two triangles which 
almost collapse to a line. This gives an intuitive understanding of why
\CP violation is small in the leading \K decays ($\d\s$ triangle) and in the leading \Bs decays ($\b\s$ triangle).
}
\end{center}
\end{figure}

The now well established \CP violation in the quark sector 
\Vera{can be}
successfully accommodated within the Standard Model (SM) of particles and fields through
the three-family Cabibbo-Kobayashi-Maskawa (CKM) quark-mixing mechanism~\cite{Cabibbo:1963,KM:1973}. 
It describes the coupling of the $W$ boson to up and down quarks and conveys the fact that the quarks with 
definite properties under charged-current weak interactions are linear combinations of the quark mass eigenstates~\cite{Beringer:2012}. 
For three families, the unitarity conditions of the quark-mixing matrix $V$ are represented by triangles in the complex plane,
as illustrated in Fig.~\ref{fig:bdUT}, and lead to 
four fundamental parameters: three magnitudes and one single irreducible phase. 
In the Wolfenstein parameterization~\cite{Wolfenstein:1983yz} we can write to ${\cal O}(\lambda_c^4)$ as
\begin{center}
\begin{eqnarray*}
\left[  \begin{array}{c}
V_{\u j} \\
V_{\c j} \\
V_{\t j} \\ 
\end{array} \right]
=
\left[  \begin{array}{ccc}
1-\lambda_c^2 & \lambda_c & A\lambda_c^3(\rho-i\eta) \\
-\lambda_c & 1-\lambda_c^2/2 & A\lambda_c^2 \\
A\lambda_c^3(1-\rho-i\eta) & -A\lambda_c^2 & 1 
\end{array} \right],
\end{eqnarray*}
\end{center}
where the index $j$ runs over \d, \s, and \b quarks,  
and $\lambda_c\approx0.226$, $A\approx0.814$, $\rho\approx0.135$ and $\eta\approx0.349$~\cite{Beringer:2012}.
Extensive tests of the CKM mechanism using all experimental 
data show a high degree of consistency~\cite{Antonelli:2009ws}. 
Historically, Kobayashi and Maskawa extended in 1973 the $2\times 2$ Cabibbo mixing 
matrix to $3\times 3$ to explain the \CP violation discovered nine years before, 
thus anticipating the existence of the third family \Vera{of quarks}, quickly
confirmed with the discovery of the $\tau$ lepton in 1975~\cite{Perl:1975} and of the fifth quark, the \b, two years later~\cite{Herb:1977}.

All local quantum field theories with Lorentz invariance and Hermiticity respect \CPT symmetry~\cite{Luders:1957,Pauli:1995}, 
in accordance with all experimental evidence~\cite{Beringer:2012}, hence there is 
\Refs{a} 
straightforward 
theoretical connection between \CP and \T violation (matter-antimatter asymmetry defines a preferred sense of time evolution).
Since the SM is 
\Vera{based on}
a quantum field theory satisfying the \CPT theorem, it follows that
the source of \CP violation 
\Vera{also requires}
of \T-violating effects.
With the complex Hermitian Lagrangian, genuine \CP phases change sign
for particles and antiparticles, so its experimental detection requires an interference experiment to 
observe the relative \CP phase between the 
\Refs{interfering} complex amplitudes.
Therefore, given the known \CP violation in weak interactions in processes involving \K and \B mesons, 
\T should also be broken in these systems.
\Vera{The question} 
is whether the expected \T asymmetry can be detected by an
experiment that, considered by itself, clearly shows a motion-reversal asymmetry independent 
\Refs{of}
\CP asymmetries and \CPT invariance.

\subsection{Experiments probing \T violation}
\label{sec:TRINPHYSICS-scenarios}

There are two main types of experiments or observables that can be used to detect directly time-reversal 
non-invariance~\cite{Wolfenstein:1999re,Sachs:1987gp,Henley:2013sja}.

\begin{figure}[htb!]
\begin{center}
 \includegraphics[width=0.35\textwidth]{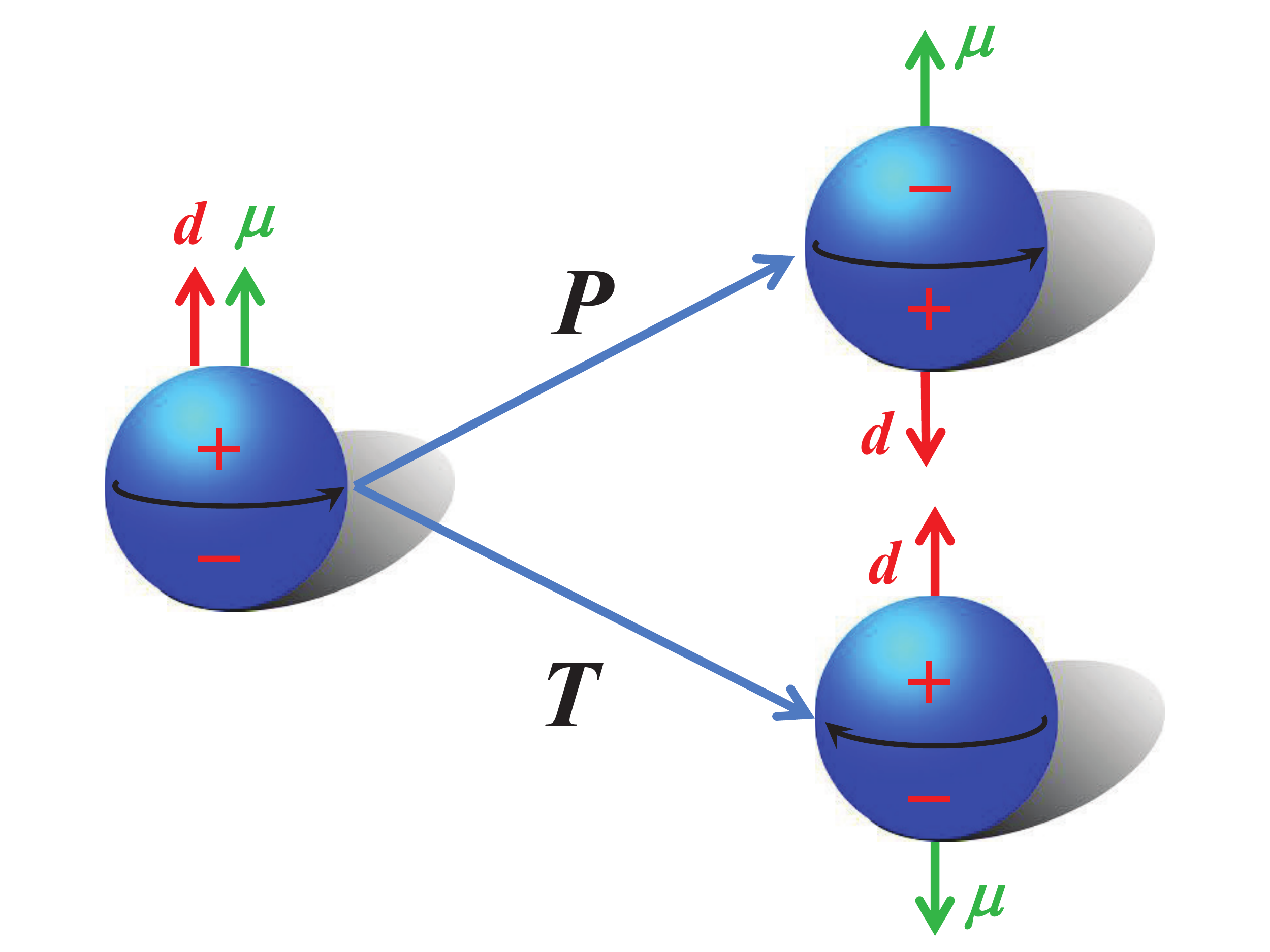}
\caption{\label{fig:edm}\captionSize A particle with spin is \Vera{represented} 
as a sphere with a spinning charge distribution. 
Its parity and time-reversed 
\Vera{images} are also shown, together with the corresponding magnetic ${\boldsymbol \mu}$ 
and electric ${\boldsymbol d}$ dipole moments.
}
\end{center}
\end{figure}

A non-zero expectation value of a \T-odd operator for a non-degenerate stationary state. This is the case
for an electric dipole moment (\edm) of a particle with spin, which is also a \P-odd, \C-even quantity, as depicted in
Fig.~\ref{fig:edm}. A parity transformation about the midplane of the sphere flips the \edm ${\boldsymbol d}$ with respect to the magnetic
dipole moment ${\boldsymbol \mu}$ (spin), 
which remains unchanged, whereas a time-reversal transformation flips ${\boldsymbol \mu}$
with respect to ${\boldsymbol d}$, which is in this case unaffected. Thus, if either parity or time reversal are good symmetries, 
the particle cannot have an \edm since one can distinguish the mirror picture from the original one.
A non-zero \edm can be generated by either strong \Vera{\T violation} 
unless it is 
\Vera{annulled}
by a Peccei-Quinn symmetry leaving the axion as remnant~\cite{Peccei:1977hh}, or by \T violation in weak interactions. 
In the SM with the CKM mechanism, a non-vanishing \edm of the neutron only appears at three-loop level. 
Hence, these experiments probe for physics beyond the SM.
To date, no signals for \edm have been found, although there are strong limits, as for the neutron
and the electron, 
\Refs{$|{\boldsymbol d}_n| < 2.9\times 10^{-26}$~e-\cm} 
and 
\Refs{$|{\boldsymbol d}_e| < 1.05\times 10^{-27}$~e-\cm~\cite{Beringer:2012}.}

Experiments involving \T-odd observables for weak decays (as well as electromagnetic transitions), constructed typically using 
odd 
products of the momentum and spin vectors of the decay products (initial and final states), are sometimes used. 
These observables are not genuine signals for \T violation since initial and final states are not interchanged, 
and in general require detailed understanding of FSI effects since they can mimic \T violation 
even with time-reversal symmetric dynamics~\cite{Wolfenstein:1999re,Sachs:1987gp,Beringer:2012}.
\Refs{In some cases, \T-odd \CP asymmetries can be built by comparing particle and antiparticle decays, without 
knowing the \CP-even FSI phases~(see for example~\cite{Gronau:2011cf} and references therein).}

\begin{figure}[htb!]
\begin{center}
  \includegraphics[width=0.35\textwidth]{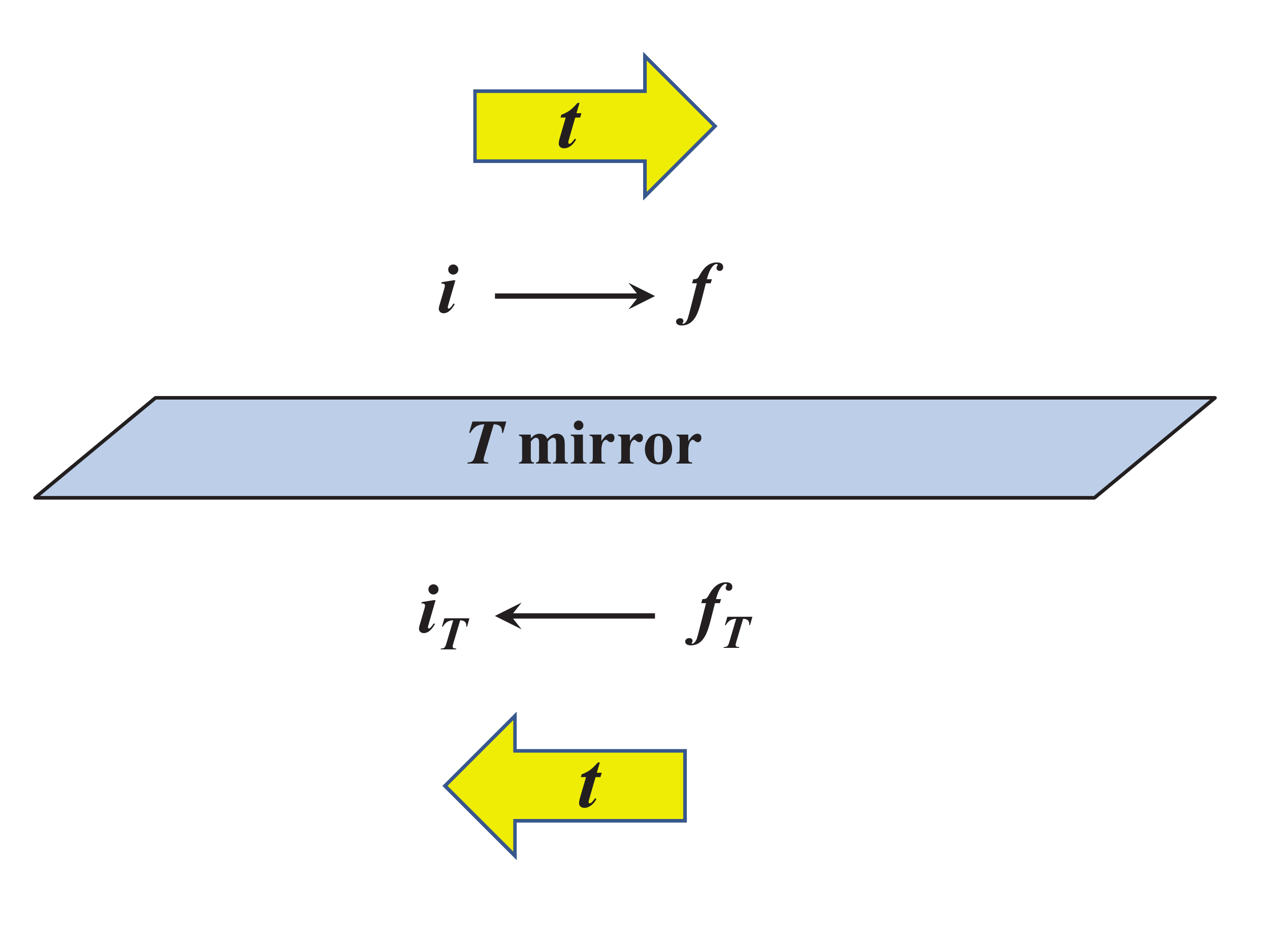}
\caption{\label{fig:atob}\captionSize 
The antiunitary property of the \T operator demands the exchange of initial and final states 
to \Refs{set up} an experiment 
\Refs{directly probing}
time-reversal symmetry in transitions. 
Motion-reversal symmetry implies 
\Auth{equal probabilities for transitions $i \rightarrow f$ and $i_\T \leftarrow f_\T$,}
once the initial conditions, namely $i$ in one case and $f_\T$ in the other, have been precisely realized. 
Here, $i_\T$ and $f_\T$ are the \T-transformed states of $i$ and $f$, respectively, with all momenta and spins reversed.
}
\end{center}
\end{figure}

We might also consider transition processes. As discussed before and illustrated in Fig.~\ref{fig:atob}, 
the antiunitary character of the \T operator 
demands the exchange of initial and final states to compare
the probabilities 
\Auth{$|\langle f |{\cal U}(t,t_0)| i\rangle|^2$} 
and 
\Auth{$|\langle U_\T i |{\cal U}(t,t_0)| U_\T f \rangle|^2$,}
once the initial conditions, namely $i$ in one case and 
\Auth{$U_\T f$} 
in the other, have been precisely realized. 
With stable particles one can consider 
\Vera{neutrino $\nu_e$ to $\nu_\mu$ mixing}~\cite{Cabibbo:1978},
but this requires high-luminosity and long-baseline neutrino facilities. 
Alternatively, 
\Vera{neutral}
\K and \B mesons, the unique 
\Vera{systems} 
\Vera{for which}
\CP violation has been detected,
are a 
\Vera{best}
choice for an experiment \Vera{probing} 
\T non-invariance.
Nevertheless, the thermodynamic irreversibility of these systems makes this option 
difficult, 
as discussed below.

\subsection{Unstable systems}
\label{sec:TRINPHYSICS-unstable}

A direct consequence of quantum dynamics is the negative-exponential time behavior of the decay of any unstable system into
two or more particles, as given by the Fermi golden rule. 
The reversal of the exponential decay law reveals that the \T transformation is not defined for a decaying state, thus
it appears that the decay prevents 
\Refs{proofs of}
motion reversal. This apparent imbalance in time
has to do with initial conditions rather than with the dynamics under time reversal; we have assumed that the initial system
is the unstable particle, but it should be formed by some process before. Hence, to address the question of time reversal
we have to choose an initial time earlier so that the production enters into the process~\cite{Sachs:1987gp}.

\begin{figure}[htb!]
\begin{center}
 \includegraphics[width=0.3\textwidth]{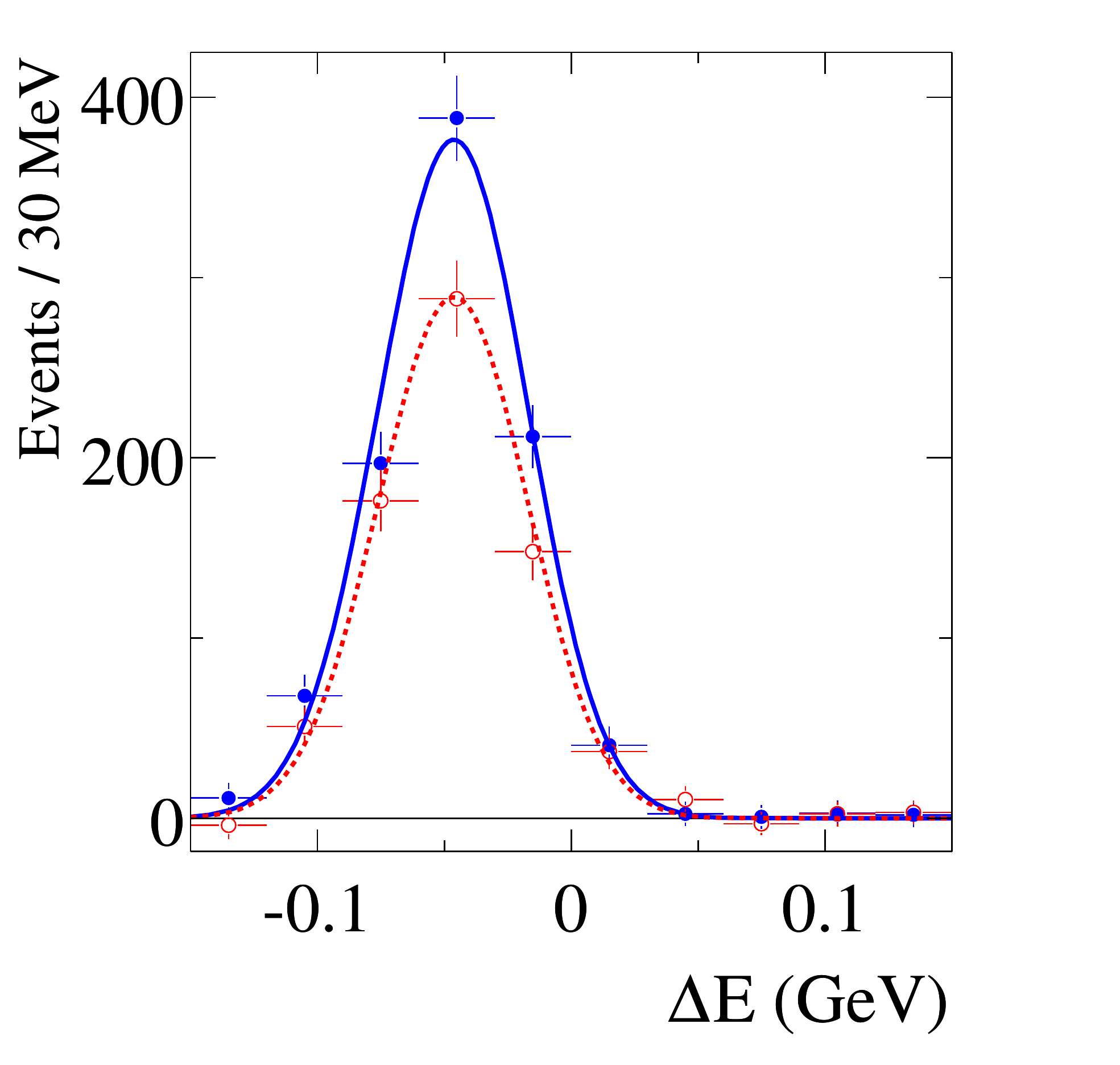}
\caption{\label{fig:kpi}\captionSize Energy distributions of $\Bz\to\Kp\pim$ 
(solid circles and solid curve) and $\Bzb\to\Km\pip$ (open circles and dashed curve) decays~\cite{Aubert:2004qm}.
Points with error bars are data and the curves are best fit projections.
The difference between the two distributions is a signature of direct \CP violation.
}
\end{center}
\end{figure}

For example, we might consider the decay of a neutral \B meson into the final state $\Kp\pim$, i.e. $\Bz\to\Kp\pim$, with rate $R_1$. 
\Refs{\CP violation is known to be large}
in this decay~\cite{Aubert:2004qm,Chao:2004mn}, 
thus we have $\Bzb\to\Km\pip$ with rate $R_2 \ne R_1$,
as 
\Refs{can} be observed in Fig.~\ref{fig:kpi}.
In the SM, this decay occurs through two different amplitudes (“penguin” and “tree”), 
\Vera{with}
different weak
phases and, in general, different strong phases.
This is a general requirement~\cite{Bernabeu:1980ke} to generate a non-vanishing interference 
for particle and antiparticle decays\footnote{The proof is 
as follows: if $A_1=a_1\exp[i(\delta_1+\phi_1)]$ and $A_2=a_2\exp[i(\delta_2+\phi_2)]$ are the two
possible amplitudes for the process $i \to f$, with $a_1$ and $a_2$ real numbers,
and $\bar i$ and $\bar f$ are the \CP-conjugate states of $i$ and $f$, respectively,
then $|A(i \to f)|^2 - |A(\bar i \to \bar f)|^2 = -4a_1a_2\sin(\Delta \delta)\sin(\Delta \phi)$
do not vanish only if both $\Delta \delta$ and $\Delta \phi$ are non-vanishing.}.
\Vera{This leads}
to a difference in the decay rates for \CP conjugate processes,
\Vera{which we refer to as}
direct \CP violation. The \B meson production could be done in electron-positron annihilation reactions
through the process $e^+ e^- \to \FourS \to \BzBzb$ at a center-of-mass energy (\cms) of $10.58$~\gev. Here only
one of the produced neutral \B mesons is analyzed for its decay into $\Kp\pim$ or $\Km\pip$, whereas
the other is not studied and might decay into any final state, say $\Bzb\to \Xbar$ or $\Bz\to \X$.
By \CPT invariance the time-reversed processes,  
$\Kp\pim\to\Bz$
and 
$\Km\pip\to\Bzb$,
would have expected rates $R_1$ and $R_2$, respectively.
However, the experiment probing motion reversal should form
the two \B mesons and the \FourS, through the chain reactions
$(\Kp\pim\to\Bz)(\Xbar\to\Bzb) \to \FourS$ 
and 
$(\Km\pip\to\Bzb)(\X\to\Bz) \to \FourS$. 
This is clearly a problem of thermodynamic irreversibility.
Besides, even if we could build such a \Refs{set up,}
strong interaction processes 
in the kaon-pion annihilation 
would completely swamp the weak interaction process responsible 
\Refs{for}
the decay.

We might consider motion reversal in the mixing, often also referred as oscillation, 
of the pseudoscalar neutral \K, \B and \D mesons.
In this case one compares the probability of a flavor eigenstate (say) \Kz transforming into a \Kzb, and vice versa.
Since the states \Kz and \Kzb are particle and antiparticle, the two transitions are connected by both \T and \CP 
\Refs{transformations.}
Even if \CPT symmetry would be broken, there exists no difference between \CP and \T in this case. 
Thus the two symmetry transformations are experimentally identical and lead to the same asymmetry.
This flavor-mixing or Kabir asymmetry~\cite{Kabir:1970} is independent of time since the two processes have identical time dependence, 
and is induced by the interference between the dispersive, $M_{12}$, and absorptive,
$\Gamma_{12}$, contributions to the mixing of \Kz and \Kzb states.
Here, $M$ and $\Gamma$
are the $2\times 2$ mass and decay Hermitian matrices of the effective Hamiltonian describing neutral meson mixing,
${\cal H}_{\rm eff}=M-i\Gamma/2$~\cite{Branco:1999fs,Bigi:2000yz},
and the index 1 (2) refers to \Bz (\Bzb) state.

\begin{figure}[htb!]
\begin{center}
 \includegraphics[angle=-90,width=0.45\textwidth]{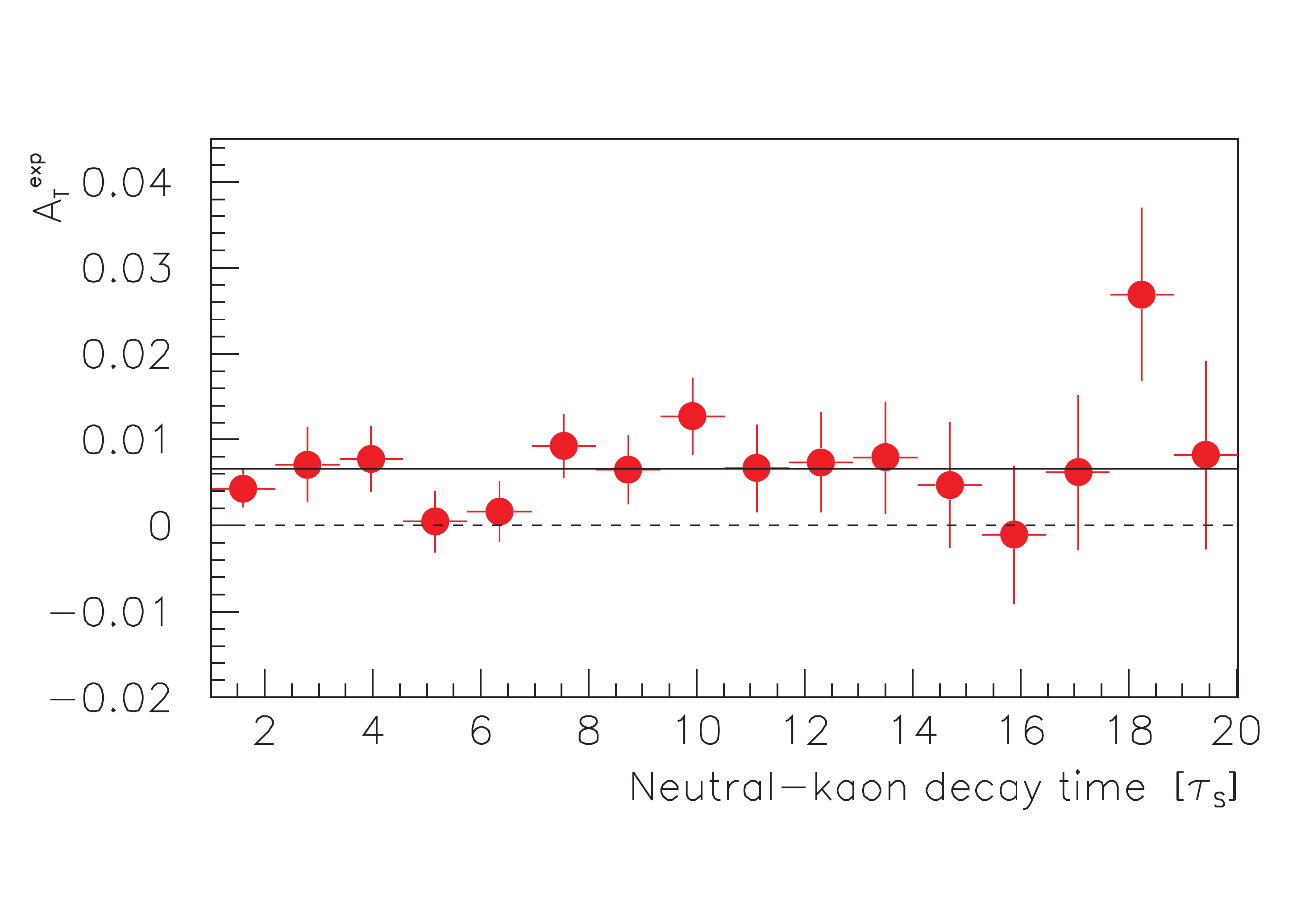}
\caption{\label{fig:CPLEAR}\captionSize The CPLEAR asymmetry versus the neutral-kaon decay time 
(in units of the \KS lifetime)~\cite{Angelopoulos:1998dv}. The solid line represents the 
\Vera{average}.
}
\end{center}
\end{figure}

Evidence at $4\sigma$ level for this 
asymmetry
was found by the CPLEAR experiment at CERN in 1998 by studying
the reaction $p\bar p \to \Kpm \pimp \Kz (\Kzb)$ in $p\bar p$ collisions~\cite{Angelopoulos:1998dv}. The strangeness
(strange and antistrange flavor content) of the \Kz and \Kzb mesons 
at production time was determined by the charge of the accompanying charged kaon. Since weak interactions do not conserve
strangeness, the \Kz and \Kzb may subsequently transform into each other. The strangeness of the neutral kaon at decay time is
determined through the semileptonic decays $\Kz\to e^+\pim\nu_e$ and $\Kzb\to e^-\pip\bar\nu_e$, respectively.
The asymmetry, shown in Fig.~\ref{fig:CPLEAR}, is effectively independent of time, and reveals a net offset with respect to zero.

\begin{figure}[htb!]
\begin{center}
\begin{tabular}{cc}
 \includegraphics[width=0.23\textwidth]{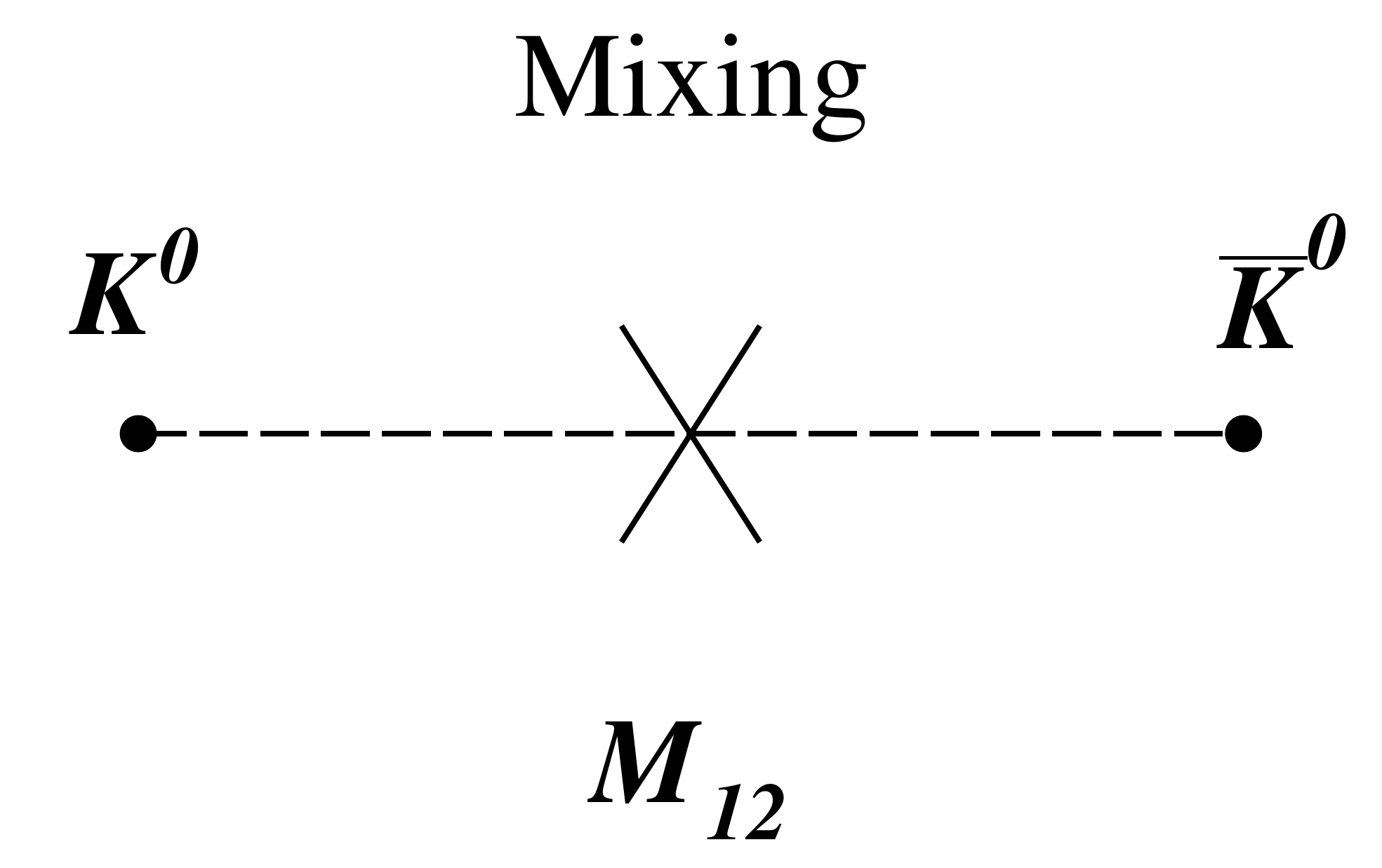} \put(-65,-10){(a)} &
 \includegraphics[width=0.23\textwidth]{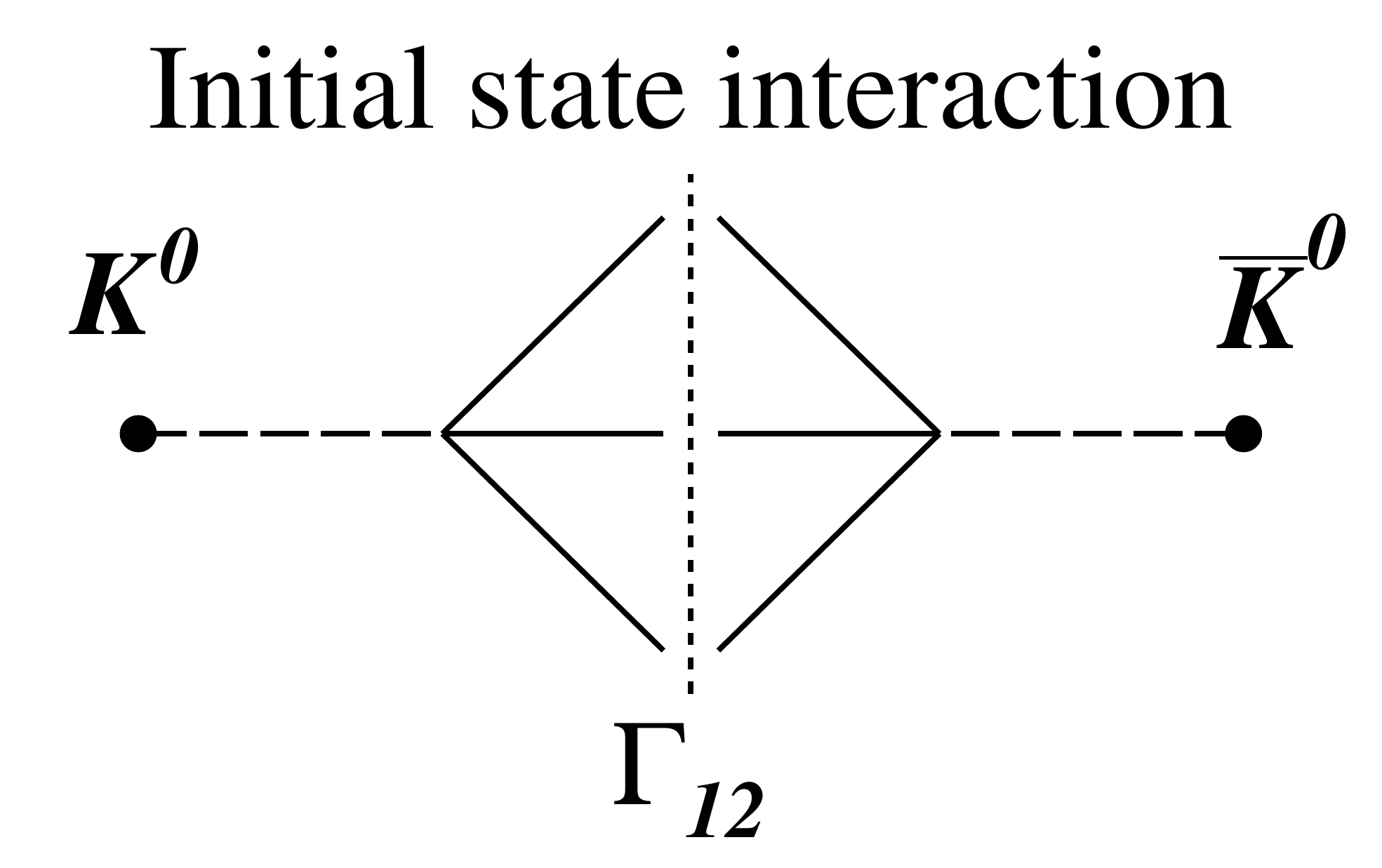} \put(-65,-10){(b)} \\
 \includegraphics[width=0.23\textwidth]{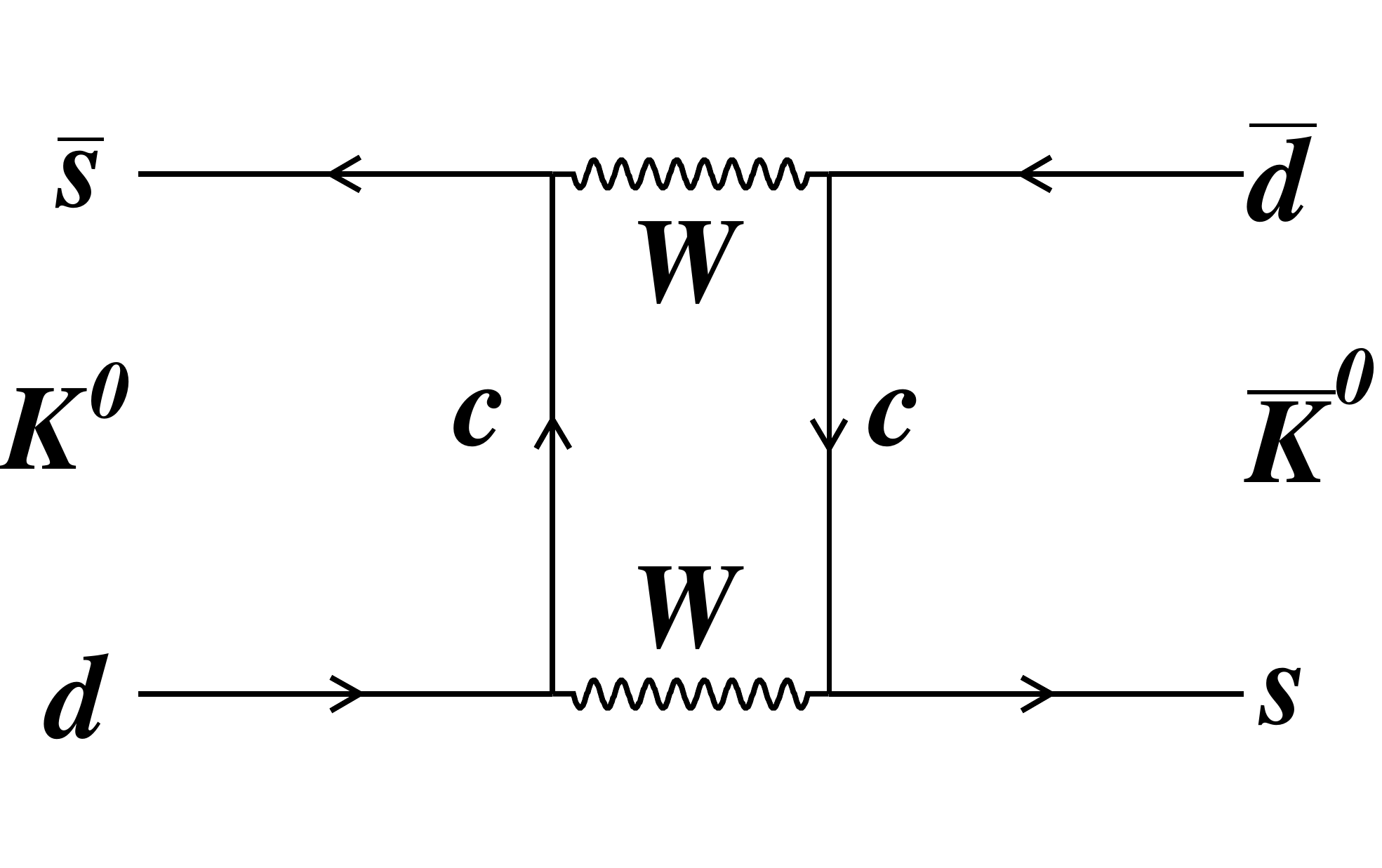} \put(-65,0){(c)} &
 \includegraphics[width=0.23\textwidth]{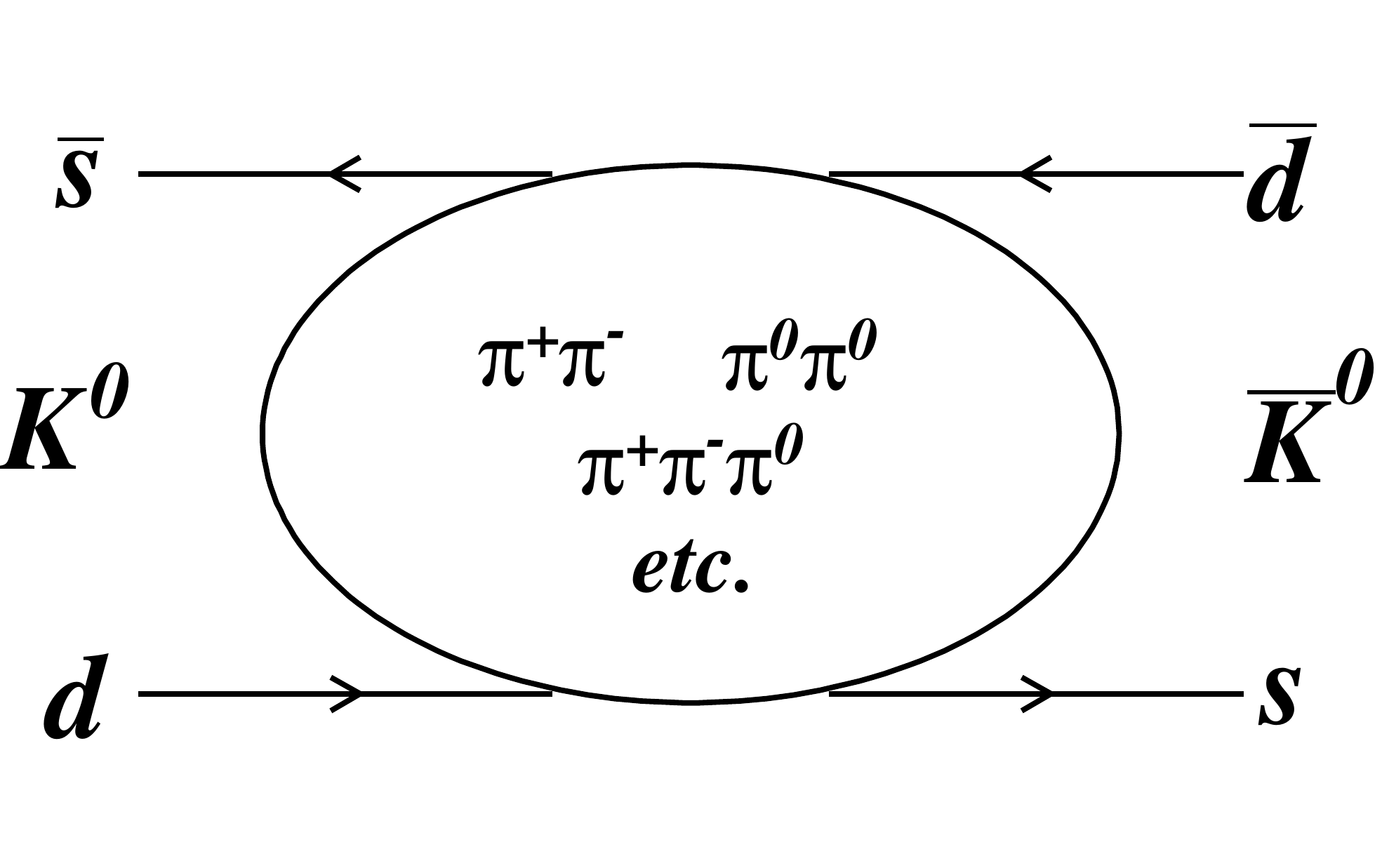} \put(-65,0){(d)}
\end{tabular}
\caption{\label{fig:Kdiagrams}\captionSize
The $\Kz\to\Kzb$ versus $\Kzb\to\Kz$ mixing asymmetry arises from the interference between
the dispersive (a) and absorptive (b) contributions to \Kz-\Kzb mixing. 
\Vera{Diagram} 
(a) involves short-distance box diagrams (c) as well as long-range interactions (d) in which the intermediate 
states are off-shell mesons.
The box diagram is matched by another where the quark triplet and the $W$ bosons are interchanged.
\Vera{Diagram} 
(b) is referred to as ``initial state interaction'' since it involves decay to intermediate on-shell states (d),
and 
it is proportional to the decay width difference between the \Kz and \Kzb mass eigenstates.
}
\end{center}
\end{figure}

The interpretation of the asymmetry relies on two main aspects. Firstly, in the framework of the 
\Vera{Weisskopf-Wigner approach~\cite{Weisskopf:1930au,Weisskopf:1930ps}}, the effect comes from the overlap (non-orthogonality) 
of the ``stationary'' \KS and \KL physical states.
Secondly, the decay plays an essential role; indeed, within the SM the dispersive and absorptive contributions 
to \Kz-\Kzb mixing (Fig.~\ref{fig:Kdiagrams})
are at leading order
proportional to the mass and decay width differences between the \Kz and \Kzb mass 
eigenstates~\cite{Branco:1999fs,Beringer:2012}, respectively.
The presence of the decay as an initial state interaction, 
essential 
to construct a
non-vanishing interference for this observable, has been argued by Wolfenstein to claim 
that this asymmetry ``is not as direct a test of time-reversal violation as one might like''~\cite{Wolfenstein:1999prl,Wolfenstein:1999re}.
In the neutral \B system, where the decay width difference between the \Bz and \Bzb mass eigenstates is negligible,
the measurement of this asymmetry has, in fact, brought negative results~\cite{Lees:2013sua,Abazov:2012hha,Aubert:2006nf,Nakano:2005jb}.
Other authors, however, have 
\Refs{argued}
that
its interpretation as a genuine signal for \T violation
does not get affected by these arguments~\cite{AlvarezGaume:1998yr,Ellis:1999xh,Gerber:2004hc}.

\section{\B factories and \CP violation}  
\label{sec:BFACTORIES}

At the asymmetric \B factories, electron and positron beams collide with high luminosity at a \cms energy 
\Vera{of}
10.58~\gev 
\Vera{corresponding to the mass of the \FourS resonance, a vector particle with $J^{PC}=1^{--}$}.
The \FourS 
\Vera{is abound state of} 
a \b and a \bbar quark, that decays 
\Vera{exclusively to}
a pair of \B and \Bb mesons. Since the mass of the \FourS is only slightly higher than twice the mass of the \B meson, 
the two \B mesons have low momenta (about 330~\mevc) and are produced almost at rest in the \FourS reference frame
with no additional particles besides those associated to the \B decays.
The energy of the electron beam is adjusted to be between twice and three times larger than that of the positrons, 
so that the \cms 
\Vera{frame}
has a Lorentz boost along the collision \Vera{axis}. 
Two \B factory colliders, \pep2 at SLAC in California and KEKB at KEK in Japan, with their corresponding detectors,
\babar~\cite{TheBABAR:2013jta,Aubert:2001tu} and Belle~\cite{Abashian:2000cg}, have been operating 
during the last \Vera{decade,} 
accumulating an integrated luminosity of data exceeding 500~\invfb and 1~\invab, \Vera{respectively.} 

The electron- and positron-beam energies have been designed to be 9.0 and 3.1~\gev at \pep2, and 8.0 and 3.5~\gev at KEKB,
thus the Lorentz boosts $\beta\gamma$ are about 0.56 and 0.42, respectively. This allows
a mean separation of the two \B mesons along the collision axis in the laboratory frame of about 250 and 200 $\mu$m.
With a typical detector resolution of 100 $\mu$m, it is 
\Vera{possible} to experimentally determine the distance \dz between 
the decay points of the two \Vera{$\B$s}
\Vera{to}
obtain the decay time difference \Vera{$\dt \approx \dz / (\beta\gamma\c)$}.
This translates into a mean
time separation of about 1.5~\ps with a resolution ranging typically between 0.6 and 0.8~\ps.

\subsection{Entangled neutral \B mesons}  
\label{sec:BFACTORIES-entangled}

The decay of the \FourS particle occurs through strong interactions, thus the system of the created pair of \B and \Bb mesons inherits 
the \FourS quantum numbers.  
About 50\% of the \BB pairs are \BzBzb, when the 
hadronization process pick ups a \ddbar quark pair, and 50\% \BpBm when it is a \uubar pair. 

Because \B and \Bb are two pseudoscalar states of a unique (complex) 
field,
Bose statistics and angular momentum conservation requires that 
the wavefunction of the \BB pair 
\Vera{be} in a P-wave, antisymmetric \CP-even state~\cite{Lipkin:1988fu},
\bea
| \Auth{\Upsilon} \rangle & = & \frac{1}{\sqrt{2}}\left[ |\Bz(t_1)\rangle|\Bzb(t_2)\rangle - |\Bzb(t_1)\rangle|\Bz(t_2)\rangle \right].
\label{eq:BBflav}
\eea
The times $t_1$ and $t_2$ in Eq.~(\ref{eq:BBflav}) do not refer to time dependence but
labels to characterize the states: state 1 (2) is labeled as the first (second) to decay, i.e. $t_1<t_2$.
The antisymmetric entanglement 
is essential: even
with \Bz-\Bzb mixing between the production time at $t=0$ and the first decay at $t=t_1$,
the state $| \Auth{\Upsilon} \rangle$ remains antisymmetric, with no trace of combinations $|\Bz\rangle|\Bz\rangle$ or $|\Bzb\rangle|\Bzb\rangle$,
and only $|\Bz\rangle|\Bzb\rangle$ and $|\Bzb\rangle|\Bz\rangle$ states appearing at any time.
\Vera{The state} of the first \B 
to decay 
at $t_1$ 
dictates 
the state of the \Vera{other} \B, 
without measuring (thus destroying) it, 
\Vera{and}
then evolves in time and \Vera{decays} 
at $t_2$. 
The antisymmetric wavefunction defined by this EPR entanglement~\cite{Einstein:1935rr,Reid:2009zz} is usually written in terms of the 
strong-interaction flavor eigenstates \Bz and \Bzb, as given by Eq.~(\ref{eq:BBflav}).
It should be noted that the basis choice 
of \Bz and \Bzb states is only a matter of convention. 
The individual state of each neutral \B is not defined in the entangled state $| \Auth{\Upsilon} \rangle$ before the first decay. 
The quantum collapse of the state at decay time $t_1$ to a flavor specific \Vera{channel} 
tags 
\Vera{the second \B}
as the 
orthogonal state.

\begin{figure}[htb!]
\begin{center}
 \includegraphics[width=0.3\textwidth]{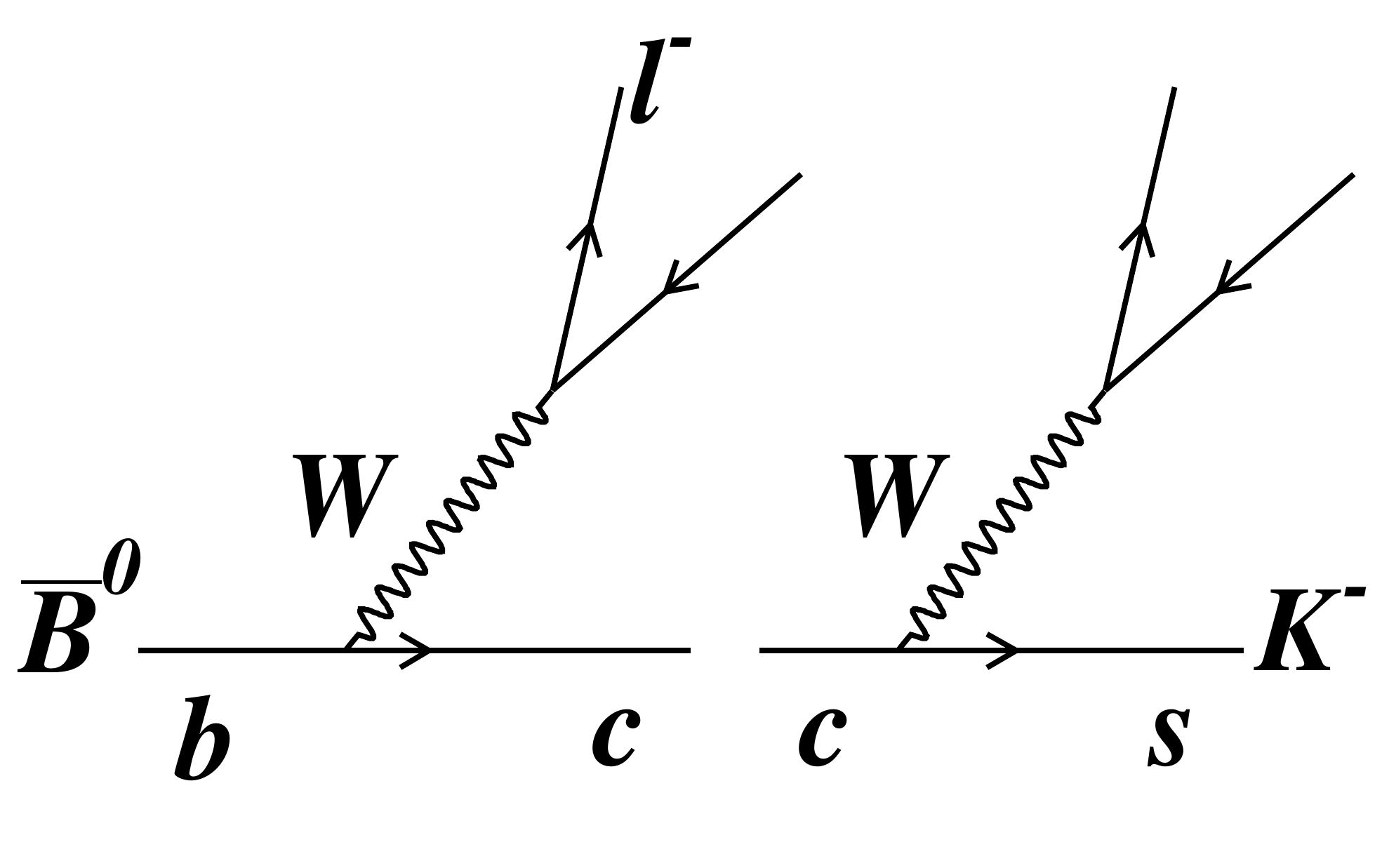}
\caption{\label{fig:BdecaySignature}\captionSize A \Bzb meson, containing a \b quark,
decays dominantly with lifetime $\sim 1.5$~\ps through a $\b\to\c$ transition. The 
\Vera{virtual}
$W^-$ gauge boson creates 
a negatively-charged prompt lepton 
\Vera{whose}
electric charge can be correlated with the 
\Vera{\b quark flavor.}
\Vera{The charge} of 
\Vera{kaons} produced through the $\b \to \c \to \s$ cascade 
\Vera{also identifies}
the flavor of the \B meson.
}
\end{center}
\end{figure}

Now, consider that the first \B decays semileptonically producing a negatively-charged prompt lepton. As shown 
in Fig.~\ref{fig:BdecaySignature}, this decay proceeds through a $\b\to\c$ transition, and the charge of the lepton is 
completely correlated with the flavor 
\Vera{of the}
\b quark, hence the parent \B is 
\Vera{a}
\Bzb at the instant 
of its decay. The anticorrelation defined by the wavefunction determines that the 
\Vera{other}
\B at that time is a \Bz, thereby it prepares (or tags) a \Bz flavor state. In quantum mechanics language this means 
that the initial state of the \B meson has been prepared or filtered as \Bz.
Neutral \B mesons start to oscillate just after their production since their mixing rate, $\dmd \approx 0.5$~\invps, is comparable 
to their 
\Vera{decay}
width, $\Gamma_d\approx 0.7$~\invps. Thus, the state prepared as \Bz will be a superposition of \Bz and \Bzb at a later time.
As 
\Refs{will} be discussed further in Secs~\ref{sec:BFACTORIES-cpv} and~\ref{sec:CONCEPT}, this quantum mechanical
superposition of states 
\Refs{is}
of key importance for the experimental exploration of \CP and, especially, time-reversal symmetries.
Analogously, with a positively-charged prompt lepton we would have a \Bzb tag. There are other signatures that
can be used, like prompt charged kaons produced through 
a $\b \to \c \to \s$ cascade, as sketched in Fig.~\ref{fig:BdecaySignature}.
In the following, we denote generically as $\ellm \Xbar$ or $\ellp \X$ any flavor-specific 
final state that can be used to identify the flavor of the decaying \B.
Note that for this procedure to work only right-sign decays should take place, i.e.
the wrong-sign decays $\Bz \to \ellm \Xbar$ and $\Bzb \to \ellp \X$ 
\Vera{do not}
occur.

\subsection{\CP violation at \B factories}  
\label{sec:BFACTORIES-cpv}

Flavor tagging based on quantum entanglement has been the basis for a decade of \CP violation physics at \B factories.
In these studies, the second \B meson, the one that is not used for flavor tagging,
is reconstructed into a \CP-eigenstate final state, for example $\jpsi\KS$, which 
has 
\Vera{\CP-odd} parity 
\Vera{(except for 0.1\%}
due to \CP violation in \Kz-\Kzb mixing~\cite{Beringer:2012}).
\Vera{The decay} of the \B meson into the flavor-eigenstate final state can occur after the decay into the \CP eigenstate; 
to account for this situation the experiments define as convention a signed decay time difference,
\bea
\dt & = & t_{\CP} - t_{\rm flavor}.
\label{Eq:deltat}
\eea
 
\begin{figure}[htb!]
\begin{center}
\includegraphics[width=0.4\textwidth]{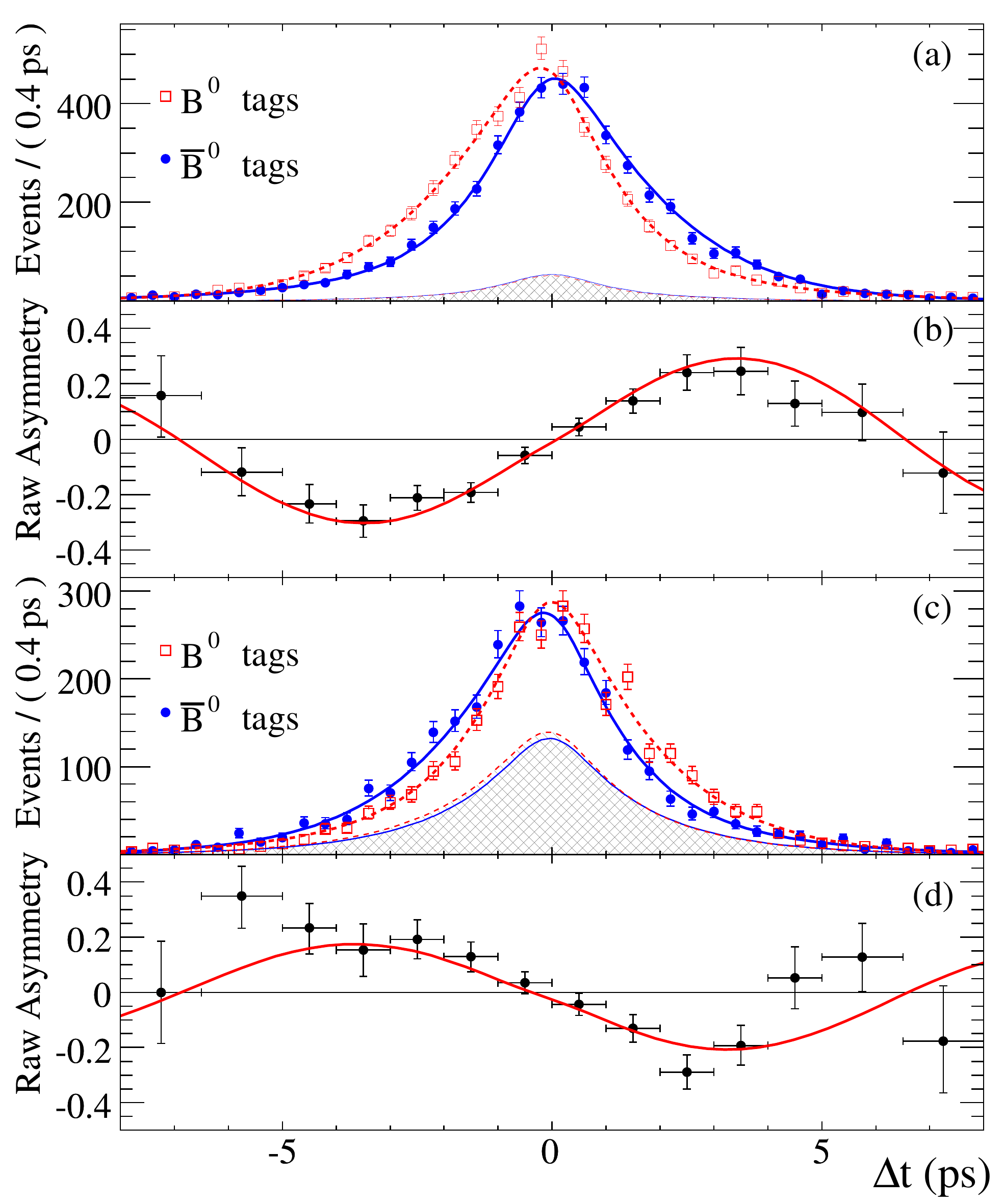}
\caption{\label{fig:CPasym-babar}\captionSize 
Flavor-tagged \dt distributions (a,c) and \CP asymmetries (b,d) from the \babar\ \Vera{experiment~\cite{Aubert:2009aw}}, for \CP-odd final
states $\jpsi\KS$, $\psi(2S)\KS$, $\chi_{c1}\KS$ and $\eta_c\KS$ (a,b), and the \CP-even final states $\jpsi\KL$ (c,d).
The solid (dashed) curves in (a) and (c) represent the best fit projections in \dt for \Bzb (\Bz) tags. 
The shaded regions represent the estimated background contributions to (a) and
(c). The curves in (b) and (d) are the fit projections of the
\CP asymmetry between \Bzb and \Bz tagged \Vera{events.} 
}
\end{center}
\end{figure}

\CP violation means any difference between the decay rate $\Bz\to\jpsi\KS$ and its \CP conjugate, $\Bzb\to\jpsi\KS$.
The search for this asymmetry has been the raison d'\^etre of the \B factories.
Experimentally everything is identical for the two processes, except the requirement of a 
$\ellm\Xbar$ or a $\ellp\X$ 
\Vera{decay of}
the \B meson used for tagging. 
The large \CP violation observed by the \B factory experiments 
in 2001~\cite{Aubert:2001nu,Abe:2001xe} follows from the comparison of the \Bz and \Bzb decay rates at equal decay proper-time 
difference, as shown in Fig.~\ref{fig:CPasym-babar}(a). 
\Vera{The difference}
between these two distributions signals \CP violation. 
\Vera{This} 
is more 
\Vera{evident}
in the time-dependent \CP asymmetry, shown in Fig.~\ref{fig:CPasym-babar}(b), 
where the difference between the two distributions is 
\Vera{normalized} to their sum,
\bea
A_{\CP,f} & = & \frac{\Gamma_{\Bzb\to f}(\dt)-\Gamma_{\Bz\to f}(\dt)}{\Gamma_{\Bzb\to f}(\dt)+\Gamma_{\Bz\to f}(\dt)}.
\label{eq:ACPdef}
\eea 
It should be noted that the notion of flavor tag adopted here as preparation of the initial state of the 
\Vera{second} \B, using the measurement of the decay of the first \B, is different from the one commonly used 
by the \B factory experiments, where tagging refers to the flavor identification of the first \B. 
The reason to adopt this convention, closer to the standard quantum mechanics language, will become more clear later.

\Vera{One}
can also use final states with different \CP parity, for example the \CP-even $f=\jpsi\KL$. Due to its opposite \CP eigenvalue,
the 
\Refs{behaviors}
of the time distributions for \Bz and \Bzb 
\Refs{are}
interchanged and the asymmetry is opposite, as observed in 
Fig.~\ref{fig:CPasym-babar}(c,d). Besides increasing the statistical power of the result, the use
of final states with opposite \CP parity provides a powerful cross-check of the \CP-violating effect.

This \CP asymmetry is induced by 
the interference of amplitudes involved
in the two possible
paths to reach the same final state $f$ from a \Bz (or its \CP conjugate), either through decay
without mixing, $\Bz \to f$,
or through mixing followed by decay, $\Bz\to\Bzb\to f$. Therefore, we usually refer to \CP violation in the interference
between decay amplitudes with and without mixing, or simply indirect \CP violation. This type of \CP symmetry breaking
can be compared to direct \CP violation (due to the interference 
\Vera{of}
different decay amplitudes) 
or \CP violation in mixing (arising from the interference 
\Vera{of}
the dispersive and absorptive contributions to mixing)~\cite{Branco:1999fs}.

The time-dependent decay rates and the \CP asymmetry can be parameterized in a model-independent way, only assuming quantum mechanics, as
\bea
\Gamma_{\Bzb(\Bz)\to f} \propto  e^{-\Gamma_d|\dt|}\times \ \ \ \ \ \ \ \ \ \ \ \ \ \ \ \ \ \ \ \ \ \ \ \ \ \ \ \ \ \ \ \ \ \ \ \  \ \ \ \nn \\
                    \Big\{ 1+(-)\big[ S_f \sin(\dmd \dt) - C_f \cos(\dmd \dt) \big] \Big\}, \ \ \ \ \ \label{eq:Gparam} \\
A_{\CP,f} = S_f \sin(\dmd \dt) - C_f \cos(\dmd \dt), \ \ \ \ \ \ \ \ \ \ \ \ \ \ \label{eq:ACPparam}
\eea
where \dmd is the mass difference
between the physical states of the neutral \B meson system, and $\Gamma_d$ is average total decay width. 
Equation~(\ref{eq:Gparam}) assumes 
a negligible difference between the decay rates of the mass eigenstates, i.e. $\dGd=0$, 
and \CP symmetry in mixing \Auth{(independently of whether \CPT and \T are or are not violated).} 
\Refs{If \CP symmetry in mixing holds, then the conditions $|q/p|=1$ and $z=0$ apply, where $q$, $p$ and $z$ are three 
complex parameters introduced to define the
eigenstates of well-defined mass and decay width (referred to as mass eigenstates) in terms of the
strong interaction (flavor) eigenstates~\cite{Beringer:2012,Branco:1999fs}. Adopting an arbitrary sign convention~\cite{Branco:1999fs}, 
\bea
| \B_L \rangle & \propto p\sqrt{1-z} | \Bz \rangle - q\sqrt{1+z} | \Bzb \rangle, \nn \\
| \B_H \rangle & \propto p\sqrt{1+z} | \Kz \rangle + q\sqrt{1-z} | \Bzb \rangle.
\label{eq:BLBHstates}
\eea
Within the Weisskopf-Wigner approach, these parameters are related to the matrix elements of the $2\times 2$ 
effective, non-Hermitian Hamiltonian ${\cal H}_{\rm eff}$ describing mixing.
Note that the real and imaginary parts of the corresponding eigenvalues are the masses and decay widths,
and their splittings are \dmd and \dGd, respectively. 
If either \CP or \CPT is a good symmetry in mixing (independent of \T), then $z=0$, and
if either \CP or \T is a symmetry of ${\cal H}_{\rm eff}$ (independent of \CPT), $|q/p|=1$. It then follows that
\CP with \CPT violation in mixing is defined by $z\ne0$,
and \CP with \T violation is through $|q/p|\ne1$.}
The curves shown in Fig.~\ref{fig:CPasym-babar} 
represent the best fit projections in \dt using Eqs~(\ref{eq:Gparam}) and (\ref{eq:ACPparam}).

The coefficients $S_f$ and $C_f$ are related to \CP violation. Within the 
\Vera{Weisskopf-Wigner approach,} these are connected with 
the fundamental parameter describing indirect \CP violation~\cite{Bigi:2000yz},
\Auth{
\bea
\lambda_f & = & \frac{q}{p}\frac{\Abar_f}{\A_f},
\label{eq:lambdaf}
\eea 
through the relations
\bea
S_f & = & 2\im\lambda_f/(1+|\lambda_f|^2),\nn \\
C_f & = & (1-|\lambda_f|^2)/(1+|\lambda_f|^2),
\eea
where 
$\A_f = \langle f | {\cal D} | \Bz \rangle$ and $\Abar_f = \langle f | {\cal D} | \Bzb \rangle$ are
the \Bz and \Bzb decay amplitudes to the final state $f$,}
with ${\cal D}$ the operator describing \Auth{the \B decay.}
\Auth{
For $f=\jpsi\KS$ and $f=\jpsi\KL$, Eq.~(\ref{eq:lambdaf}) becomes
\bea
\lambda_f & = & \eta_f \frac{q}{p}\frac{\Abar}{\A}\frac{p_\K}{q_\K},
\label{eq:lambdaf_2}
\eea 
where $\A = \langle \jpsi\Kz | {\cal D} | \Bz \rangle$, $\Abar = \langle \jpsi\Kzb | {\cal D} | \Bzb \rangle$,
and $\eta_f = -1$($+1$) for $f=\jpsi\KS$($\jpsi\KL$) is} 
associated to the \CP parity of the final state.
The factor $p_\K/q_\K$ 
arises from \Kz-\Kzb mixing, 
essential for the interference because \Bz and \Bzb decay into $\jpsi\Kz$ and $\jpsi\Kzb$, 
but not into $\jpsi\Kzb$ and $\jpsi\Kz$, respectively.

Assuming that the amplitude $\A_f$ can be described by a single weak phase, 
the two contributions to \CP violation in the parameter $\lambda_f$, \CP with \T violation and \CP with \CPT violation,
can be identified easily by separating it into modulus and phase, $\lambda_f = |\lambda_f| \exp(i\phi_f)$~\cite{Schubert:2014cva}.
\CPT invariance in the decay requires $|\Abar_f/\A_f|=1$~\cite{Lee:1957qq}. 
For $|q/p|=1$, 
experimentally well verified~\cite{Lees:2013sua,Abazov:2012hha,Aubert:2006nf,Nakano:2005jb},
it follows that $|\lambda_f|=1$. 
\T invariance 
\Auth{in the time evolution followed by decay}
requires $\phi_f =0$ or $\pi$, i.e. $\im\lambda_f = 0$~\cite{Enz:1965tr}.
Instead, if $\A_f$ is the sum of two (or more) amplitudes with non-vanishing weak and strong phase differences between the 
two amplitudes, then we have $|\Abar_f/\A_f| \ne 1$, even if ${\cal D}$ is \CPT symmetric. 
Thus, if $|\Abar_f/\A_f| = 1$ then we
have either both \CPT symmetry in decay and a single amplitude, 
or an unlikely cancellation of \T and \CPT violation in decay amplitudes.

\begin{figure}[htb!]
\begin{center}
\begin{tabular}{cc}
\multicolumn{2}{c}{ \includegraphics[width=0.23\textwidth]{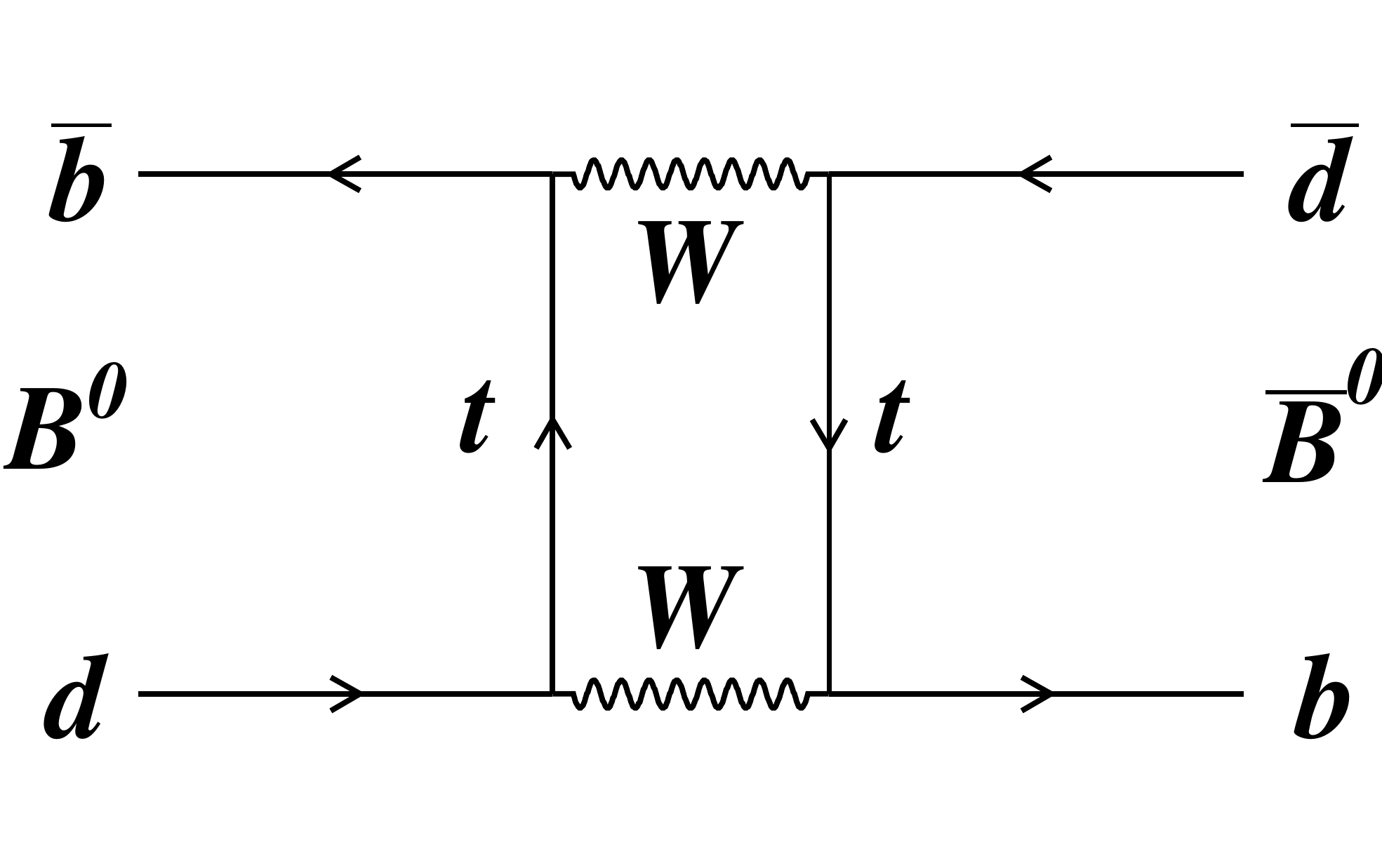} } \put(-125,0){(a)} \\
 \includegraphics[width=0.23\textwidth]{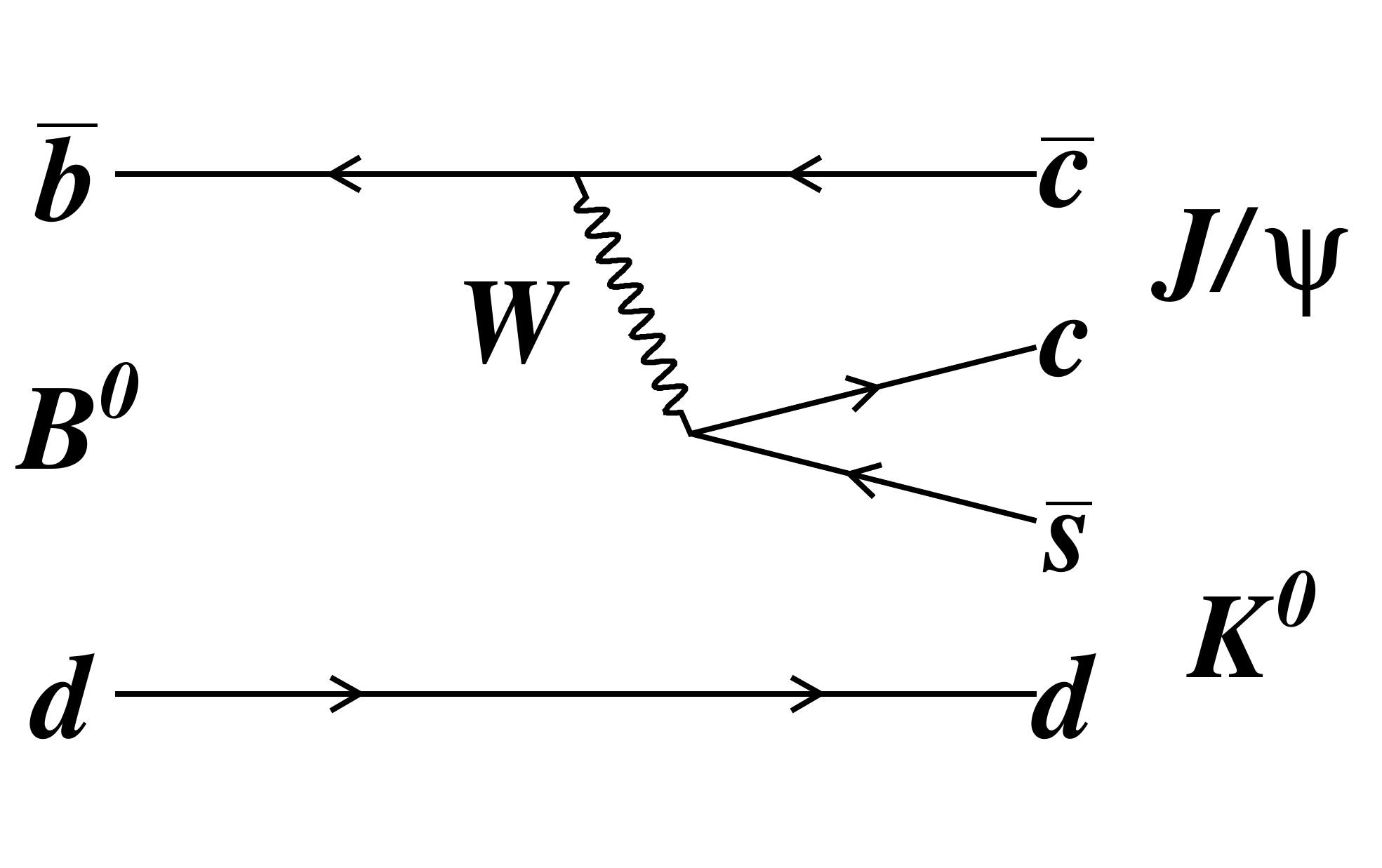} \put(-65,0){(b)} &
 \includegraphics[width=0.23\textwidth]{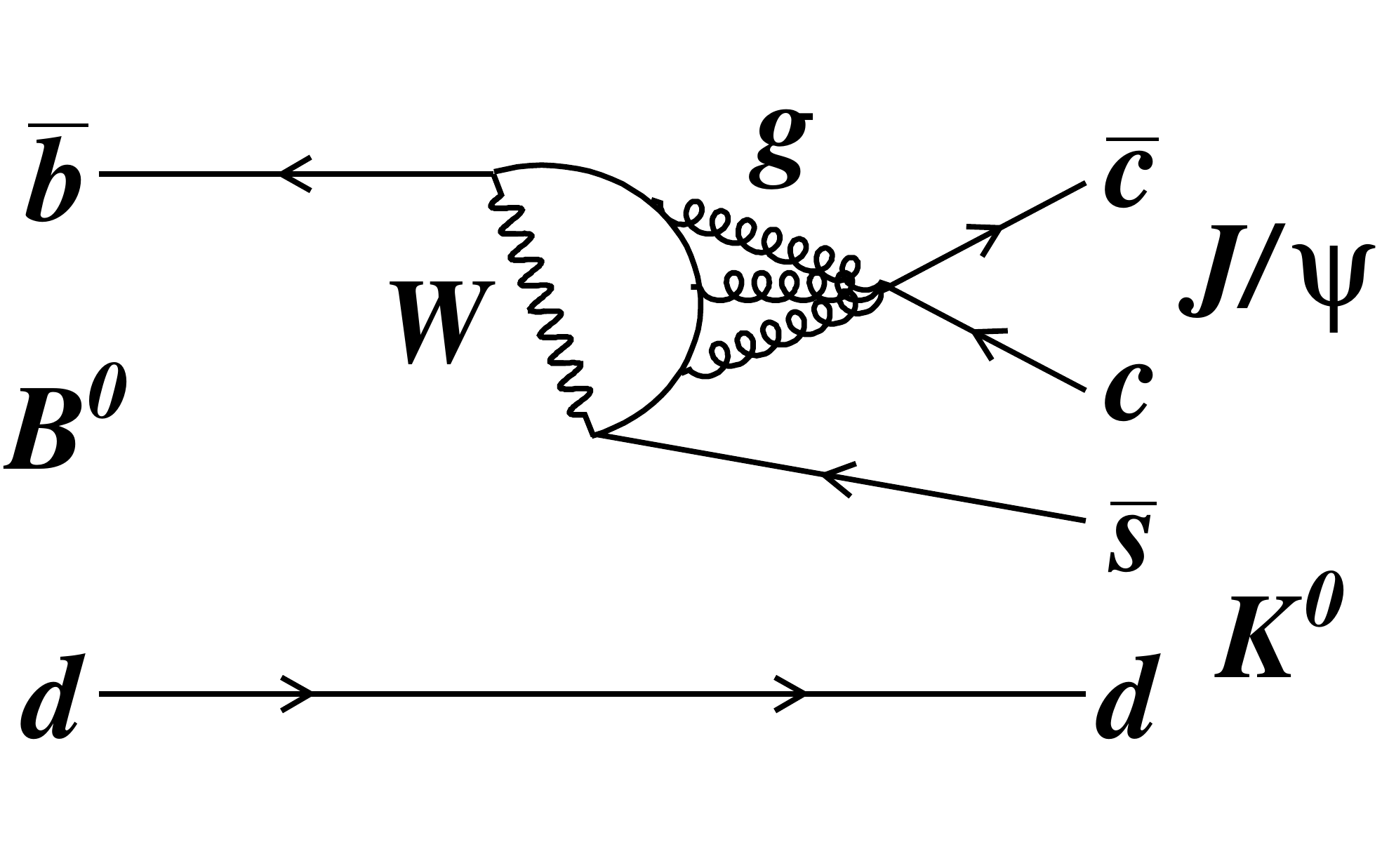} \put(-65,0){(c)}
\end{tabular}
\caption{\label{fig:Bdiagrams}\captionSize
(a) Box diagram corresponding to the SM short-distance contributions to \Bz-\Bzb mixing.
This contribution is matched by a diagram where the quark triplet and the $W$ bosons are interchanged.
(b) Tree and (c) \Auth{penguin SM diagrams} for the $\Bz\to\ccbar\Kz$ decay.
In the Wolfenstein parameterization the tree amplitude is ${\cal O}(\lambda_c)$, with $\lambda_c\approx 0.226$, 
whereas the penguin contribution is ${\cal O}(\lambda_c^2)$ and has the same weak phase. 
}
\end{center}
\end{figure}

In the SM, \Bz-\Bzb mixing is dominated by the box diagrams shown in Fig.~\ref{fig:Bdiagrams}(a), 
leading to $q/p \approx V_{\t\b}^* V_{\t\d}/V_{\t\b} V_{\t\d}^*$~\cite{Branco:1999fs,Bigi:2000yz}.
For the final state $f=\jpsi\KS$, the \B decay is dominated by the 
$\b\to\ccbar\s$ tree amplitude in Fig.~\ref{fig:Bdiagrams}(b), 
followed by \Kz-\Kzb mixing (Fig.~\ref{fig:Kdiagrams}).
Therefore, from Eq.~(\ref{eq:lambdaf_2}) it follows
\bea
\lambda_f & = & \eta_f \frac{V_{\t\b}^*V_{\t\d}}{V_{\t\b}V_{\t\d}^*} \frac{V_{\c\b}V_{\c\s}^*}{V_{\c\b}^*V_{\c\s}} \frac{V_{\c\s}V_{\c\d}^*}{V_{\c\s}^*V_{\c\d}}. 
\eea
This leads to $C_f = 0$ and $S_f = \im\lambda_f = - \eta_f \sin 2\beta$, where 
$\beta \equiv {\rm arg}[-(V_{\c\d}V_{\c\b}^*)/(V_{\t\d}V_{\t\b}^*)]$~\cite{Beringer:2012} is 
the angle between the \t and \c sides of the
$\b\d$ unitarity triangle, illustrated in Fig.~\ref{fig:bdUT}.
In the Wolfenstein parameterization, $\beta \equiv -{\rm arg} V_{\t\d}$.
Thus, the same magnitude is expected for the \CP-even and \CP-odd modes up to 
\Refs{a}
small correction due to \CP violation in \Kz-\Kzb oscillations.
Higher-order (penguin) amplitude contributions have either the same weak phase -Fig.~\ref{fig:Kdiagrams}(c)-
or are CKM suppressed, so that these predictions apply within the SM up to ${\cal O}(\lambda_c^4)$,
where $\lambda_c\approx 0.226$ in the Wolfenstein parameterization~\cite{Wolfenstein:1983yz,Beringer:2012}.
The deviation due to the penguin contributions with different CKM phase has been estimated to be 
smaller than 1\%~\cite{Boos:2004xp,Boos:2006vq}. 
\Refs{Physics beyond the SM could modify the phase of $q/p$ (thus $S_f$)
and $C_f$, although large effects are unlikely to be generated in the latter
due to the dominance of the tree amplitude in decay~\cite{Chiang:2009ev}.}

\section{Time-reversal experiment concept}  
\label{sec:CONCEPT}

We might be tempted to interpret these \CP violation results as evidence for time-reversal non-invariance.
The experimental study has been performed invoking $\dGd=0$ 
\Auth{and \CP invariance in mixing ($|q/p|=1$ and $z=0$),} 
although the good agreement between the data points and curves observed in Fig.~\ref{fig:CPasym-babar} hints that 
the effects 
\Auth{of possible deviations}
from these assumptions
are well below the statistical sensitivity of the current data. 
Indeed, dedicated studies have shown the good agreement 
\Auth{of the results}
with and without making these 
assumptions~\cite{Aubert:2003hd,Aubert:2004xga}. Moreover, the results are consistent with 
\Refs{$C_f = 0$ and $S_f \ne 0$, and therefore, according to Eq.~(\ref{eq:lambdaf}), with
$|\Abar_f/\A_f|=1$ for $|q/p|=1$ and $\im\lambda_f \ne 0$.}
Within the 
\Vera{Weisskopf-Wigner approach}
these values are compatible with \CP with \CPT symmetry in decay, 
and \CP with \T violation in the interference
of decay with and without mixing~\cite{Schubert:2014cva,Fidecaro:2013gsa}.
This is not, however, the question of interest here, but to \Refs{set up} an experiment capable to 
demonstrate by itself motion reversal non-invariance 
between states that are not \CP conjugate to each other, as discussed previously.

\subsection{Entangled neutral \B mesons revisited}  
\label{sec:CONCEPT-entangled}

The solution~\cite{Banuls:1999aj,Banuls:2000ki,Wolfenstein:1999re,Quinn:2009,Bernabeu:2012ab} arises from the quantum mechanical properties 
imposed by the EPR entanglement~\cite{Einstein:1935rr,Reid:2009zz} between the two neutral \B mesons produced in
the \FourS resonance decay. Just as one \B meson in the entangled pair is prepared in the \Bzb or \Bz states at the time when the
other \B is observed as a \Bz or \Bzb by a decay into $\ellp\X$ or $\ellm\Xbar$, respectively, the first decay of one \B into 
the final states $\jpsi\KS$ or $\jpsi\KL$
prepares the other \B into well defined, orthogonal linear combinations of \Bz and \Bzb states.
In fact, this idea offers the opportunity to explore separately time reversal, \CP and \CPT symmetries, selecting appropriately 
different transitions defined by different decay channels.

For the entangled state of the two mesons produced by the \FourS decay, the individual state of each neutral \B meson
is not defined before its collapse as a filter imposed by the observation of the decay. Thus, the state $| \Auth{\Upsilon} \rangle$ in
Eq.~(\ref{eq:BBflav}) can be written in terms of any pair of orthogonal states of the individual \B mesons, i.e. a linear 
combination of \Bz and \Bzb, that we denote \Bplus, and its orthogonal state, \Bminus,
\bea
| \Auth{\Upsilon} \rangle & = & \frac{1}{\sqrt{2}}\left[ |\Bplus(t_1)\rangle|\Bminus(t_2)\rangle - |\Bminus(t_1)\rangle|\Bplus(t_2)\rangle \right]. \ \ \ \ \ 
\label{eq:BplusBminusEPR}
\eea
As in Eq.~(\ref{eq:BBflav}), the time evolution (including mixing) preserves only 
$|\Bplus\rangle|\Bminus\rangle$ and $|\Bminus\rangle|\Bplus\rangle$ terms.

Neglecting \CP violation in \Kz-\Kzb mixing, which holds within ${\cal O}(10^{-3})$~\cite{Beringer:2012}, and 
assuming that $\Bz\to\jpsi\Kz$ and $\Bzb\to\jpsi\Kzb$, the normalized states
\bea
|\Bplus\rangle  & = & {\cal N} \left( |\Bz\rangle + \frac{\A}{\Abar} |\Bzb\rangle \right),\nn \\
|\Bminus\rangle & = & {\cal N} \left( |\Bz\rangle - \frac{\A}{\Abar} |\Bzb\rangle \right),
\label{eq:BplusBminus}
\eea
where ${\cal N}=|\Abar|/\sqrt{|A|^2+|\Abar|^2}$,
have the property that the former decays into $\jpsi\KL$, 
but not into $\jpsi\KS$, and the latter into $\jpsi\KS$, 
but not into $\jpsi\KL$~\cite{Lipkin:1988fu,Bernabeu:2012ab}.

The proof is as follows. 
\Auth{Adopting the same sign convention} as in 
Eq.~(\ref{eq:BLBHstates})~\footnote{\Auth{The \KL state is experimentally found to be the} heavier state.},
and assuming \CPT invariance in 
kaon mixing, we have
\bea
| \KS \rangle & = {\cal N}_\K \left( p_\K | \Kz \rangle -q_\K | \Kzb \rangle \right), \nn \\
| \KL \rangle & = {\cal N}_\K \left( p_\K | \Kz \rangle +q_\K | \Kzb \rangle \right),
\label{eq:KSKLstates}
\eea
with ${\cal N}_\K=1/\sqrt{|p_\K|^2+|\q_\K|^2}$. In the absence of wrong-strangeness \B decays, 
$\langle \jpsi\Kzb | {\cal D} | \Bz \rangle = \langle \jpsi\Kz | {\cal D} | \Bzb \rangle = 0$,
it is straightforward to show that
\bea
\langle \jpsi\KS(\KL)|{\cal D}|\Bplus\rangle  & = & {\cal N} {\cal N}_\K \left[ p_\K\A -(+) q_\K\A \right],\nn\\
\langle \jpsi\KL(\KS)|{\cal D}|\Bminus\rangle & = & {\cal N} {\cal N}_\K \left[ p_\K\A -(+) q_\K\A \right].~~~~~
\eea
Assuming now \CP invariance in \Kz-\Kzb mixing, we have $p_\K = q_\K$ and the above expressions yield
\bea
\langle \jpsi\KL|{\cal D}|\Bplus\rangle = \langle \jpsi\KS|{\cal D}|\Bminus\rangle & = & 2 p_\K {\cal N} {\cal N}_\K  \A,\nn\\
\langle \jpsi\KS|{\cal D}|\Bplus\rangle = \langle \jpsi\KL|{\cal D}|\Bminus\rangle & = & 0,
\eea
hence \Bplus cannot decay into $\jpsi\KS$ and \Bminus cannot into $\jpsi\KL$.
Note that using Eq.~(\ref{eq:lambdaf}) and inverting Eq.~(\ref{eq:KSKLstates}) it is straightforward to obtain
Eq.~(\ref{eq:lambdaf_2}) and the relation $\lambda_{\jpsi\KL}=-\lambda_{\jpsi\KS}$, 
as 
\Auth{introduced}
in Sec.~\ref{sec:BFACTORIES-cpv}. The fact that the two 
amplitude ratios
differ \Auth{only} by a minus sign is of key importance for the definition of the \Bplusminus states.
According to the definition in Eq.~(\ref{eq:BplusBminus}), these states 
are well 
behaved
under \Bz and \Bzb rephasing,
although 
\Refs{they}
are not 
\CP eigenstates independent of the flavor of the decay channel
due to arbitrary quark phases and possible deviations from $|\A/\Abar| = 1$.

The observation of the decay to the $\jpsi\KS$ ($\jpsi\KL$) final state at time $t_1$ generates an automatic transfer of information to the
(still living) partner meson. Hence, the other \B meson at that time is in a \Bplus (\Bminus) state because this is the state
that cannot decay into $\jpsi\KS$ ($\jpsi\KL$).
We call the quantum preparation of the initial state at $t_1$, using the filter imposed by the observation at $t_1$, a ``\Bplus (\Bminus) tag''.
Sometimes it is referred to as a ``\CP tag''~\cite{Banuls:1998mg} because it is defined through decays into \CP-eigenstate final states.
The \Bplus or \Bminus states prepared by entanglement at $t_1$ evolve in time until they are observed at some later 
time $t_2$ in a decaying final state filtering another linear combination state.
For convenience, we restrict to flavor-specific final states $\ellm\Xbar$ and $\ellp\X$ that can be used to filter,
assuming no wrong-sign \B decays, the state of the \B meson at $t_2$ as \Bzb and \Bz, respectively.

Initial states \Bz and \Bzb can be prepared similarly, as already discussed in Sec.~\ref{sec:BFACTORIES-entangled}.
The observation of the decay to $\ellm\Xbar$ ($\ellp\X$) at time $t_1$ dictates that the other \B meson 
at that time is in a \Bz (\Bzb) state because this state cannot decay into $\ellm\Xbar$ ($\ellp\X$). As before,
this requires assuming the absence of wrong-sign \B decays, i.e. $\Bz\to\ellm\Xbar$ and $\Bzb\to\ellp\X$ 
\Vera{do not}
occur.  
We refer 
\Refs{to}
the quantum preparation of this initial state at $t_1$ as ``\Bz (\Bzb) tag''.
The \Bz or \Bzb states prepared by entanglement at $t_1$ evolve in time until they are observed at some later 
time $t_2$ in a decaying final state filtering a given \Bz and \Bzb linear combination state, 
which we 
select
\Bplus or \Bminus by observing the appropriate decay channels.

By virtue of the EPR correlation of Eq.~(\ref{eq:BplusBminusEPR}), 
\Bplus and \Bminus have been defined as states orthogonal to the states, denoted as $\BminusT$ and $\BplusT$,
defined through the filter imposed by the observation at $t_1$ of the decay into $\jpsi\KS$ and $\jpsi\KL$, respectively,
i.e. $\langle \Bplus | \BminusT \rangle = \langle \Bminus | \BplusT \rangle = 0$.
The exchange of initial and final states with identical boundary conditions required by motion reversal 
imposes that the basis of tagging states $(\Bplus,\Bminus)$ must be identical to the 
basis of states filtered by decay $(\BplusT,\BminusT)$. This condition is met when \Bplus and \Bminus 
(and thus \BplusT and \BminusT) are orthogonal to each other.
It is then straightforward to prove that $\langle \Bminus | \Bplus \rangle = 0$ if $|\A/\Abar|=1$. 
For this to hold, the decay amplitude $\A$ should have only one weak phase,
as it is expected to apply at 1\% level or better for $\Bz\to\jpsi\Kz$ and $\Bzb\to\jpsi\Kzb$, 
and have \CPT symmetry in the decay amplitude (see Sec.~\ref{sec:BFACTORIES-cpv}).

\begin{figure*}[htb!]
\begin{center}
\includegraphics[width=0.8\textwidth]{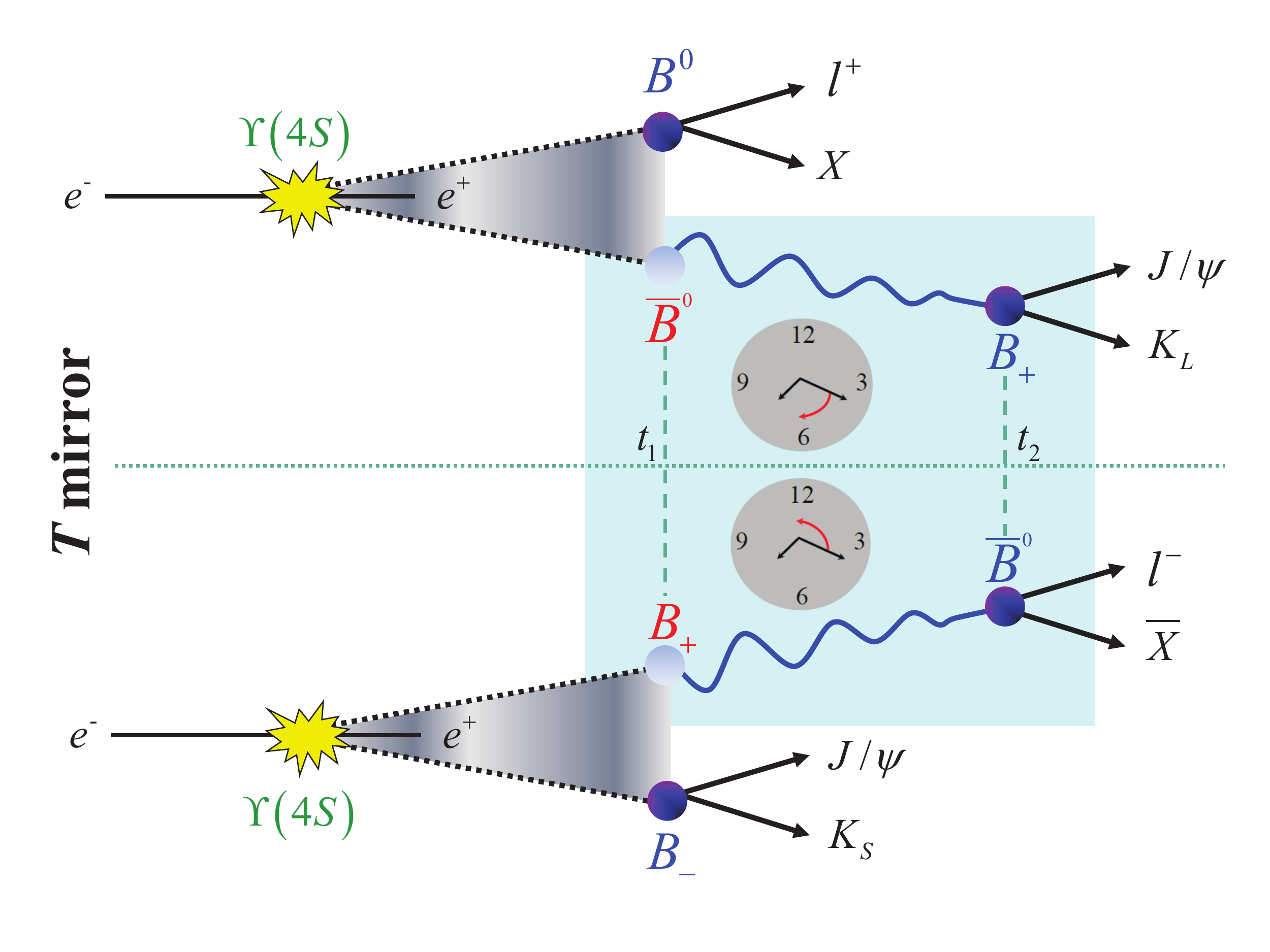}
\caption{\label{fig:method}\captionSize 
\Vera{Basic concept} 
of the time-reversal experiment. 
Electron-positron collisions at the asymmetric \B factory produce \FourS resonances, 
\Vera{which decay via}
strong interaction 
\Vera{to}
an entangled pair of \B mesons.
When one \B meson decays at $t_1$, the identity of the other is ``tagged'' without measuring it specifically. 
In the top panel, the \B meson observed to decay to the final state $\ellp\X$ at $t_1$ 
transfers information to the 
\Vera{other}
meson and dictates that it is in a \Bzb state.
This surviving meson tagged as \Bzb is \Vera{observed} 
at $t_2$
\Vera{to decay} into a final 
state $\jpsi\KL$ that filters the \B meson to be in a \Bplus state, a linear combination of \Bz and \Bzb states. 
This case corresponds to a transition $\Bzb \to \Bplus$. To study time reversal we have to compare
the rate at which this transition occurs to the rate of the time-reversed transition, $\Bplus \to \Bzb$ \Vera{(bottom panel).} 
}
\end{center}
\end{figure*}

To clarify the foundations of the time-reversal experiment, consider the case illustrated in Fig.~\ref{fig:method}.
In the top panel, the first \B decays into a $\ellp\X$ final state
(for example a semileptonic final state producing a positively-charged prompt lepton or an hadronic final state
with a \Vera{positively-charged} 
kaon). This final state filters the quantum-mechanical state of the 
other \B at time $t_1$ 
as \Bzb.
The surviving meson tagged as \Bzb is observed later at $t_2$ 
to decay into a final state $\jpsi\KL$ that filters the \B meson to be in a \Bplus state.
Therefore, the event class $(\ellp\X,\jpsi\KL)$, encapsulating a time ordering,
undergoes a transition $\Bzb \to \Bplus$ in the elapsed time $t=t_2-t_1$.
This transition has, according to the convention in Eq.~(\ref{Eq:deltat}), 
a decay time difference $\dt=t>0$.
As shown in the bottom panel, \T transformation demands to \Refs{set up} the conditions for 
the mirrored transition $\Bplus \to \Bzb$. The \Bplus tag requires
the first \B to decay into the $\jpsi\KS$ final state, whereas the surviving meson tagged at $t_1$
as \Bplus has to be observed at $t_2$ through its decay into a $\ellm\Xbar$ flavor specific final state.
The time-ordered event class is now $(\jpsi\KS,\ellm\Xbar)$,
which according to Eq.~(\ref{Eq:deltat}) yields a decay time difference $\dt<0$.
Thus and from now on, ``$\dt>0$'' (``$\dt<0$'') will refer to the time ordering in an event class 
with a flavor- (\CP-) eigenstate decay channel appearing first.
Note that this non-trivial \T transformation is not defined by the ``\dt reversal'' obtained just by 
flipping the time ordering of the decay channels for a given event class.
For this to be achieved, EPR entanglement together with the availability of both flavor- and \CP-eigenstate
decay modes to filter appropriate neutral \B states,
has been key for the quantum preparation of the \B-meson initial state in the
transition and its motion-reversed version; the problem of particle instability has been thus 
\Vera{avoided.}

\subsection{Transitions among \B meson states}  
\label{sec:CONCEPT-transitions}

\Vera{For}
the four initial states \Bz, \Bzb, \Bplus, and \Bminus 
\Vera{prepared}
by entanglement, 
and the four final states \Bz, \Bzb, \Bplus, and \Bminus 
\Vera{available through}
decay, 
it is possible to
construct 
\Vera{eight} transitions and their corresponding time-ordered event classes $(f_1,f_2)$
given in Table~\ref{TAB:TRV-EVTCLASSESvsTRANSITIONS}~\cite{Bernabeu:2012ab}. 
In this notation, the final state $f_1$ is observed at 
time $t_1$ and the final state $f_2$ is observed at time $t_2 = t_1 + t$, where $t>0$ is the elapsed time. 
The matching between event classes and transitions applies under the assumption of absence of wrong-strangeness and wrong-sign \B decays,
\CP invariance in \Kz-\Kzb mixing, and $|\Abar/\A|=1$. 
Recently an extended discussion, including wrong-strangeness and wrong-sign \B decays, has been presented~\cite{Applebaum:2013wxa}.

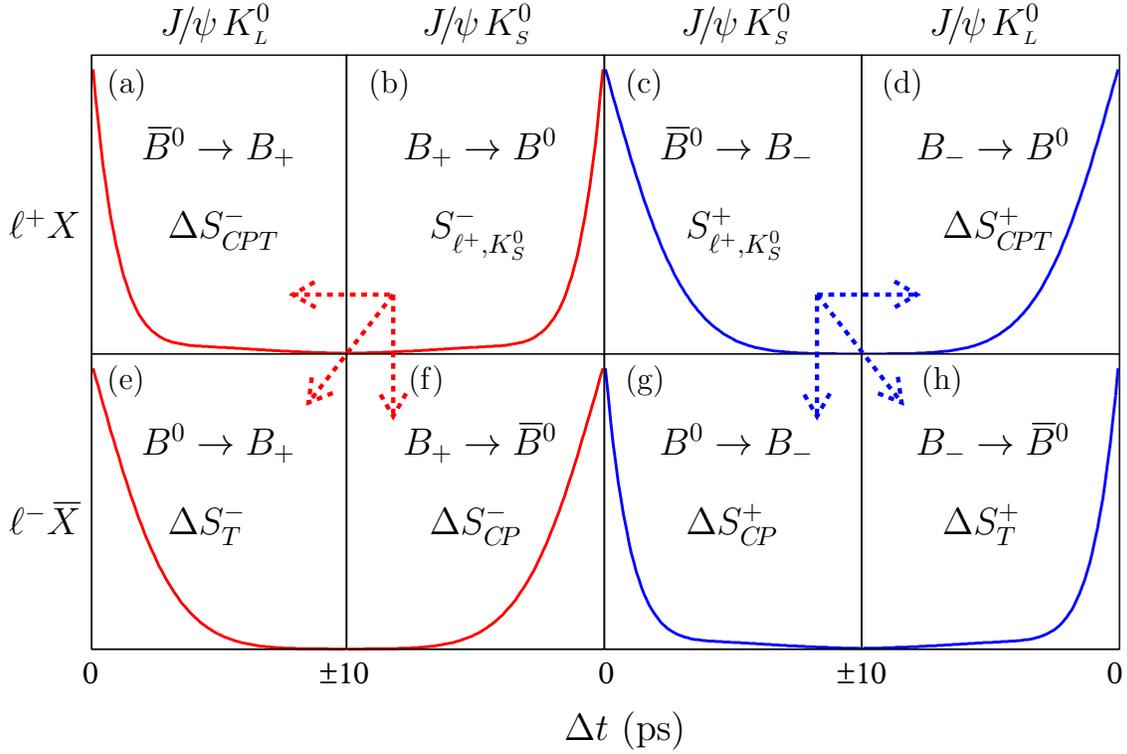
\begin{figure*}[htb!]
\begin{center}
 \includegraphics[width=0.8\textwidth]{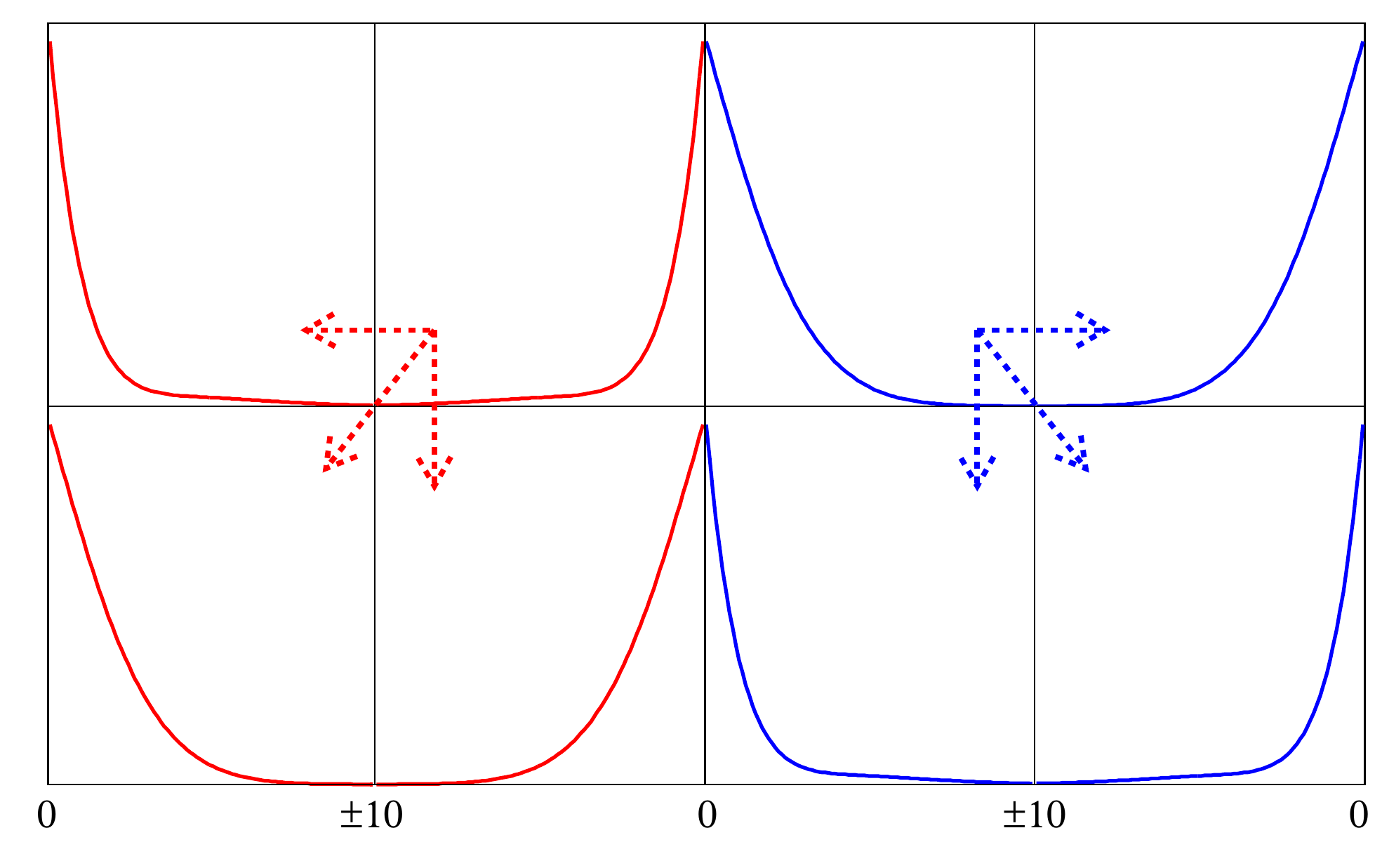}
\put(-215,-10){\Large $\dt$ (ps)}
\put(-388,235){\large (a)}
\put(-375,210){\Large $\Bzb \to \Bplus$}
\put(-365,180){\Large \Red{$\Delta S^-_{\CPT}$}}
\put(-289,235){\large (b)}
\put(-276,210){\Large $\Bplus \to \Bz$}
\put(-266,180){\Large \Red{$S^-_{\ellp,\KS}$}}
\put(-192,235){\large (c)}
\put(-179,210){\Large $\Bzb \to \Bminus$}
\put(-169,180){\Large \Blue{$S^+_{\ellp,\KS}$}}
\put(-95,235){\large (d)}
\put(-82,210){\Large $\Bminus \to \Bz$}
\put(-72,180){\Large \Blue{$\Delta S^+_{\CPT}$}}
\put(-388,123){\large (e)}
\put(-375,98){\Large $\Bz \to \Bplus$}
\put(-365,68){\Large \Red{$\Delta S^-_{\T}$}}
\put(-274,123){\large (f)}
\put(-276,98){\Large $\Bplus \to \Bzb$}
\put(-266,68){\Large \Red{$\Delta S^-_{\CP}$}}
\put(-192,123){\large (g)}
\put(-179,98){\Large $\Bz \to \Bminus$}
\put(-169,68){\Large \Blue{$\Delta S^+_{\CP}$}}
\put(-80,123){\large (h)}
\put(-82,98){\Large $\Bminus \to \Bzb$}
\put(-72,68){\Large \Blue{$\Delta S^+_{\T}$}}
\put(-368, 257){\Large $\jpsi\KL$}
\put(-269, 257){\Large $\jpsi\KS$}
\put(-171, 257){\Large $\jpsi\KS$}
\put(-77, 257){\Large $\jpsi\KL$}
\put(-425,180){\Large $\ellp\X$}
\put(-425,68){\Large $\ellm\Xbar$}
\caption{\label{fig:processes}\captionSize 
\Refs{Expected time-dependent probability distributions for the eight time-ordered event classes given in 
Table~\ref{TAB:TRV-EVTCLASSESvsTRANSITIONS}, shown as a planar map. They are identified by the flavor-specific 
($\ellm \Xbar$, $\ellp \X$) and \CP-eigenstate ($\jpsi\KS$, $\jpsi\KL$) decay products and the time ordering.
The dashed arrows in the left (red) and right (blue) panels indicate the time-ordered event classes connected by the three symmetries, 
\CP (vertical), \T (oblique), and \CPT (horizontal), independently of each other. 
The two panels (red, blue) of $t$-reverse decay channels are unconnected by the symmetries.
}
}
\end{center}
\end{figure*}

\Vera{The observation} of an asymmetry between the 
\Auth{probabilities}
for transitions on the 
right and left panels of Table~\ref{TAB:TRV-EVTCLASSESvsTRANSITIONS},
\begin{eqnarray} 
|\langle\Bzb|{\cal U}(t)|\Bminus\rangle|^2  - |\langle\Bminus|{\cal U}(t)|\Bzb\rangle|^2,~~ & \nn\\ 
|\langle\Bminus|{\cal U}(t)|\Bz\rangle|^2  - |\langle\Bz|{\cal U}(t)|\Bminus\rangle|^2,~~ & \nn\\ 
|\langle\Bzb|{\cal U}(t)|\Bplus\rangle|^2  - |\langle\Bplus|{\cal U}(t)|\Bzb\rangle|^2,~~ & \nn\\ 
|\langle\Bplus|{\cal U}(t)|\Bz\rangle|^2  - |\langle\Bz|{\cal U}(t)|\Bplus\rangle|^2,~~ &  
\label{eq:Tcomparisons}
\end{eqnarray}
\Vera{will}
be an unambiguous demonstration of motion reversal in time evolution
of states
that are not \CP conjugate to each other.
\Auth{Here, ${\cal U}(t)$ is the time-evolution operator determined by the effective Hamiltonian ${\cal H}_{\rm eff}$.
Note that the notation has been changed in comparison to Sec.~\ref{sec:TRINPHYSICS-cm-qm} since $t$ denotes now
the elapsed time.}
We immediately realize that each of the four independent time-reversal comparisons in Table~\ref{TAB:TRV-EVTCLASSESvsTRANSITIONS}
uses a pair of time-ordered event classes involving four different final states at different times, 
$\ellp\X$ and $\ellm\Xbar$ at times $t_1$ (or $t_2$) and $t_2$ (or $t_1$),
and $\jpsi\KS$ and $\jpsi\KL$ at times $t_2$ (or $t_1$) and $t_1$ (or $t_2$), respectively.
Therefore, 
\Vera{time-reversal tests require}
the comparison of 
\Auth{probabilities}
for event classes with opposite 
time ordering (opposite \dt sign) and different observed
flavor- and \CP-eigenstate final states, as illustrated in Fig.~\ref{fig:processes}~\cite{Bernabeu:2012ab}.

With the same approximations, differences like
\bea
|\langle\Bz|{\cal U}(t)|\Bminus\rangle|^2  - |\langle\Bminus|{\cal U}(t)|\Bzb\rangle|^2,
\eea
or
\bea
|\langle\Bminus|{\cal U}(t)|\Bz\rangle|^2  - |\langle\Bminus|{\cal U}(t)|\Bzb\rangle|^2,
\eea
probe \CPT and \CP symmetry, respectively, in the time evolution of \Bz-\Bzb transitions.
For each of these cases there are also four independent asymmetries, although clearly not all 
comparisons involving time reversal, \CP, and \CPT transformations are independent.
Whereas \CP transformation requires the comparison 
\Vera{of}
pairs of time-ordered event classes
with different flavor-specific final states, but common \CP-eigenstate final state and same \dt sign,
\CPT demands a common flavor-specific final state, different \CP-eigenstate final states, and opposite \dt sign
(see Fig.~\ref{fig:processes})~\cite{Bernabeu:2012ab}.
Therefore, the time-reversal case 
\Vera{is}
the most challenging.

\begin{table}[htb!]       
\renewcommand{\arraystretch}{1.4}
\begin{center}
\caption{\label{TAB:TRV-EVTCLASSESvsTRANSITIONS}\captionSize \Auth{Time-ordered event classes $(f_1,f_2)$} and 
their corresponding transitions between \B meson states.
The matching between event classes and transitions applies under the assumption of absence of wrong-strangeness and wrong-sign \B decays, 
a single weak amplitude, i.e. $|\Abar/\A|=1$, and no \CP violation in \Kz-\Kzb mixing.
The event classes and transitions on the right panel \Vera{(Time reversed)} are the time-reversed versions of those in the left panel \Vera{(Reference).}
}
       \begin{tabular}{ cc  cc } \hline \hline	
\multicolumn{2}{c}{Reference} & \multicolumn{2}{c}{Time reversed} \\ \hline
         Event class & Transition\phantom{+} & Event class & Transition\phantom{+} \\ \hline
 $(\ellp\X,\jpsi\KL)$ & $\Bzb \to \B_+$  &  $(\jpsi\KS,\ellm\Xbar)$ & $\B_+\to \Bzb$ \\
 $(\jpsi\KS,\ellp\X)$ & $\B_+\to \Bz$    &  $(\ellm\Xbar,\jpsi\KL)$ & $\Bz \to \B_+$ \\ 
 $(\ellp\X,\jpsi\KS)$ & $\Bzb \to \B_-$  &  $(\jpsi\KL,\ellm\Xbar)$ & $\B_-\to \Bzb$ \\
 $(\jpsi\KL,\ellp\X)$ & $\B_-\to \Bz$    &  $(\ellm\Xbar,\jpsi\KS)$ & $\Bz \to \B_-$  \\ [0.05in] \hline \hline
       \end{tabular}
    \end{center}	
  \end{table}

\section{\babar\ time-reversal analysis}  
\label{sec:ANALYSIS}

\Vera{The \babar\ detector} recorded an integrated luminosity of 
518~\invfb of data, of which 424~\invfb were taken at a \cms energy corresponding to the 
\Vera{mass of the}
\FourS resonance, 
28 and 13.6~\invfb around the \ThreeS and \TwoS resonances (10.36 and 10.02~\gev), 4~\invfb above the \FourS,
and 48~\invfb below the resonances [44~\invfb at a \cms energy 40~\mev below the \FourS]~\cite{TheBABAR:2013jta}. 
These luminosities correspond to about $470\times 10^6$~\BB, $690\times 10^6$~\ccbar, 
and $500\times 10^6$~$\tau^+\tau^-$ pairs, and $121\times 10^6$~\ThreeS and $99\times 10^6$~\TwoS resonances.
For the time-reversal analysis~\cite{Lees:2012kn} \babar\ used all \BB and \FourS off-resonance data.

\begin{figure*}[htb!]
\begin{center}
 \includegraphics[width=0.75\textwidth]{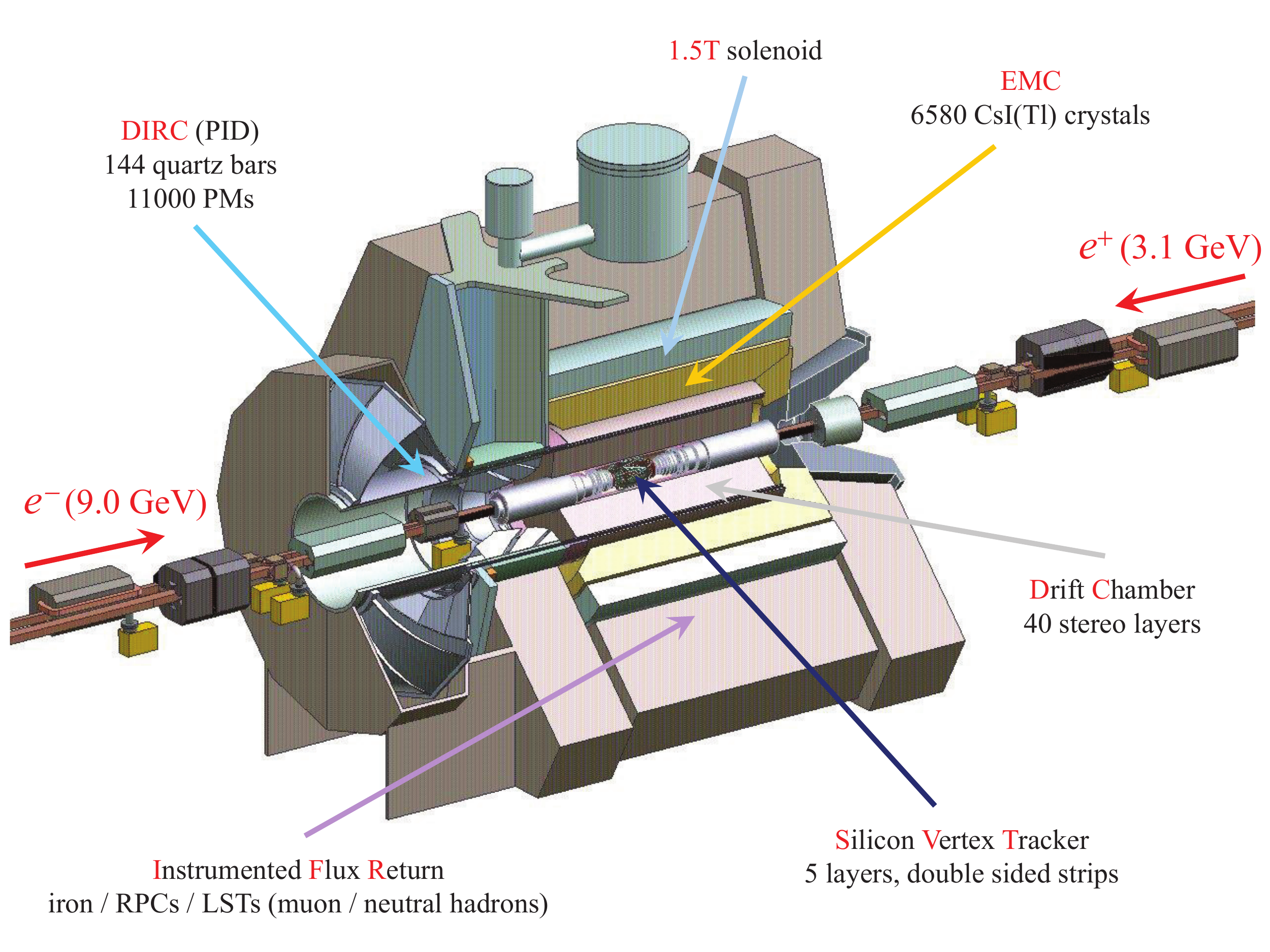}
\caption{\label{fig:babardetector}\captionSize A sketch of the \babar\ detector. The five subsystems and the solenoid
magnet providing the 1.5 T magnetic field are indicated. 
Source: \babar\ Collaboration.
}
\end{center}
\end{figure*}

\begin{figure}[htb!]
\begin{center}
 \includegraphics[width=0.48\textwidth]{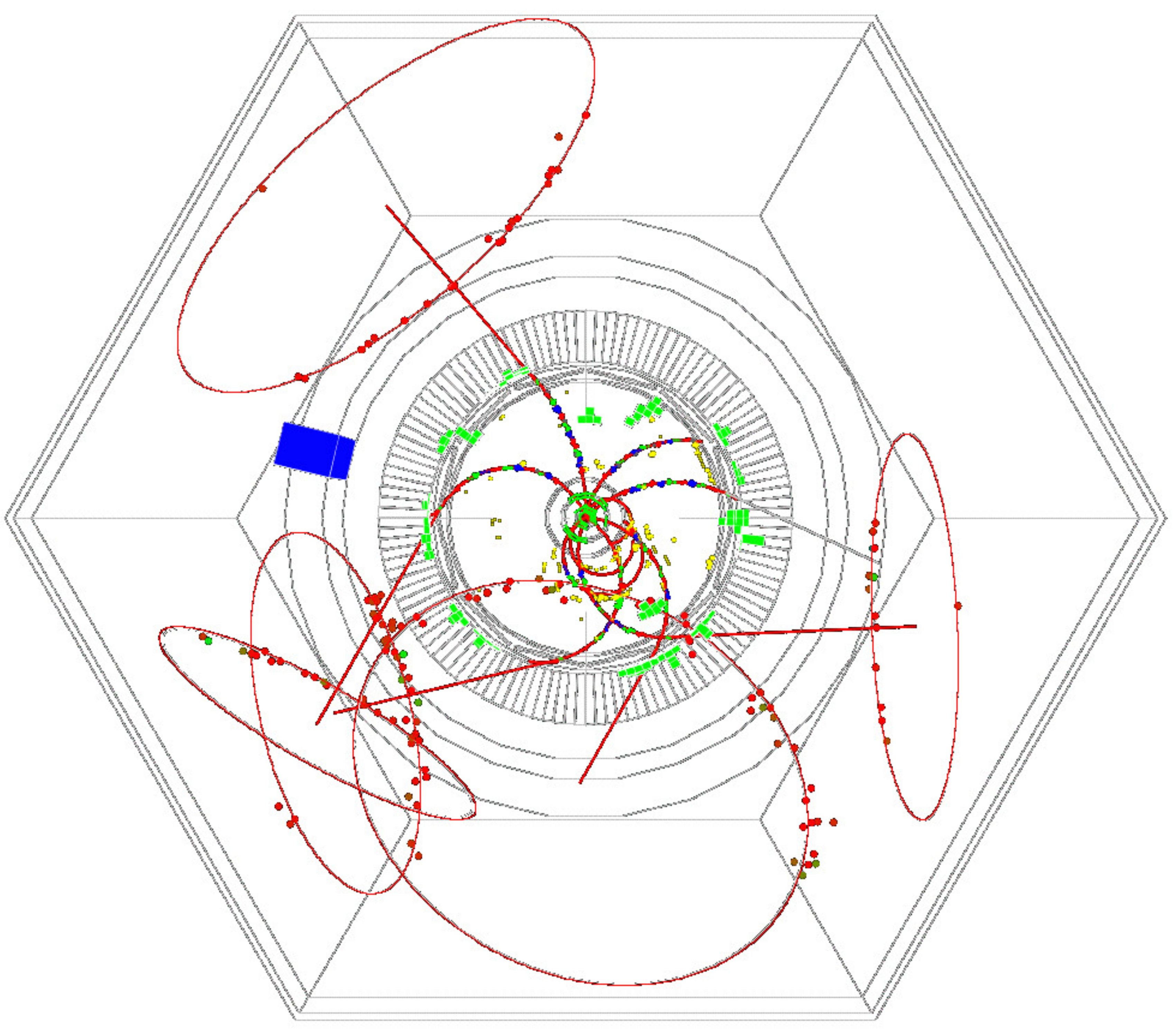}
\caption{\label{fig:eventdisplay}\captionSize Transverse view of the computer reconstruction of a \BB event in the \babar\ detector. 
The display shows the decay products of the two \B mesons as curved tracks (in red) in the central region of the detector (SVT and DC). 
Some particles 
\Vera{deposit} energy in the EMC calorimeter (green and blue blocks). Particle identification 
is assisted by measuring the Cherenkov rings \Vera{(DIRC).}
Source: \babar\ Collaboration.}
\end{center}
\end{figure}

The \babar\ detector, sketched in Fig.~\ref{fig:babardetector}, 
\Vera{was}
designed as a general purpose detector 
for $e^+e^-$ annihilation physics~\cite{Aubert:2001tu}. 
Surrounding the interaction point is a five-layer, double-sided Silicon Vertex Tracker (SVT), which measures the
\Vera{angles and}
impact parameters of charged particle \Vera{tracks.}
A 40-layer Drift Chamber (DC) surrounds the SVT and provides measurements of the momenta for charged particles.
Charged hadron identification is achieved through measurements of 
\Vera{energy-loss} in the tracking system
and the Cherenkov angle obtained from a Detector of Internally Reflected Cherenkov light (DIRC). 
A CsI(Tl) electromagnetic calorimeter (EMC) provides photon detection, electron identification, and $\piz$ reconstruction. 
These components are 
\Vera{inserted inside}
a solenoid magnet, which provides a 1.5 T magnetic field. 
\Vera{The flux} return of the magnet (IFR) is instrumented with resistive plate chambers and
limited streamer tubes in order to 
\Vera{detect}
muons and 
\Vera{$\KL$s~\cite{TheBABAR:2013jta}.} 
Figure~\ref{fig:eventdisplay} illustrates the transverse view of the computer reconstruction of a \BB event.
The detector 
\Vera{performance}
has been very stable over the nine years of operation, 
\Vera{supporting a broad flavor physics program.}

Although the \CP-violation and time-reversal analysis concepts are fundamentally different, the 
experimental analyses 
\Vera{are} closely connected to each other. Reconstruction algorithms, 
event selection criteria, calibration techniques and descriptions of the background composition
are common~\cite{Aubert:2009aw}. However, 
\Vera{the selected signal events for the time-reversal measurement require somewhat different treatment.}

\subsection{Event reconstruction and selection}  
\label{sec:ANALYSIS-selection}

In addition to $\jpsi\KS$, $\psi(2S)\KS$ and $\chi_{c1}\KS$ decay modes are also considered; all these three final states
have \CP-odd parity and are hereafter denoted generically as $\ccbar\KS$.
The $\ccbar$ particle states are reconstructed in the decay channels $\jpsi,\psi(2S)\to e^+ e^-, \mu^+ \mu^-$, $\psi(2S)\to\jpsi\pip\pim$,
and $\chi_{c1}\to\jpsi\gamma$.
Whereas \KS mesons decay near the collision point and are reconstructed through their decays into $\pip\pim$ and $\piz\piz$
(the latter only for $\jpsi\KS$), the long-living \KL mesons pass through the tracking systems undetected and 
are reconstructed 
\Vera{by}
their hadronic interactions in the EMC and/or the IFR and the
\Vera{\jpsi} in the \B decay. 
The background rejection relies on vetoes to specific \B decay modes, and on angular and event shape variables
to suppress continuum events arising from $e^+e^- \to \qqbar$, $\q=\u,\d,\s$ reactions.
These variables exploit the different topology of $\qqbar$ and \BB events, jet-like for the former and spherical for the latter, 
consequence of the large mass difference between light and \b quarks.
 
\begin{figure}[htb!]
\begin{center}
\begin{tabular}{cc}
 \includegraphics[width=0.23\textwidth]{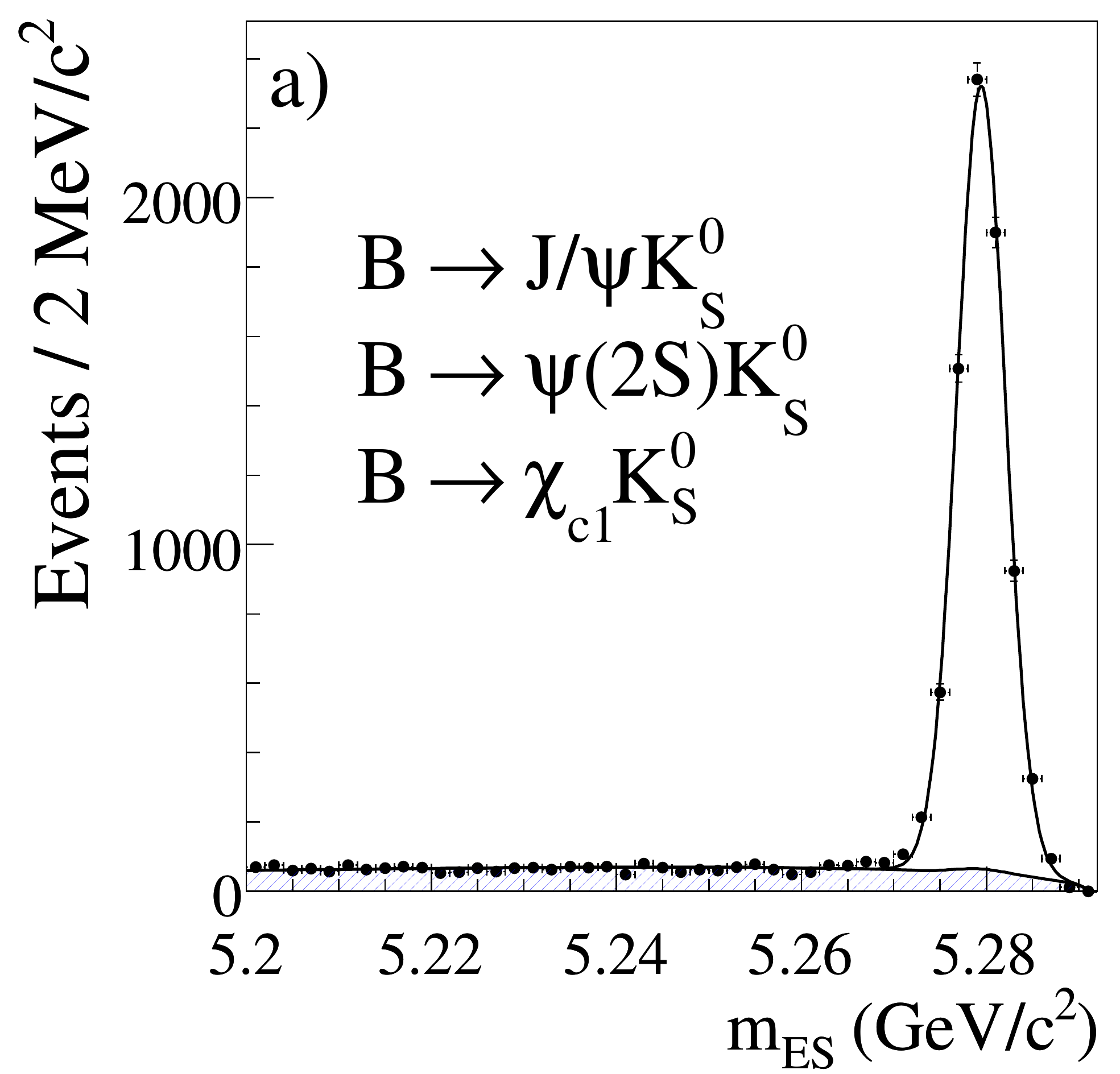} &
 \includegraphics[width=0.23\textwidth]{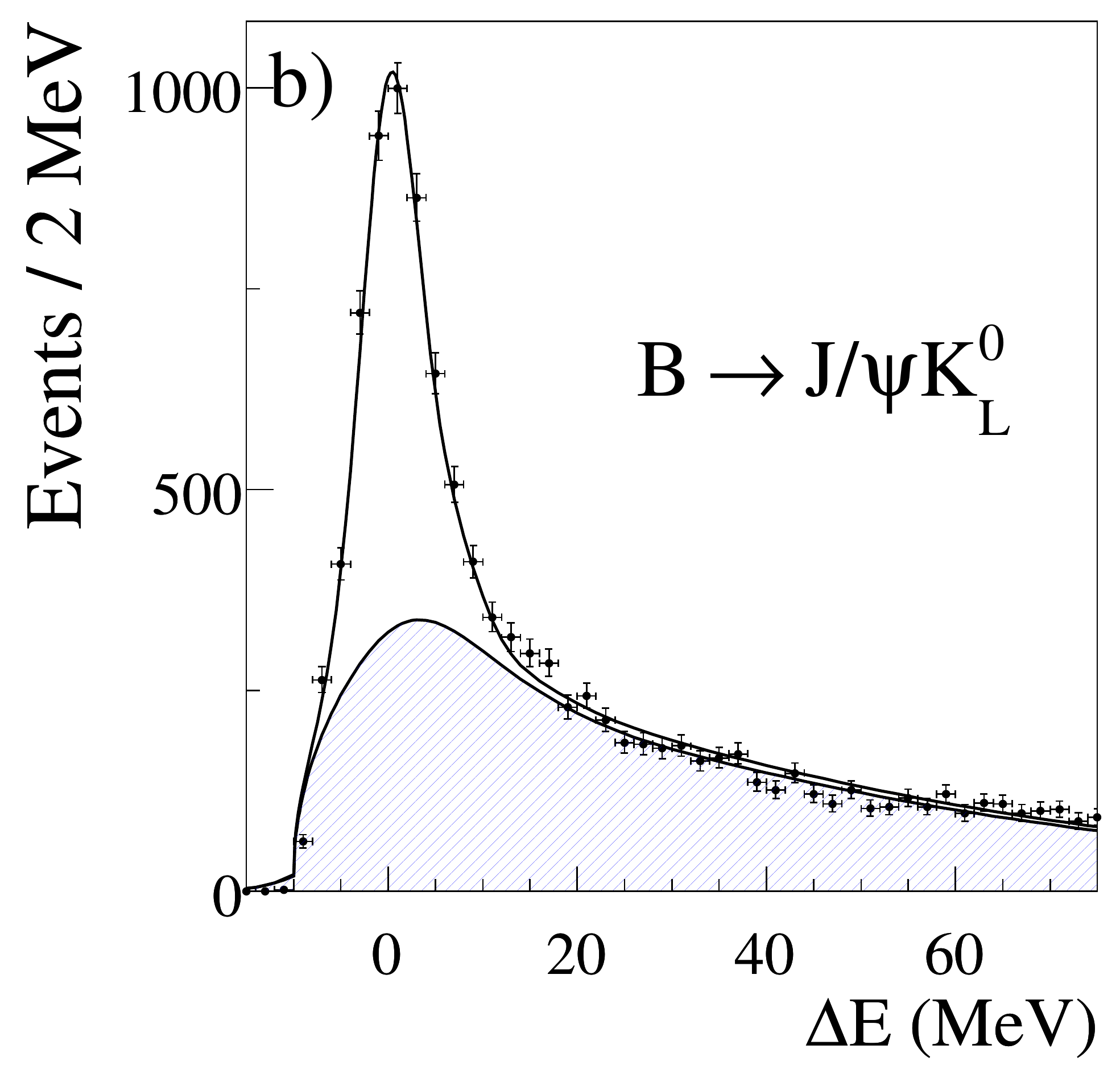} \\
\end{tabular}
\caption{\label{fig:sample}\captionSize (a) \mes and (b) \de distributions for the final sample of neutral \B decays reconstructed in 
the $\ccbar\KS$ and $\jpsi\KL$ final states, respectively. The shaded regions represent the estimated background contributions.}
\end{center}
\end{figure}

\B meson candidates decaying to $\jpsi\KL$ are characterized by the energy difference
$\de = E_\B^* - E_{\rm beam}^*$ between the \B energy and the beam energy in the $e^+e^-$ \cms frame,
\Vera{while for the}
$\ccbar\KS$ final states 
the beam-energy substituted mass, 
$\mes = \sqrt{(E_{\rm beam}^*)^2-|{\mathbf p}_\B^*|^2}$, \Vera{is used,}
where ${\mathbf p}_\B^*$ is the \B momentum in the \cms frame.
These two kinematic variables are based on the fact that \B mesons are produced almost at rest in \cms frame 
and the beam energies are precisely known. 
Figure~\ref{fig:sample} shows the \mes and \de data distributions for the final sample
\Vera{of about}
7800 $\ccbar\KS$ signal events with purities ranging from 87 to 96\%, and 
5800 $\jpsi\KL$ signal events with purities around 60\%.

Since the mass of the \FourS is only slightly higher than twice the mass of the \B meson, no additional particles are \Vera{produced.}
Thus, the flavor identity of the \B meson
not associated with the reconstructed $\ccbar\KS$ or $\jpsi\KL$ final state
is 
\Vera{determined}
from the remaining particles in the event,
on the basis of the charges of prompt leptons and kaons, pions from $\Dstarp\to\Dz\pip$ decays, and high-momentum
charged particles. The notation $\ellm\Xbar$ ($\ellp\X$) introduced previously denotes all 
these inclusive final states that identify the flavor of the \B as \Bzb (\Bz). 
In practice, these flavor identity signatures are combined using a neural network 
\Refs{whose}
output is used
to divide the events into six hierarchical, mutually exclusive flavor categories of increasing misidentification probability.
\Vera{The proper} time difference between the decay of the 
two \B mesons, \dt, 
\Vera{is measured by the}
separation of the two decay vertices along the $e^+e^-$ collision axis and the known
boost, as discussed in Sec.~\ref{sec:BFACTORIES}.

Besides the \CP eigenstate final states, 
other
high-statistics, flavor-specific final states like $\Bz\to\D^{*-}\pip$ and $\Bz\to\jpsi\Kstarz[\to\Kp\pim]$ 
(and their \CP conjugates)
are reconstructed, 
\Vera{and}
are used for calibrating 
the \dt
resolution and the 
flavor misidentification probability.
Charged \B meson decays like $\Bpm\to\jpsi\Kpm,\psi(2S)\Kpm,\jpsi\Kstarpm$ are likewise detected and used for 
\Vera{systematic}
checks.

\subsection{Signal data treatment and results}  
\label{sec:ANALYSIS-fitting}

Assuming $\dGd =0$, the time dependence of each of the eight transitions depicted in Fig.~\ref{fig:processes} can be
parameterized in a model-independent way as
\begin{eqnarray}
\frac{g^\pm_{\alpha,\beta}(t)}{e^{-\Gamma_d t}} \propto 
1+S^\pm_{\alpha,\beta}\sin(\deltamd t) + C^\pm_{\alpha,\beta}\cos(\deltamd t), 
\label{eq:Gparam2}
\end{eqnarray} 
where the lower indices $\alpha=\ellp, \ellm$ and $\beta = \KS, \KL$ stand for the final reconstructed decay modes
$\ellp\X,~\ellm\Xbar$ and $\ccbar\KS,~\jpsi\KL$, respectively,  
and the upper indices encapsulate the time ordering, ($+$) for \Bz or \Bzb tagged states and ($-$) for \Bplus or \Bminus tagged states.
This expression is analogous to Eq.~(\ref{eq:Gparam}) with the distinction that is associated to eight pairs 
$(S^\pm_{\alpha,\beta},C^\pm_{\alpha,\beta})$
instead of only one, and $t>0$.

\begin{figure*}[htb!]
\begin{center}
\begin{tabular}{ccc}
 \includegraphics[width=0.25\textwidth]{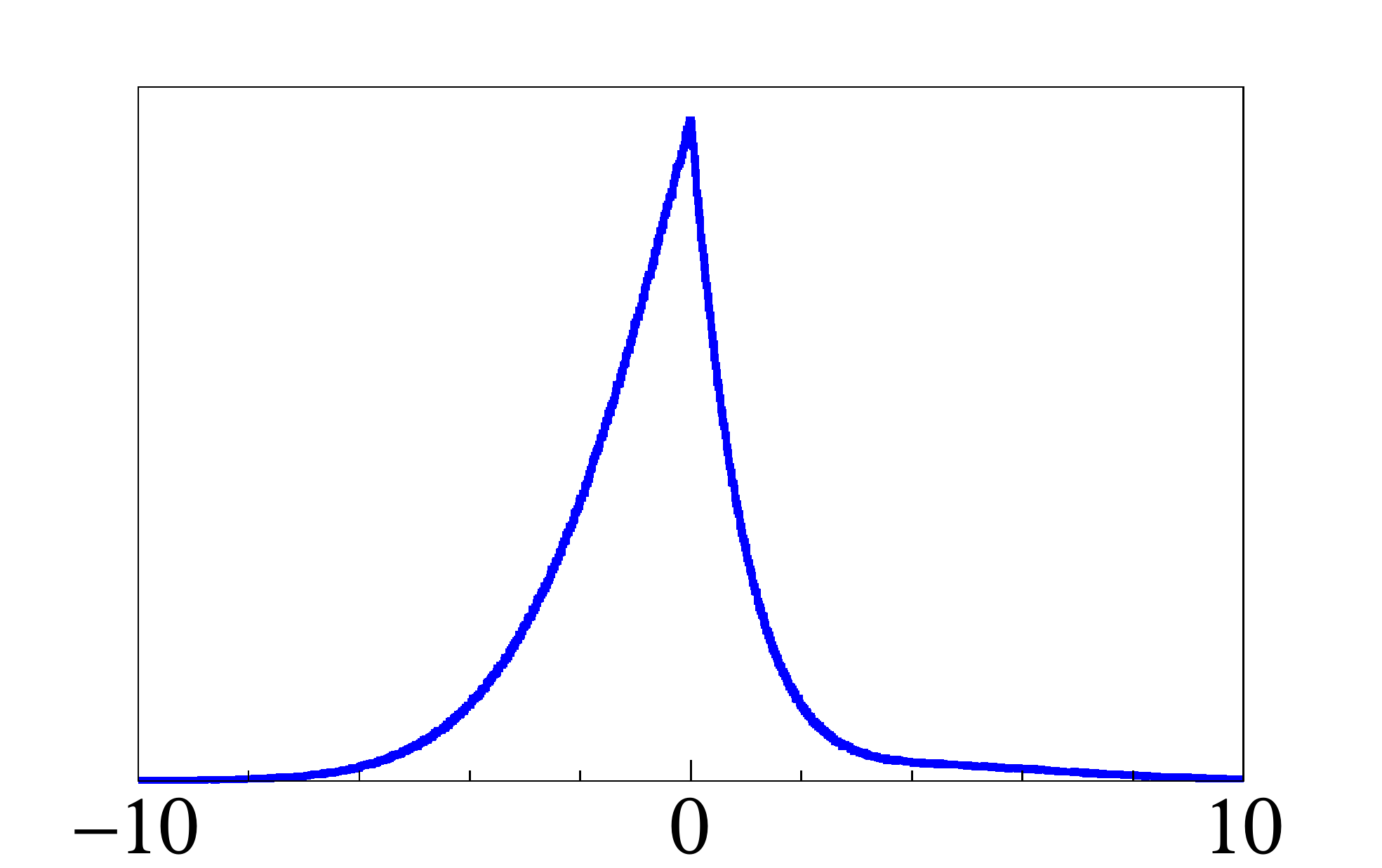} &
 \includegraphics[width=0.25\textwidth]{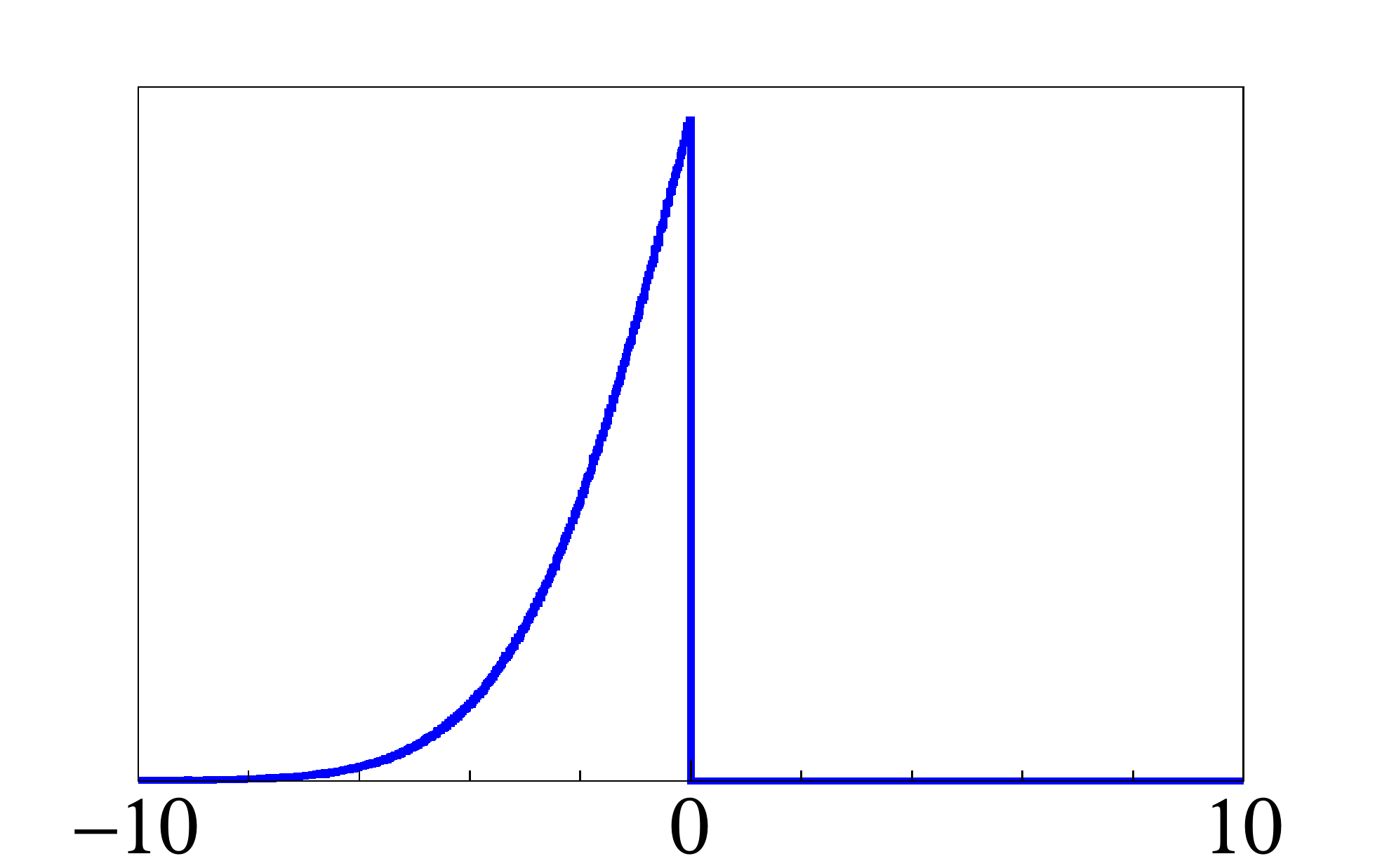} \put(-93,76){\Large $\Bminus \to \Bz$} &
 \includegraphics[width=0.25\textwidth]{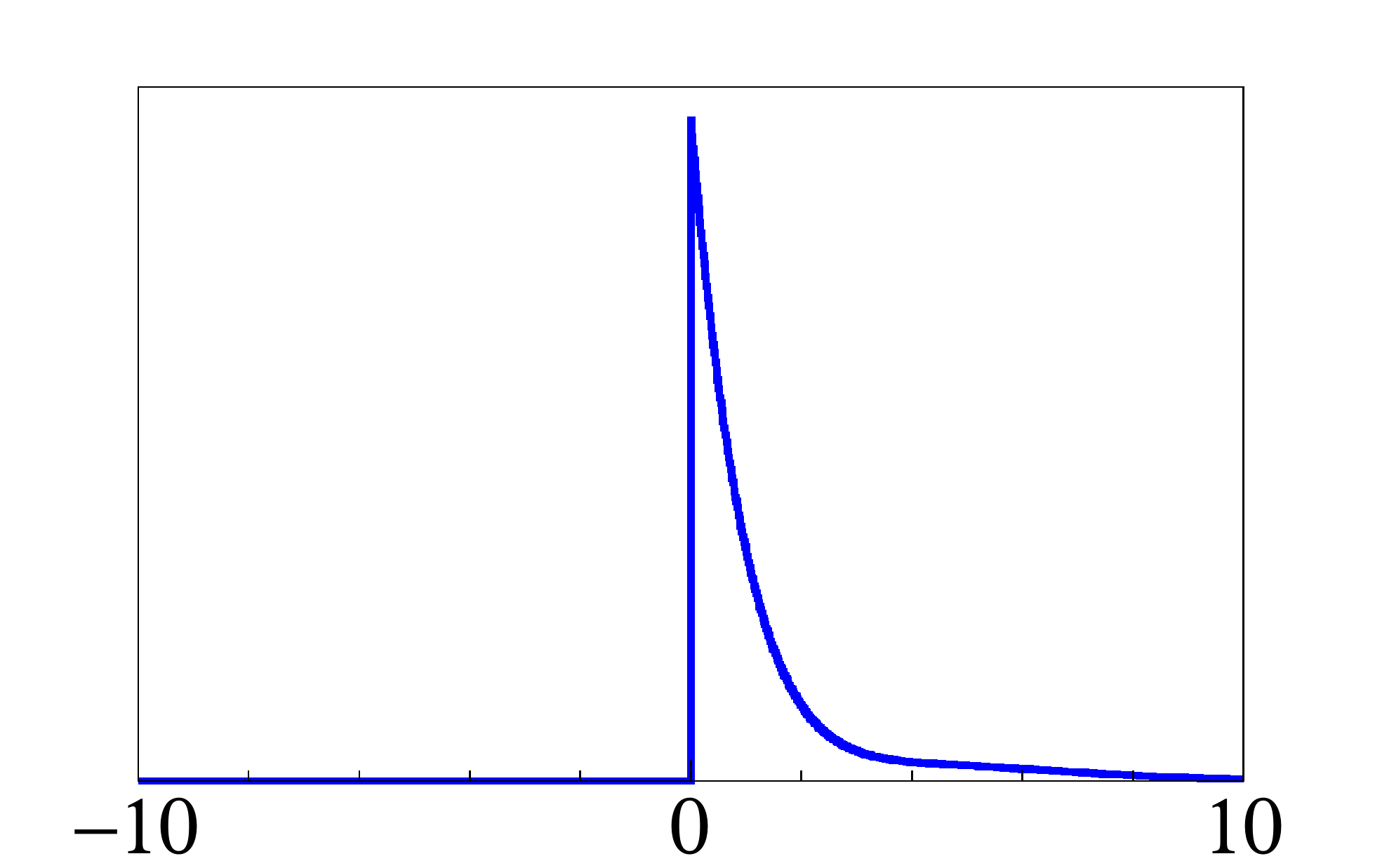} \put(-93,76){\Large $\Bzb \to \Bplus$}\\
 \includegraphics[width=0.25\textwidth]{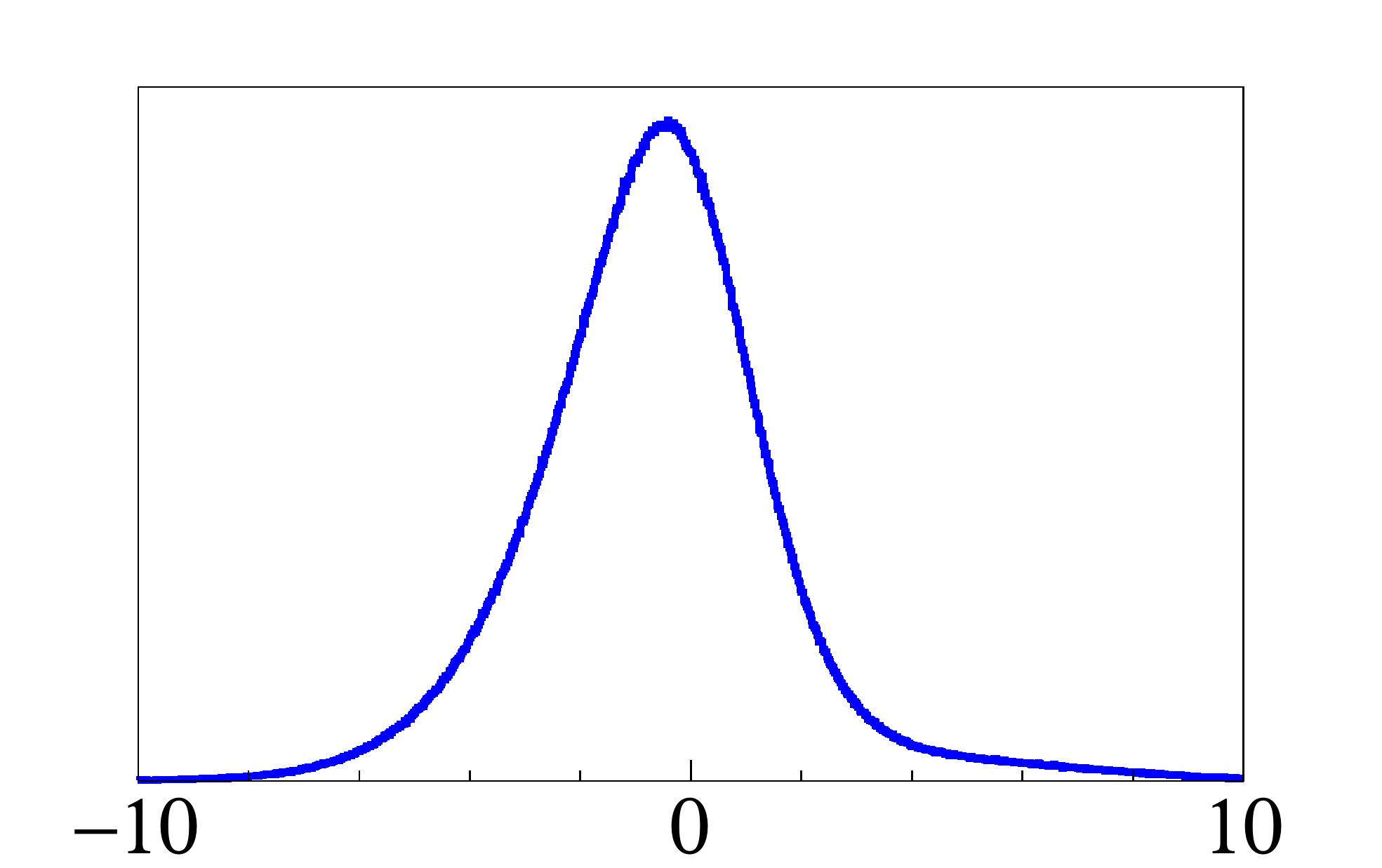} &
 \includegraphics[width=0.25\textwidth]{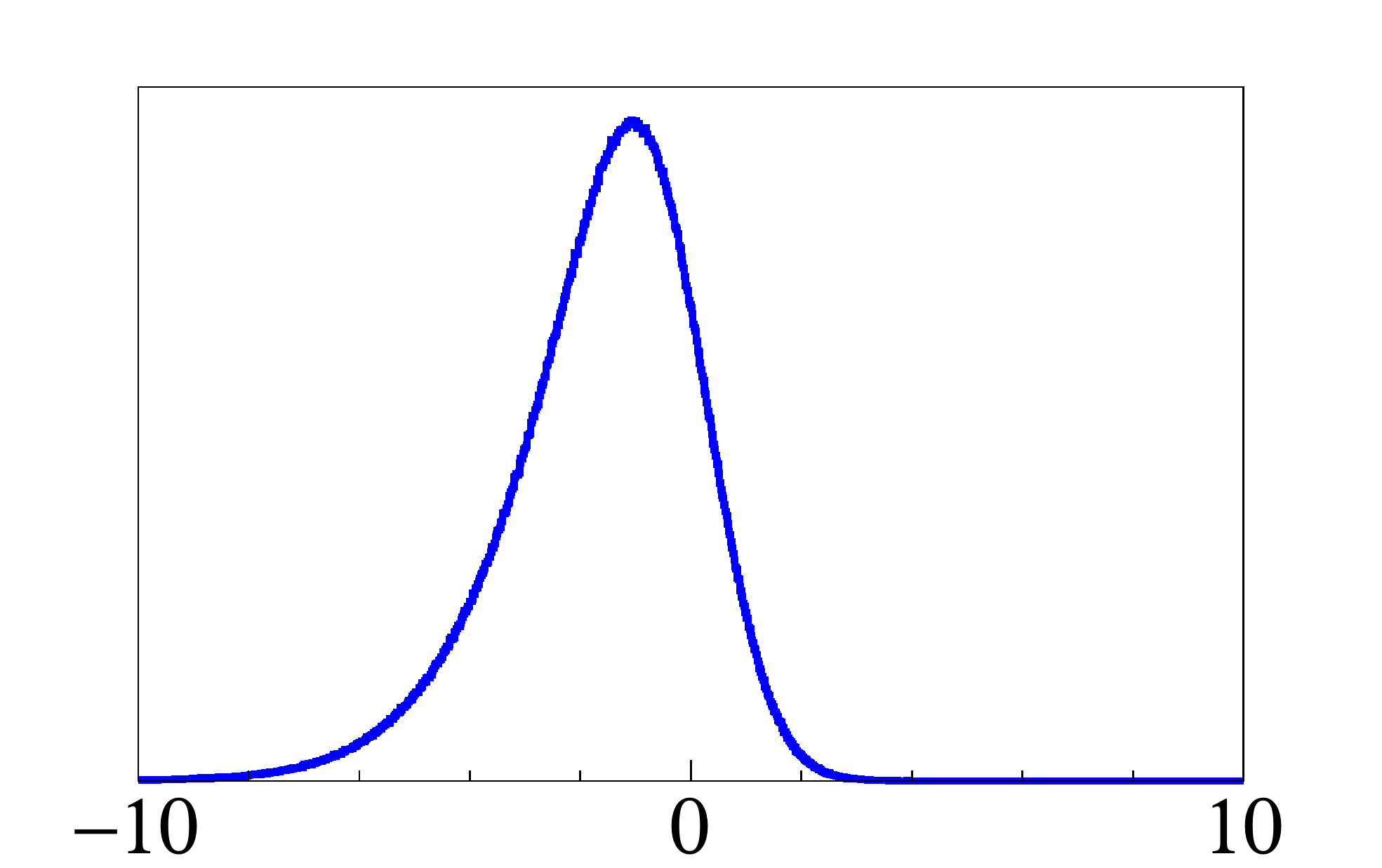} &
 \includegraphics[width=0.25\textwidth]{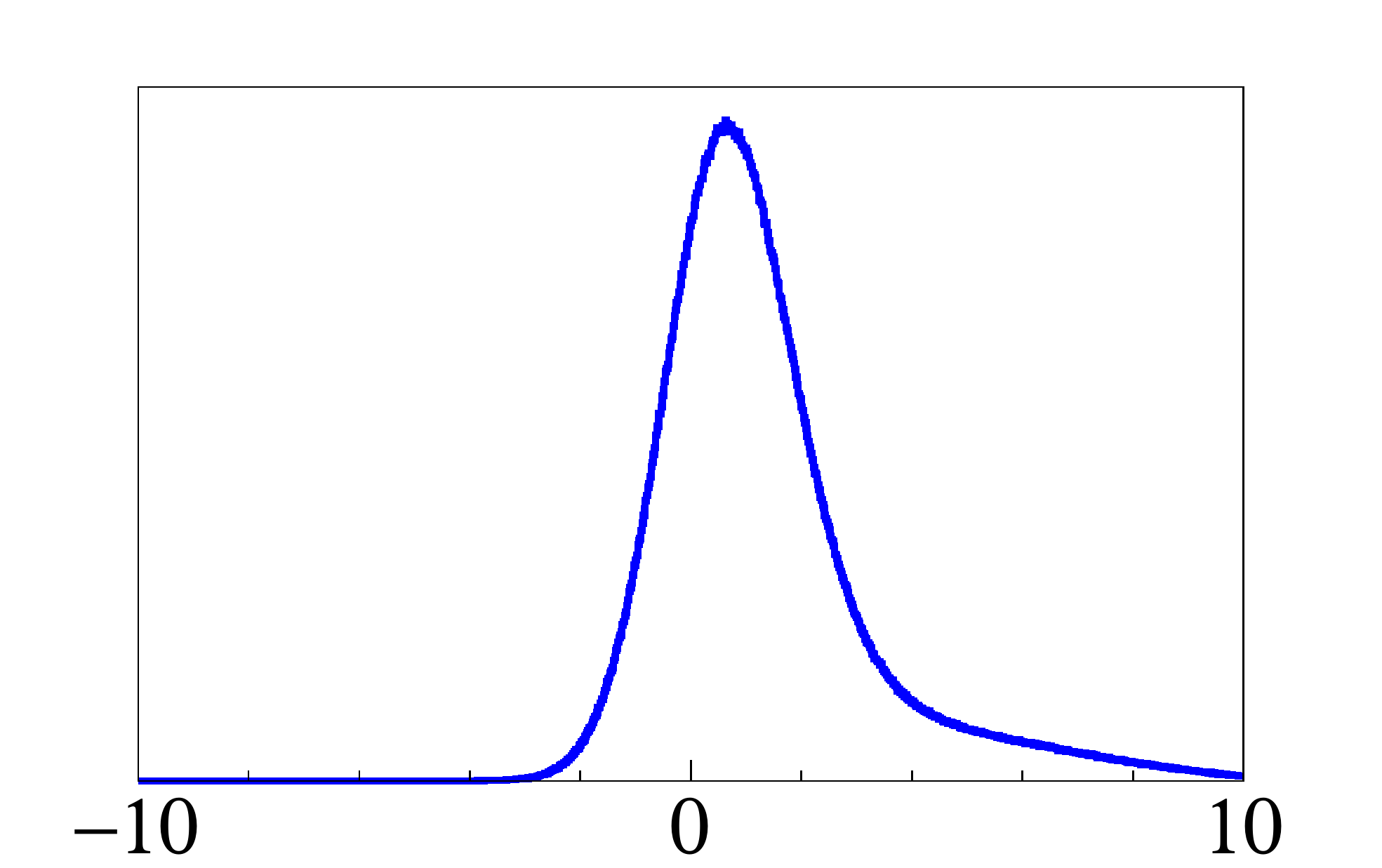} \\
\end{tabular}
 \put(-220,-95){\Large \dt (ps)}
 \put(-400,10){\begin{rotate}{90}{\large Perfect}\end{rotate}}
 \put(-390,10){\begin{rotate}{90}{\large resolution}\end{rotate}}
 \put(-400,-75){\begin{rotate}{90}{\large Detector}\end{rotate}}
 \put(-390,-75){\begin{rotate}{90}{\large resolution}\end{rotate}}
 \put(-269, 35){\LARGE =} \put(-138, 35){\LARGE +} 
 \put(-269,-50){\LARGE =} \put(-138,-50){\LARGE +} 
\caption{\label{fig:unfolding}\captionSize 
\Vera{Unfolding of the measured \dt distributions:} 
the reconstructed \dt distribution (bottom left) 
\Vera{is}
unfolded to disentangle the
\dt resolution function and the true \dt distribution (top left) used to define 
the \Bplusminus (true $\dt<0$, top middle) and flavor (true $\dt>0$, top right) tags.
The \dt resolution mixes true \Bplusminus or flavor tags with fake flavor or \Bplusminus tags
(bottom middle and bottom right), respectively.
The case shown here corresponds to a signal sample in which the two \B mesons have been reconstructed in the decay modes 
$\jpsi\KL$ and $\ellp\X$. The event classes \Vera{are} 
$(\jpsi\KL,\ellp\X)$ and $(\ellp\X,\jpsi\KL)$, 
corresponding to transitions $\Bminus \to \Bz$ and $\Bzb\to\Bplus$, respectively.
}
\end{center}
\end{figure*}
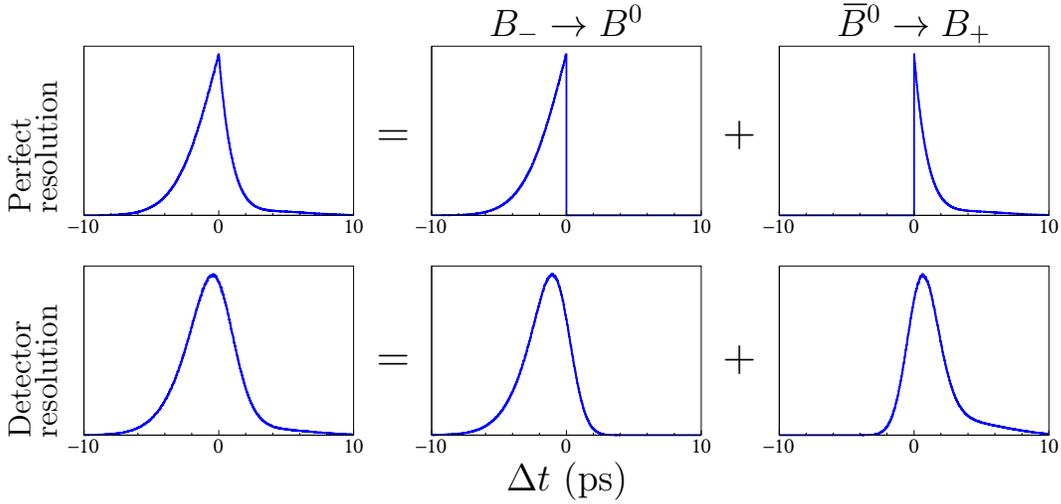

The pairs \Vera{of parameters} $(S^\pm_{\alpha,\beta},C^\pm_{\alpha,\beta})$ are determined
by a maximum likelihood fit to the measured \dt distributions of the four signal samples 
in which one \B meson is reconstructed in a $\ccbar\KS$ or $\jpsi\KL$ decay mode,
and the flavor content of the other \B is identified through a $\ellm\Xbar$ or $\ellp\X$ decay mode. 
Neglecting time resolution, the elapsed time between the first and second decay is $t=\deltat$
for flavor tags, and $t=-\deltat$ for \Bplusminus tags.
As illustrated in Fig.~\ref{fig:unfolding}, time resolution mixes events with positive and negative 
true \dt, i.e. a true event class $(\ellp\X,\jpsi\KL)$,
corresponding to a $\Bzb\to\Bplus$ transition, 
might appear reconstructed as $(\jpsi\KL,\ellp\X)$, corresponding to a $\Bminus\to\Bz$ transition,
and vice versa. 
To determine separately the coefficients for event classes with true positive \dt (flavor tag) or true negative \dt (\Bplusminus tag),
it is necessary to unfold the time ordering and the \dt resolution. This
is accomplished by using a signal probability-density-function for the four distributions of the form
\begin{eqnarray} 
{\cal H}_{\alpha,\beta}(\deltat) & = & g^+_{\alpha,\beta}(\dttrue)H(\dttrue) \otimes{\cal R} (\delta t; \sigma_{\dt}) + \nonumber \\
                              &  & g^-_{\alpha,\beta}(-\dttrue)H(-\dttrue) \otimes{\cal R}(\delta t; \sigma_{\dt}), \nonumber \\
\end{eqnarray} 
where 
\dttrue 
is the signed difference of proper times between the two \B decays in the 
limit of perfect \dt resolution,
$H$ is the Heaviside step function, ${\cal R}(\delta t;\sigma_{\dt})$ is the resolution 
function, with $\delta t = \dt-\dttrue$,
and $\sigma_{\dt}$ is the estimate of the \dt uncertainty obtained by the reconstruction algorithms. 
This unfolding procedure, which requires good \dt resolution and excellent \Refs{knowledge} of the resolution function, 
especially in the low $|\dt|$ region, is of critical importance to resolve the time ordering and represents
in practice the main experimental challenge of 
\Vera{this}
time-reversal
\Vera{analysis~\cite{Lees:2012kn}} 
in comparison 
to the associated \CP violation analysis~\cite{Aubert:2009aw}.

From the eight pairs of signal coefficients, reported in the left panel of Table~\ref{tab:results}, 
\Vera{one}
might construct two sets ($\pm$) 
of three pairs each of independent asymmetry parameters, $(\DeltaSpmT, \DeltaCpmT)$, $(\DeltaSpmCP, \DeltaCpmCP)$, 
and $(\DeltaSpmCPT, \DeltaCpmCPT)$, as illustrated in Fig.~\ref{fig:processes} and also shown in the right panel of Table~\ref{tab:results}. 
There are multiple choices to define the asymmetry parameters but for convenience 
\Vera{two sets} \Auth{(panels in the first and second columns, and in the third and fourth 
columns of Fig.~\ref{fig:processes})} of asymmetries related to 
each of the three discrete symmetries
\Vera{are chosen.} 
The reference transitions are taken by convention
$\Bplus\to\Bz$ and $\Bzb\to\Bminus$, corresponding to event classes $(\ccbar\KS,\ellp\X)$ and $(\ellp\X,\ccbar\KS)$, respectively
(see Fig.~\ref{fig:processes} and Table~\ref{tab:results}).
This choice has the advantage that the breaking of time-reversal symmetry would directly manifest itself through any nonzero value of 
\DeltaSpmT 
or any difference between \DeltaSpmCP and \DeltaSpmCPT.

  \begin{table*}[htb!]
  \renewcommand{\arraystretch}{1.4}
    \begin{center}
      \caption{\label{tab:results}\captionSize Measured values of the $S_{\alpha,\beta}^\pm$ and $C_{\alpha,\beta}^\pm$ coefficients,
and of the asymmetry parameters, defined as differences among coefficients for 
symmetry-transformed transitions as depicted in Fig.~\ref{fig:processes}~\cite{Lees:2012kn}.
The first uncertainty is statistical and the second systematic.
The lower indices \ellm, \ellp, \KS, and \KL stand for reconstructed decay modes that 
identify the \B meson state as \Bzb, \Bz and \Bminus, \Bplus, respectively,
and the upper indices encapsulate the time ordering, ($+$) when the decay to \ellm and \ellp occurs first and ($-$) otherwise.
The asymmetry parameters $\DeltaSpmT$, $\DeltaCpmT$ and the differences
$\DeltaSpmCP-\DeltaSpmCPT$, $\DeltaCpmCP-\DeltaCpmCPT$ are all motion-reversal violating.
The first column refers to the index labelling in Fig.~\ref{fig:processes}.}       
\begin{tabular}{ c c  c  c  c  c}	\hline \hline	
    \multicolumn{2}{c}{Transition}       & Coefficient & Result & Asymmetry parameter & Result \\ \hline
(b) & $\B_+\to\Bz  $ & \SmBzKs  & $-0.66\pm0.06\pm0.04$ & Reference \\ 
    &                & \CmBzKs  & $-0.05\pm0.06\pm0.03$ & Reference \\ [0.05in] 
(e) & $\Bz\to\B_+  $  & \SpBzbKl & $\phm0.51\pm0.17\pm0.11$ & $\DeltaSmT$   &  $\phm1.17\pm 0.18\pm 0.11$ \\ 
    &                 & \CpBzbKl & $-0.01\pm0.13\pm0.08$    & $\DeltaCmT$   &  $\phm0.04\pm 0.14\pm 0.08$ \\ [0.05in] 
(a) & $\Bzb\to\B_+ $  & \SpBzKl  & $-0.69\pm0.11\pm0.04$    & $\DeltaSmCPT$ &  $-0.03\pm 0.13\pm 0.06$ \\ 
    &                 & \CpBzKl  & $-0.02\pm0.11\pm0.08$    & $\DeltaCmCPT$ &  $\phm0.03\pm 0.12\pm 0.08$ \\ [0.05in] 
(f) & $\B_+\to\Bzb $ & \SmBzbKs & $\phm0.67\pm0.10\pm0.08$  & $\DeltaSmCP$  &  $\phm1.33\pm 0.12\pm 0.06$\\ 
    &                & \CmBzbKs & $\phm0.03\pm0.07\pm0.04$  & $\DeltaCmCP$  &  $\phm0.08\pm 0.10\pm 0.04$\\ [0.05in] 
\hline
(c) & $\Bzb\to\B_-$   & \SpBzKs  & $\phm0.55\pm0.09\pm0.06$ & Reference \\
    &                 & \CpBzKs  & $\phm0.01\pm0.07\pm0.05$ & Reference \\ [0.05in] 
(h) & $\B_-\to\Bzb$  & \SmBzbKl & $-0.83\pm0.11\pm0.06$    & $\DeltaSpT$ & $-1.37\pm 0.14\pm 0.06$ \\ 
    &                & \CmBzbKl & $\phm0.11\pm0.12\pm0.08$ & $\DeltaCpT$ & $\phm0.10\pm 0.14\pm 0.08$ \\ [0.05in] 
(d) & $\B_-\to\Bz $  & \SmBzKl  & $\phm0.70\pm0.19\pm0.12$ & $\DeltaSpCPT$ & $\phm0.16\pm 0.21\pm 0.09$ \\ 
    &                & \CmBzKl  & $\phm0.16\pm0.13\pm0.06$ & $\DeltaCpCPT$ & $\phm0.14\pm 0.15\pm 0.07$ \\ [0.05in] 
(g) & $\Bz\to\B_- $  & \SpBzbKs & $-0.76\pm0.06\pm0.04$    & $\DeltaSpCP$  & $-1.30\pm 0.11\pm 0.07$\\ 
    &                & \CpBzbKs & $\phm0.08\pm0.06\pm0.06$ & $\DeltaCpCP$  & $\phm0.07\pm 0.09\pm 0.03$\\ [0.05in] 
\hline \hline
      \end{tabular}
    \end{center}	
  \end{table*}

Systematic uncertainties in the measurement of the coefficients and asymmetry parameters in Table~\ref{tab:results}
are dominated by the knowledge of the \dt resolution and background composition of event classes containing $\jpsi\KL$ 
final states, and any possible deviation of the experimental procedure observed in detailed Monte Carlo simulations~\cite{TheBABAR:2013jta}. 
Besides, the experiment has performed a number of cross-checks, based on both simulated and data control samples, 
to assess the robustness of the results. Of special relevance is the check performed using 
event classes where the neutral \B meson reconstructed in the $\ccbar\KS$ or $\jpsi\KL$ final states is replaced 
by a charged \B decaying to $\ccbar\Kpm$ and $\jpsi\Kstarp$, respectively.
\Vera{It is found that}
all coefficients and parameters, \Vera{shown} in 
Table~\ref{tab:results}, are consistent with zero.

\subsection{Interpretation of results and significance}  
\label{sec:ANALYSIS-results}

The ability of 
\Vera{this}
analysis to describe the 
\Vera{data} 
\Vera{can be assessed by}
visualizing the 
rate differences between the transitions and their time-reversed conjugates in Table~\ref{TAB:TRV-EVTCLASSESvsTRANSITIONS}. 
The inequalities among 
\Auth{probabilities}
are better determined in the form of asymmetries along the lines of Eq.~(\ref{eq:ACPdef}).
For transition $\Bzb\to\B_-$ (first entry in Table~\ref{TAB:TRV-EVTCLASSESvsTRANSITIONS}),
\begin{eqnarray} 
A_T(\deltat) & = & \frac {{\cal H}^-_{\ellm,\KL}(\deltat)-{\cal H}^+_{\ellp,\KS}(\deltat)}
                         {{\cal H}^-_{\ellm,\KL}(\deltat)+{\cal H}^+_{\ellp,\KS}(\deltat)},
\end{eqnarray} 
where ${\cal H}^\pm_{\alpha,\beta}(\dt)={\cal H}_{\alpha,\beta}(\pm\dt) H(\dt)$. 
With this construction, $A_T(\dt)$ is defined only for positive \dt values. 
Neglecting reconstruction effects, 
\begin{eqnarray} 
A_{\T}(t) & \approx & \frac{\DeltaSpT}{2} \sin(\dmd t) + \frac{\DeltaCpT}{2} \cos(\dmd t).
\end{eqnarray} 
The three other asymmetries, corresponding to the last three entries in Table~\ref{TAB:TRV-EVTCLASSESvsTRANSITIONS},
are constructed analogously and have the same time dependence, with $\DeltaSpT$ replaced by \DeltaSmT, $\DeltaSmCP-\DeltaSmCPT$,
and $\DeltaSpCP-\DeltaSpCPT$, respectively, and equally for $\DeltaCpT$.

\begin{figure}[htb!]
\begin{center}
\begin{tabular}{c}
 \includegraphics[width=0.4\textwidth]{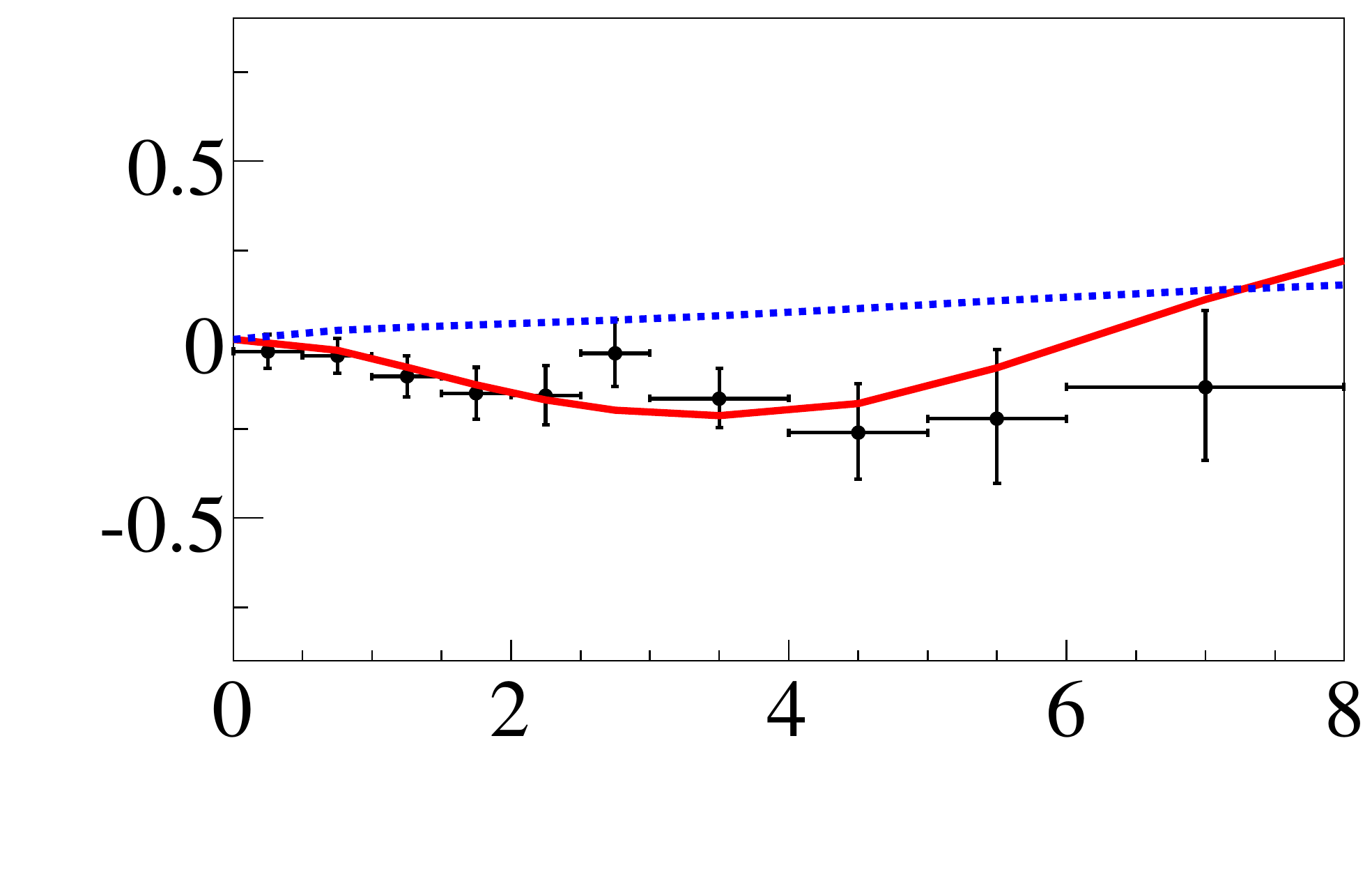} \put(-115,113){\Large $\Bzb\rightarrow\Bminus$} \\
 \includegraphics[width=0.4\textwidth]{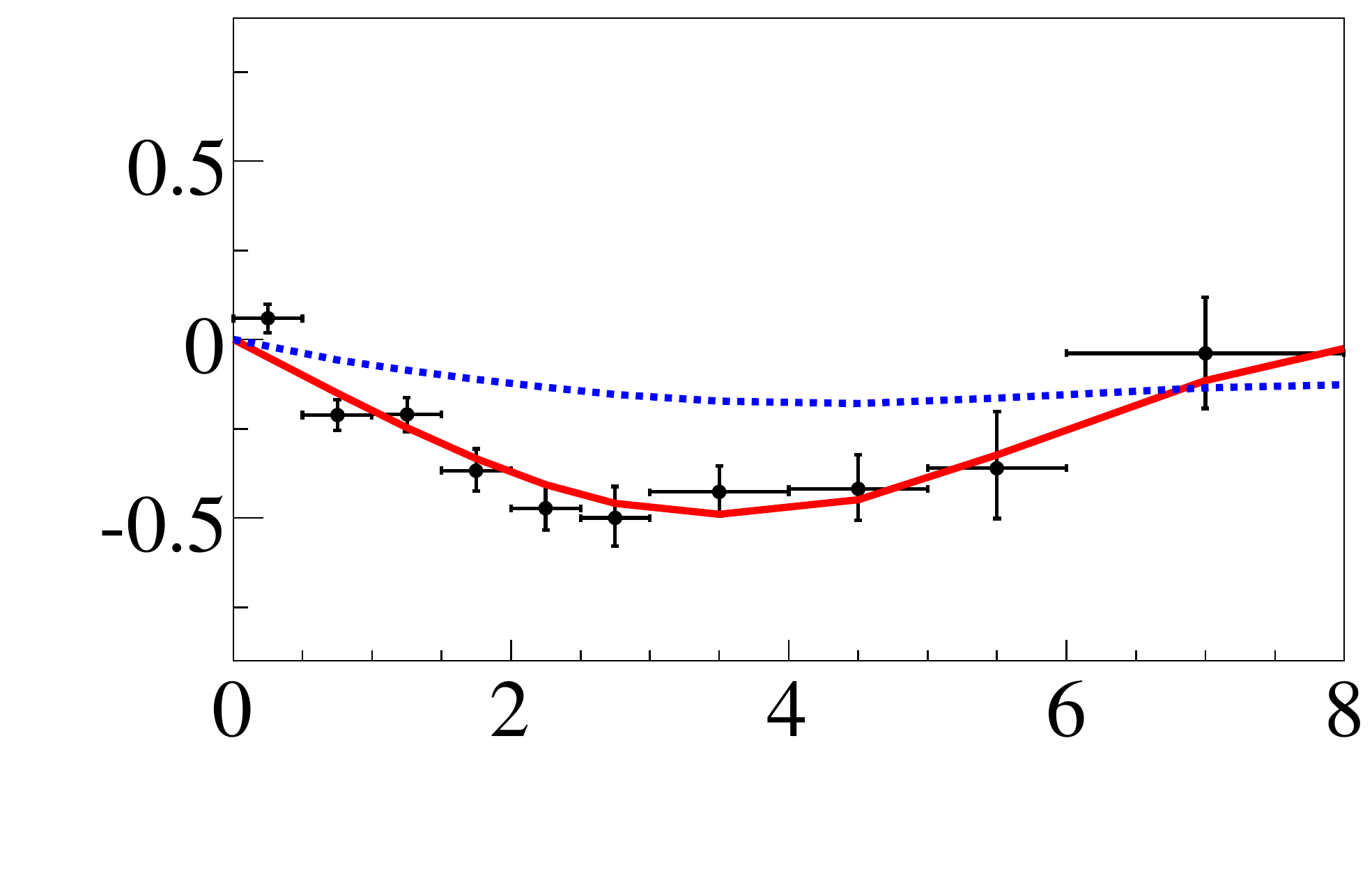} \put(-115,113){\Large $\Bminus\rightarrow\Bz$}  \\
 \includegraphics[width=0.4\textwidth]{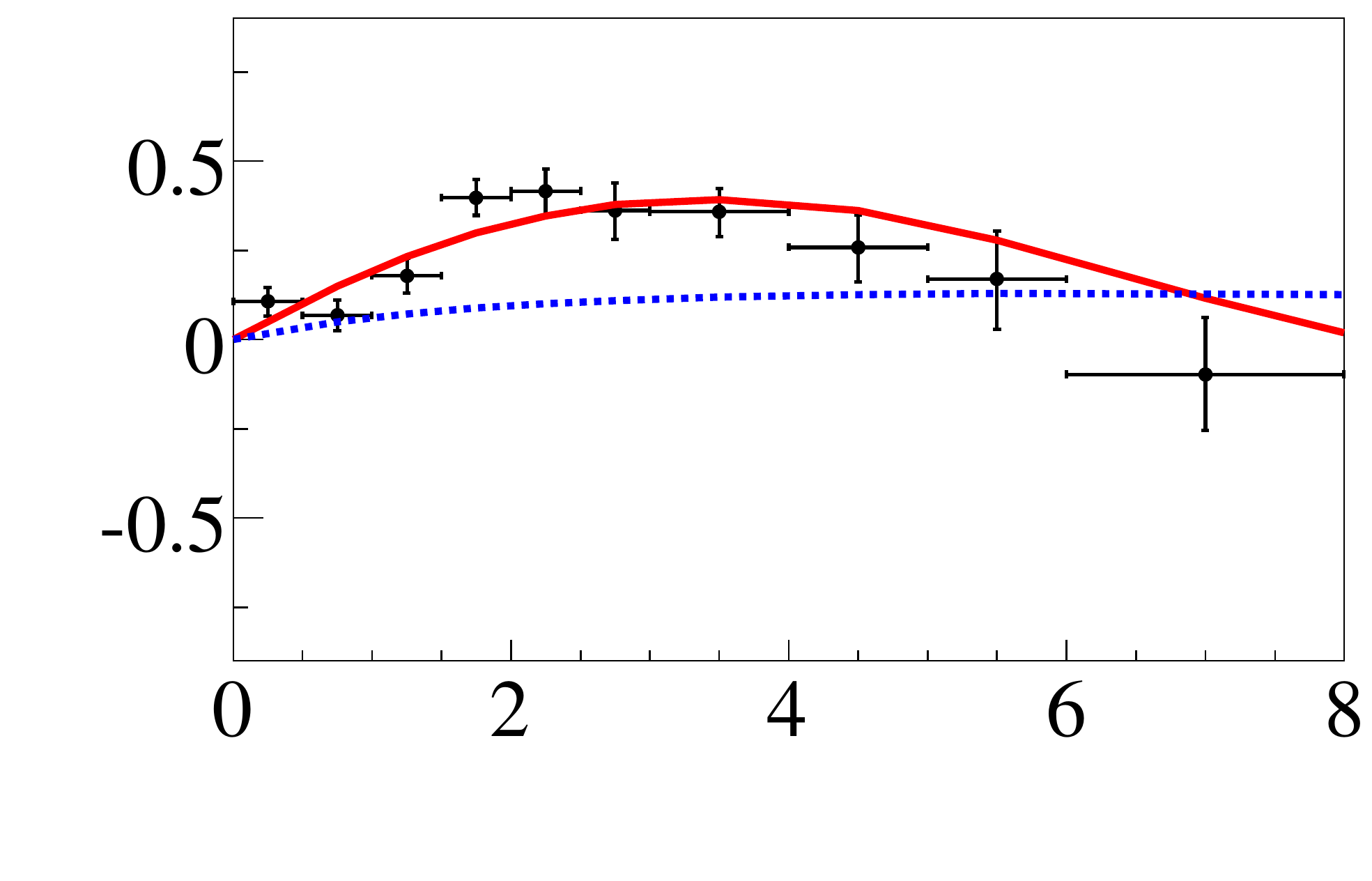} \put(-115,113){\Large $\Bzb\rightarrow\Bplus$} \\
 \includegraphics[width=0.4\textwidth]{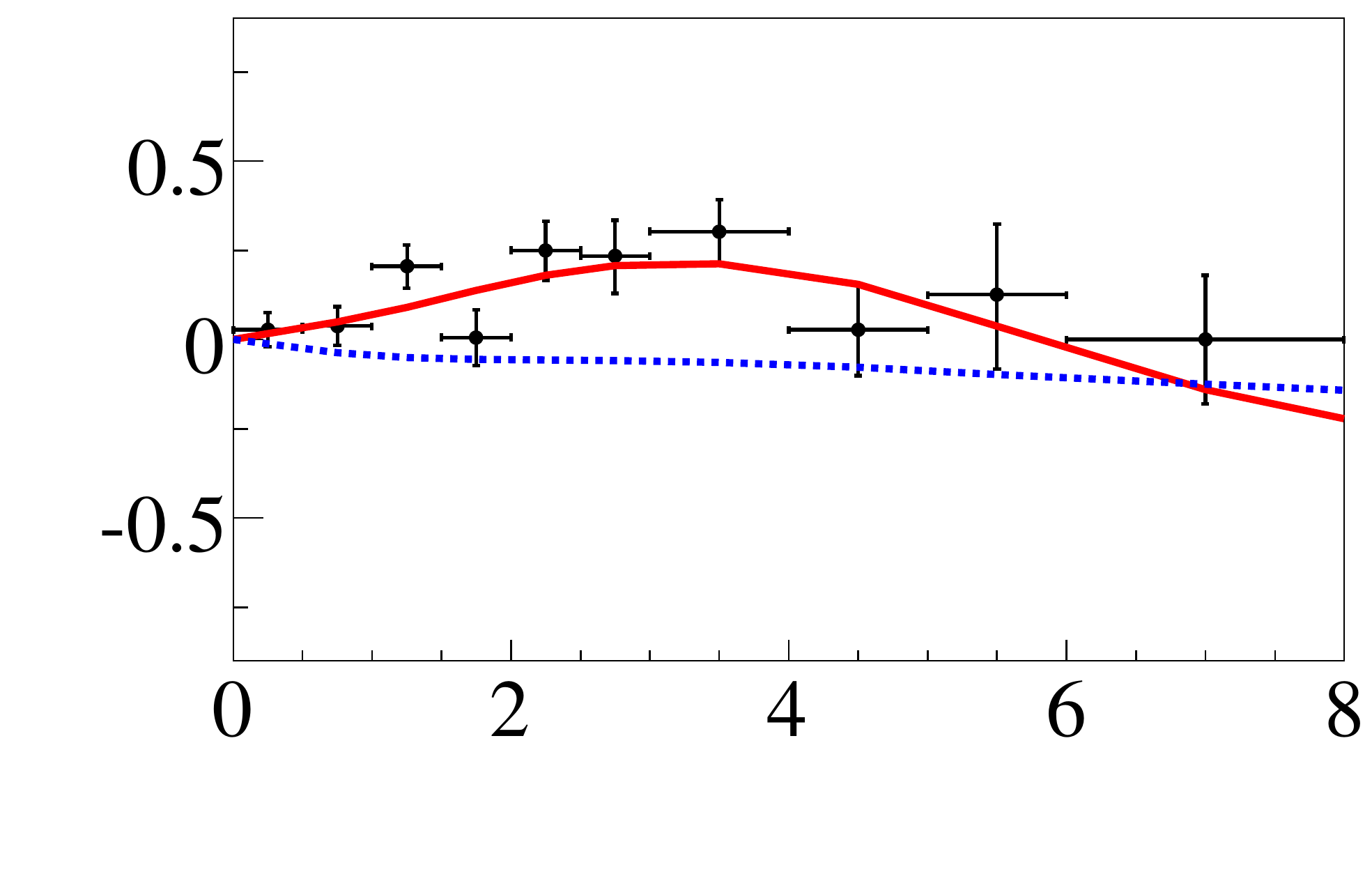} \put(-115,113){\Large $\Bplus\rightarrow\Bz$} 
     \put(-110,0){\Large \dt (ps)}
     \put(-200,80){ \begin{rotate}{90}{\Large \hskip5cm ASYMMETRY $A_T$}\end{rotate} } \\
\end{tabular}
\caption{\label{fig:AT}\captionSize The four independent asymmetries measured by the \babar\ experiment~\cite{Lees:2012kn}
between transitions (from top to bottom)
$\Bzb\rightarrow\Bminus$,
$\Bminus\rightarrow\Bz$,
$\Bzb\rightarrow\Bplus$,
$\Bplus\rightarrow\Bz$,
and their time-reversed versions. 
The points with error bars represent the data, the red solid and dashed blue curves represent the 
projections of the best fit results with and without time-reversal violation, respectively.
}
\end{center}
\end{figure}

Figure~\ref{fig:AT} shows the four time-reversal asymmetries constructed in this way.
The data 
\Refs{are}
well described by the red solid curves, which represent the projection of the best
fit to the eight $(S^\pm_{\alpha,\beta},C^\pm_{\alpha,\beta})$ pairs, as reported in Table~\ref{tab:results}.
These curves deviate significantly from the dashed blue curves, which represent the fit projection 
\Vera{for}
time-reversal \Vera{invariance,} 
i.e. $\DeltaSpmT=0$, $\DeltaCpmT=0$, $\DeltaSpmCP=\DeltaSpmCPT$, and 
$\DeltaCpmCP=\DeltaCpmCPT$. The fact that the dashed blue curves are not identically zero is a consequence of 
experimental effects, in particular the asymmetry with respect to $\delta t=0$ of the time resolution function.

All eight $C^\pm_{\alpha,\beta}$ in Table~\ref{tab:results} are compatible with zero, therefore the $|\Abar/\A|=1$ condition
discussed in Sec.~\ref{sec:CONCEPT-entangled} is validated within errors and the association between event classes and 
transitions in Table~\ref{TAB:TRV-EVTCLASSESvsTRANSITIONS} is confirmed.
It 
\Vera{follows} that the observation $\DeltaSpmT \ne 0$ and $\DeltaSpmCP \ne \DeltaSpmCPT$ in Table~\ref{tab:results} is an
unambiguous, direct detection of time-reversal violation in the time evolution of neutral \B mesons, obtained through 
\Vera{motion} reversal in transitions that are not \CP conjugate to each other. The violation of time-reversal symmetry is also clearly seen 
through the large differences between the red solid and dashed blue curves in all four asymmetries 
shown in Fig.~\ref{fig:AT}.

\begin{figure}[htb!]
\begin{center}
 \includegraphics[width=0.45\textwidth]{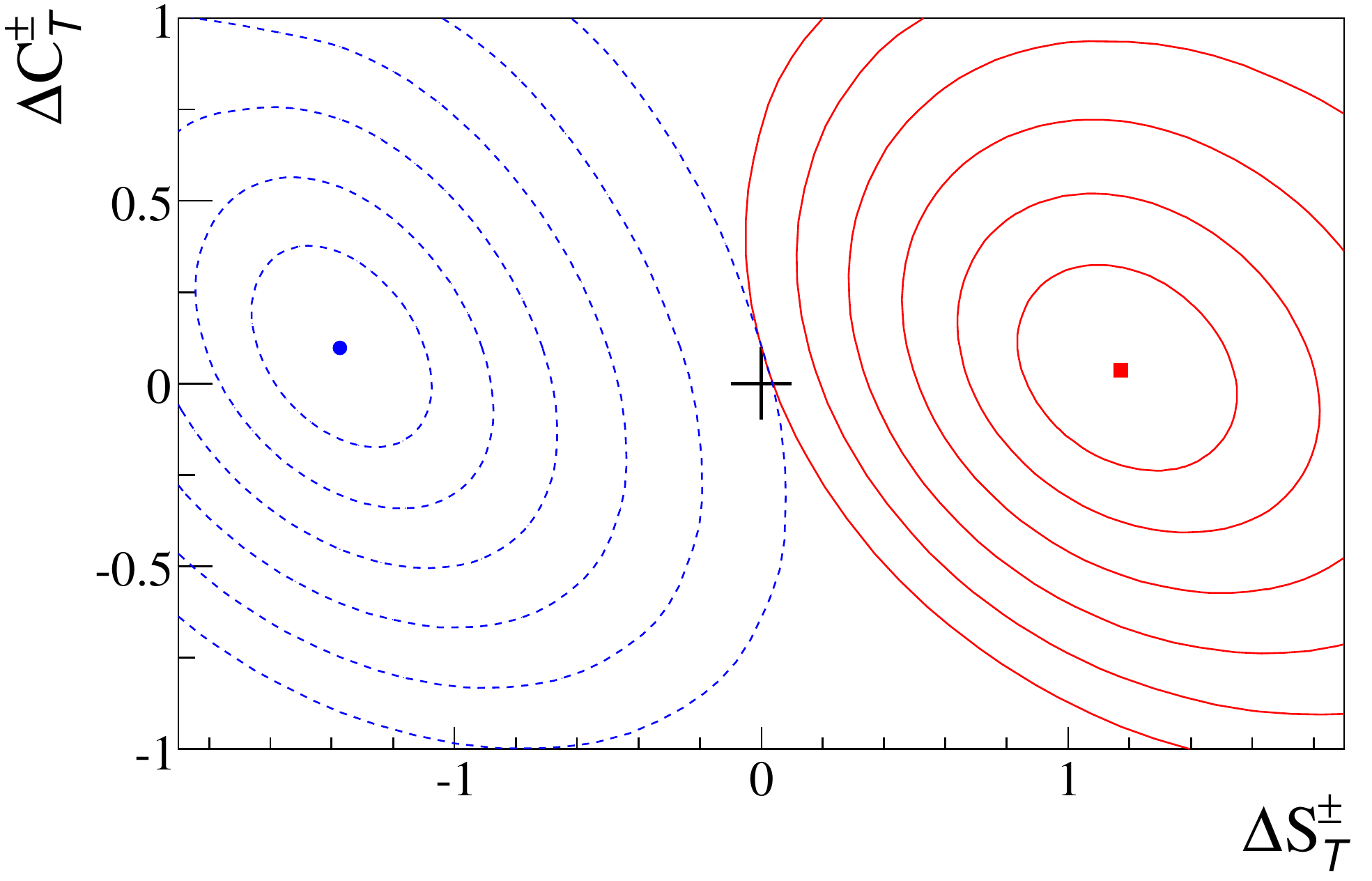}
\caption{\label{fig:scanT}\captionSize 
Central values (blue point and red square) and two-dimensional $p$-value contours
(for $p=0.317$, $4.55\times10^{-2}$, $2.7\times10^{-3}$, $6.3\times10^{-5}$, $5.7\times10^{-7}$, and $2.0\times10^{-9}$)
for the \T-asymmetry parameters 
$(\DeltaSpT,\DeltaCpT)$ (blue dashed curves) and $(\DeltaSmT,\DeltaCmT)$ (red solid curves),
as reported by \babar~\cite{Lees:2012kn}.
The \T-invariance point is shown as a $+$ sign. 
}
\end{center}
\end{figure}

The significance of the observed time-reversal violation is evaluated on the basis of changes in
log-likelihood value $\ln {\cal L}$ with respect to the maximum (\twoDLL).
The difference \twoDLL between the fit without \T violation and the best fit is 226,
including systematic uncertainties. Assuming Gaussian statistics, this corresponds 
to a significance of $14\sigma$, evaluated from the upper integral at \twoDLL of the $\chi^2$ probability 
distribution for 8 degrees of freedom~\footnote{Eight is the difference in the number of fit constraints with and 
without \T violation.} ($p$-value)~\cite{Beringer:2012}.
Figure~\ref{fig:scanT} shows $p$-value contours calculated from the change \twoDLL in two dimensions for the 
\T-asymmetry parameters $(\DeltaSpT,\DeltaCpT)$ and $(\DeltaSmT,\DeltaCmT)$. 
In the two cases cases we can observe that the \T invariance point is excluded at $6\sigma$ level.
The difference \twoDLL for fits assuming \CPT or \CP symmetry are 5 and 307, corresponding to
$0.3\sigma$ and $17\sigma$, consistent with \CPT invariance and \CP violation, respectively. 
These values, combined with those in Table~\ref{tab:results}, are compatible with \CP violation 
as due to time-reversal violation and \CPT invariance.
The 
\Vera{larger} 
significance of \CP violation is because the comparison of 
\Auth{probabilities}
for event classes 
e.g. $(\ellm\Xbar,\ccbar\KS)$ and $(\ellp\X,\ccbar\KS)$ has a higher statistical and systematic 
significance than the comparison of e.g. $(\ellm\Xbar,\ccbar\KS)$ and $(\ellm\Xbar,\ccbar\KL)$.

Following the discussion in Sec.~\ref{sec:CONCEPT-entangled} and assuming \CP conservation in mixing 
(i.e \CP with \T invariance, and \CP with \CPT symmetry in mixing), all eight pairs of coefficients $S^\pm_{\alpha,\beta}$ and $C^\pm_{\alpha,\beta}$
are related,
\bea
S & = & \phantom{-}S^+_{\ellp,\KS} =           -S^+_{\ellm,\KS} =           -S^-_{\ellp,\KS} = \phantom{-}S^-_{\ellm,\KS} = \nn \\
  &   &           -S^+_{\ellp,\KL} = \phantom{-}S^+_{\ellm,\KL} = \phantom{-}S^-_{\ellp,\KL} =           -S^-_{\ellm,\KL} , \nn \\
C & = & \phantom{-}C^+_{\ellp,\KS} =           -C^+_{\ellm,\KS} = \phantom{-}C^-_{\ellp,\KS} =           -C^-_{\ellm,\KS} = \nn \\
  &   & \phantom{-}C^+_{\ellp,\KL} =           -C^+_{\ellm,\KL} = \phantom{-}C^-_{\ellp,\KL} =           -C^-_{\ellm,\KL}.\ \ \ \ \ \ \
\eea
In the SM, the eight $S^\pm_{\alpha,\beta}$ coefficients are measurements of $S=\sin 2\beta \approx 0.7$. The results 
in Table~\ref{tab:results} lead to a mean value 
$S=0.686\pm0.029$,
which is consistent with the value obtained 
from the \CP violation investigation based on the same data~\cite{Aubert:2009aw}. Analogously, the measurement of the eight 
$C^\pm_{\alpha,\beta}$ coefficients result in a mean value of $0.022\pm0.021$, consistent with both the previous \CP analysis and zero.

\section{Conclusion}  
\label{sec:CONCLUSION}

The arrow of time in 
systems with large number of degrees of freedom is a thermodynamic property 
of entropy associated to the irreversibility of boundary conditions. 
However, 
\Refs{time's arrow} is not related to the question of time-reversal symmetry in the fundamental laws of physics.
Only two physical systems in nature, the unstable \K and \B mesons, have a relatively large expected breaking of time-reversal 
symmetry; in these systems \CP violation has been observed
and no experimental data contradicts the \CPT theorem. Therefore, \K and \B mesons are a 
\Vera{best}
choice for an experiment 
detecting directly time-reversal non-invariance. 

The main principle for a direct detection of time-reversal violation in transitions is the exchange of initial and \Vera{final states.}
A unique opportunity arises from the quantum-mechanical properties imposed by the EPR entanglement between the two 
neutral \B mesons produced in the \FourS resonance decay at \B factories. 
\Vera{The observation} of the first \B decaying into the flavor eigenstates $\ellp\X$ or $\ellm\Xbar$, 
or the \CP eigenstates $\ccbar\KS$ or $\jpsi\KL$, \Vera{prepares (tags)} the initial state of the other \B as \Bzb, \Bz, \Bplus, or \Bminus, respectively. 
The initial states tagged by entanglement are filtered at a later time through its decay into a \CP- or a flavor-eigenstate decay mode.
The four \B meson states appearing as initial and final states make possible to build eight different transition 
\Auth{probabilities}
for the 
time evolution of the neutral \B meson.
In appropriate combinations, time-reversal, \CP, and \CPT symmetries can be analyzed separately through four time-dependent
asymmetries, which can be expressed in terms of certain asymmetry parameters.

The \babar\ experiment has measured the four time-reversal asymmetries and extracted the corresponding asymmetry parameters.
The results show a 
\Vera{highly significant}
departure from motion reversal symmetry.
A precise exchange of initial and final states is needed to interpret the results 
as direct detection of time-reversal non-invariance. This requires the absence of both wrong-strangeness 
($\Bz\nrightarrow\ccbar\Kzb$ and $\Bzb\nrightarrow\ccbar\Kz$) and 
wrong-sign ($\Bz\nrightarrow\ellm\Xbar$ and $\Bzb\nrightarrow\ellp\X$) \B decays, \CP invariance 
in \Kz-\Kzb mixing, and $|\Abar/\A|=1$ (a single decay amplitude and \CPT symmetry in the $\Bz\to\ccbar\Kz$ decay amplitude).
All these effects are small and have been either accounted for in the systematic uncertainties (the second),
directly demonstrated in the experimental analysis and incorporated in the uncertainties (the fourth), 
or neglected since their impact is well below the statistical sensitivity [${\cal O}(10\%)$] according to measurements 
[the first, ${\cal O}(0.1\%)$] or SM expectations [the third, ${\cal O}(1\%)$].

\Vera{Time-reversal} and \CP symmetry breakings are seen in two 
separate observations (the states involved in the transitions are not \CP conjugate to each other)
and the asymmetries are time dependent with only a $\sin(\dmd\dt)$ term, of order ${\cal O}(10^{-1})$, and are induced by the interference
of decay amplitudes with and without mixing. The corresponding measurement of the weak phase from time-reversal asymmetries
match those from \CP asymmetries, therefore the observed \T and \CP violations balance to each other, supporting \CPT invariance 
in the time evolution of \B mesons.
This is in contrast to the flavor-mixing asymmetry in \Kz-\Kzb transitions measured
by CPLEAR, where \CP and \T transformations 
are identical and the asymmetry is time independent, 
of order ${\cal O}(10^{-3})$, and is produced by the interference between the dispersive and absorptive contributions to \Kz-\Kzb mixing.

The concept of direct detection of time-reversal violation in transitions might be 
\Vera{extended to include}
systematic tests 
using pairs of \B and \D mesons created 
in the decay of the \FourS and $\psi(3770)$ resonances~\cite{Bevan:2013rpr},
as well as pairs of \K mesons from the $\phi(1020)$~\cite{Bernabeu:2012nu}.
In the latter \Vera{case,} there are important differences triggered by a non-vanishing decay width difference, 
the non-orthogonality of the \KL and \KS states,
and the small effects expected within the SM.
This opens the possibility to embark upon a complete time-reversal (and \CPT) violation program in weak
interactions to probe possible new physics contributions in tree and loop decays at future high-luminosity 
flavor factories, Belle II at SuperKEKB~\cite{Abe:2010gxa,Aushev:2010bq} and KLOE-2 at DA$\Phi$NE~\cite{AmelinoCamelia:2010me}.

The main limitation of the method is associated with the identification of the appropriate decay 
channels used to filter the states of the time-reversed transition. \Vera{Specifically,} 
\Refs{it is necessary to identify}
pairs of decay channels that project into meson states orthogonal to each other
This orthogonality condition is satisfied by conjugate flavor eigenstate decay 
channels ($\ellm\Xbar$,$\ellp\X$) and by \CP eigenstates of opposite \CP parity with the 
same flavor content ($\ccbar\KS$,$\jpsi\KL$). 
The fact that the \B decay amplitudes to $\ccbar\KS$ and $\jpsi\KL$ 
are given by the same diagram followed by \Kz-\Kzb mixing (needed to make possible the interference)
is essential for the definition of the states \Bminus and \Bplus.
The precise implementation of these restrictions imposes the requirements summarized above.
An alternative time-reversal asymmetry based on EPR entanglement open to any pair of decay channels has been recently 
\Vera{suggested~\cite{Bernabeu:2013qea}, though}
the connection between the experiment and the time-reversal observables 
requires some theoretical input.

\begin{acknowledgments}
\Auth{We thank V. Luth and F. J. Botella for carefully reading the manuscript and providing helpful comments,
and A. J. Bevan, A. Di Domenico, N. E. Mavromatos, T. Nakada, V. Rubakov, K. R. Schubert and P. Villanueva-P\'erez
for enlightening discussions on the subject.}
We are grateful for the support from MINECO No. FPA2010-21549-C04, FPA2011-23596
and Generalitat Valenciana No. PROMETEO 2010/056, 2013/017, GVISIC 2012/020 (Spain).

\end{acknowledgments}

\bibliography{rmp-trv}

\end{document}